%% file: main.tex
\documentclass[10pt]{article}
\usepackage{graphicx}
\usepackage{epstopdf, epsfig}

\usepackage[utf8]{inputenc} 
\usepackage[T1]{fontenc}    
\usepackage[dvipsnames]{xcolor}
\usepackage{amsmath,amsbsy,amssymb}
\usepackage{authblk}
\usepackage[numbers]{natbib}

\newcommand\Rey{\mbox{\textit{Re}}}  

\usepackage[]{geometry}
\begin{document}

\title{Direct numerical simulation of cambered airfoil aerodynamics at Re\,=\,20,000}

\author[1]{Bjoern F. Klose}
\author[2]{Geoffrey R. Spedding}
\author[1]{Gustaaf B. Jacobs%
 \thanks{Corresponding author. Email: gjacobs@sdsu.edu}}

\affil[1]{Department of Aerospace Engineering, San Diego State University, San Diego, CA 92182, USA}
\affil[2]{Department of Aerospace and Mechanical Engineering, University of Southern California, Los Angeles, CA 90007, USA}

\maketitle

\begin{abstract}
A comprehensive and detailed overview of the flow topology over a cambered NACA 65(1)-412 airfoil at \Rey{}\,=\,$2\times10^4$ is presented for angles of attack ranging from 0$^\circ$ to 10$^\circ$ using high-order direct numerical simulations.
It is shown that instabilities bifurcate the flow and cause it to change at a critical angle of attack from laminar separation without reattachment over a laminar separation bubble at the trailing edge to a bubble at the leading edge.
The transition of the flow regimes is governed by the Karman vortex shedding of the pressure side boundary layer at the trailing edge, Kelvin-Helmholtz instabilities within the separated shear layer on the suction side, as well as three-dimensional instabilities of elliptic flow within the vortex cores and hyperbolic flow in the shear layer between subsequent Karman vortices.
As the suction side shear layer transitions and reattaches, the interaction of the two and three-dimensional instabilities results in three-dimensional tubular structures and large-scale turbulent puffs.
The formation and shifting of the laminar separation bubble defines the far-wake topology several chord-lengths behind the airfoil and is accompanied by a sudden increase of the lift force and decrease in the drag that underscores the sensitive nature of low-Reynolds number airfoil aerodynamics.
Lift and drag polars are presented for direct numerical simulations, wind tunnel experiments, and simplified numerical procedures where incorrect prediction of the force coefficients is caused by the failure to correctly model the low-pressure region at the trailing edge that is caused by the time-dependent generation of the Karman vortices.
\end{abstract}

\section{Introduction} \label{introduction}
\input{content/introduction}

\section{Computational formulation} \label{gov_eqn}
\input{content/gov_eqn}

\section{Setup} \label{setup}
\input{content/setup}

\section{Results and Discussion} \label{results}
\input{content/results3D}

\section{Concluding Remarks} \label{conclusion}
\input{content/conclusion}

\medskip
\noindent
Declaration of Interests. The authors report no conflict of interest.

\section*{Acknowledgments}
We gratefully acknowledge funding by the Air Force Office of Scientific Research under FA9550-16-1-0392 of the Flow Control Program and from Solar Turbines. 
The authors thank the Department of Defense for computational time on the DoD HPC.
\appendix
\section{Parameter study: 2D simulations} \label{appendix_2D}
\input{content/results2D}

\clearpage
\bibliographystyle{unsrtnat}
\bibliography{main}

\end{document}

%% file: content/introduction.tex
Most practical aerodynamics occurs at chordwise Reynolds numbers (\Rey{}\,=\,$Uc/\nu$, where $U$ is a flight speed, $c$ is a chord length and $\nu$ is the kinematic viscosity) of $10^6$ or more, and classical inviscid theories and numerical procedures yield satisfactory estimates of lift.  Drag is more difficult to account for but there are a number of standard methods for making reasonable estimates, even on complex geometries (e.g. \citep{anderson:10, destarac:04}).  However, there is a growing number of cases (turbine blade elements at altitude, wind turbines, small-scale autonomous aircraft) where Re is not necessarily large.  In particular there is a range $10^4$\,$\leq$\,\Rey{}\,$\leq$\,$10^5$ where Re is small enough so that the viscous boundary layer almost never remains attached, and large enough so the resulting shear layer can readily destabilise and generate complex flows, even leading to turbulence.  In that case, as long ago noted by \citet{Lissaman83}, airfoil and wing performance is almost entirely dictated by the propensity for separation of the initially laminar boundary layer.

\subsection{Separation, reattachment and instabilities}

Because airfoils have a suction and a pressure side, the boundary layers on the upper and lower surfaces are governed by different forces and dynamics.  In the Re regime described above (which we denote as moderate Reynolds number), the boundary layer on the pressure side of the airfoil commonly remains laminar and does not separate upstream of the trailing edge.  On the suction side, the loss of momentum through viscous effects and an adverse pressure gradient downstream of the suction peak forces the laminar boundary layer to separate from the airfoil.  Flow reattachment then occurs if the separated shear layer transitions to turbulence far enough upstream from the trailing edge so the momentum from the outer flow can be transported to the surface, re-energising the now-turbulent boundary layer which can more robustly follow the surface contour.  The region of recirculating flow between the separation and the reattachment points is called a laminar separation bubble (LSB) and has been a subject of research for many years in experiment, theory and computation (e.g. \citet{Horton68, stewartson:70, smith:79, AS00, scheichl:08, JSS08, burgmann:08a, burgmann:08b}).

The term laminar separation bubble is somewhat misleading because the reattachment that makes the bubble possible can appear well-defined in a time-averaged sense, but the physical mechanisms are complex, three-dimensional and highly unsteady.  The relevant instabilities are both two- and three-dimensional.  Two dimensional instabilities include the shedding of vortices behind the airfoil, similar to those seen in bluff body wakes \citep{WS86, Williamson96}, together with Kelvin-Helmholtz (K-H) instabilities and the formation of vortices within the separated shear layer itself.  The two-dimensional vortices themselves then are subject to three-dimensional modes related to elliptic instabilities that lead to deformation of the vortex core \citep{Williamson96a, Kerswell02}, or Crow instabilities in counter-rotating vortex pairs, which result in the local, long-wave displacement of the vortex wake \citep{Crow70, LLW16}. Three-dimensional flow is induced within the shear layer between successive vortices, where hyperbolic streamlines drive a hyperbolic instability that results in the formation of the braid loops \citep{JSS08, MLR13}.  Once a three-dimensional instability is established, the process of loop generation is self-sustaining through induction from the previous vortex, as described by \citet{Williamson96a}.

\citet{JSS08} investigated the extent to which a combination of elliptic and hyperbolic instabilities can drive self-sustained turbulence in the LSB on a NACA 0012 at \Rey{}\,=\,$5\times 10^4$.  Inside an LSB, the mean flow at the surface is reversed, leading upstream, and the magnitude of the reverse flow can be used as an indicator of stability in the bubble.  Although the study found reverse flow levels of 15.3\%, thought to be at the lower limit of 15\% -- 20\% required for an absolute instability in a LSB \citep{AS00}, linear stability analysis on the time-averaged flow profile did not yield an absolute instability, and three-dimensional modes cannot be overlooked.  The possible significance of three-dimensional instability modes agrees with the observation by \citet{Theofilis11}, who reports that reverse flow levels of $\mathcal{O}$(10\%) are sufficient to sustain it.  \citet{MLR13} elaborate on the instabilities in LSBs and argue that the elliptic and hyperbolic instabilities may both occur at fundamental and subharmonic frequencies of the vortex shedding and that the simultaneous occurrence of several instabilities results in the rapid disintegration of the spanwise vortices.  The interaction of Karman and LSB vortices in the wake of the SD 7003 airfoil was investigated by \citet{DLR16}, who showed that the frequencies of the Karman vortices and the shedding from a suction-side LSB are locked in and the LSB shedding occurs at a subharmonic frequency.

The comparative fragility of the laminar boundary layer (which can also support two-dimensional Tollmien-Schlichting waves) and the numerous two- and three-dimensional instability modes, and their interactions, leads to significant challenges in flow measurement, prediction and control, with great sensitivity to both environmental and surface geometry details.

This highly sensitive nature of the transitional flows in and around the LSB is first associated with the receptivity of the separated shear layer in an adverse pressure gradient to instabilities, external disturbances, and feedback mechanisms \citep{JSS08, JSS10}.  Though they present challenges, the exquisite sensitivities can also be an opportunity in flow control strategies that deliberately exploit them, for example through synthetic jets \citep{glezer02, suzuki04, visbal11, BKJH20} or through acoustic excitation \citep{yang:13a, yang:14}.  Because such devices can be coupled with the inherent instabilities of the base flow to achieve global modifications of the flow structure \citep{glezer02}, even with low amplitude input energies, an understanding of the naturally occurring instabilities behind the flow transition is of practical importance.
For a more comprehensive review on control of low-Reynolds number flows, we refer the reader to the review by \citet{CS11}.

\subsection{Challenges in experiment and computation}

The rich set of dynamics available within one chord length of airfoils and wings at moderate Re has resulted \emph{inter alia} in a broad range of conflicting results for ostensibly similar conditions.  In \citep{selig:95} the discrepancies that emerged for \Rey{}\,$\leq$\,$10^5$ between lift-drag polars of the Eppler 387 airfoil measured at different facilities was a clear sign that not all aspects of the flows were equivalent, and \citet{TSS17} showed that similar discrepancies could be found in the existing literature for the NACA 0012 airfoil.  The sensitive dynamics of the LSB and its evolution was at the heart of some (but not all) of these differences, and it was clear that a repeatable result could come only from more precise conditions.  The extreme sensitivity to LSB dynamics was not restricted to a small class of specialised airfoils and will be found over most airfoil geometries with greater than 10\% thickness at these \Rey{}.  In comparing experiment with available computational results, it was also clear that the extreme computational effort required to capture the flow dynamics was not widely available, and a parametric study, for example over a range of angles of attack, $\alpha$, was impractical.

The existing high-fidelity simulations of airfoil flows have been primarily carried out on the canonical, symmetric NACA 0012 or on the thin, cambered SD 7003, which is a popular profile for low-Reynolds number operations as it allows for a thin and stable LSB over a range of $\alpha$ \citep{SD89}.  Even when direct numerical simulations (DNS) with no modeling coefficients are feasible, there remain issues concerning the two- vs. three-dimensional domain, the time resolution and length of simulation, and then extrapolation from limited and specific $\alpha$.  

\citet{SJL05} computed the flow about a NACA 0012 for $\alpha$\,=\,$4^\circ$ and \Rey{}\,=\,$10^5$ using a finite difference scheme that was second order accurate in time, and sixth order in space.  The spanwise extent of the computational domain, $b$, was $0.1c$, and Mach number, $M$\,=\,0.2.  Three-dimensional modes were found to grow rapidly, originating from the trailing edge but affecting the otherwise two-dimensional KH modes near separation.  The upstream propagation of the downstream modes was conjectured to be possible through pressure waves.
\citet{JSS08} performed a DNS (fourth order in space and time) on the NACA 0012 for $\alpha$\,=\,$5^\circ$ and \Rey{}\,=\,$5 \times 10^5$ (with $b$\,=\,$0.2c$, $M$\,=\,0.4), using a volume forcing to promote transition to turbulence, which was then found to be self-sustaining.  The time-averaged flow field was not absolutely unstable but amplification of three-dimensional modes in the large-scale two-dimensional shed vortices could be convected upstream, depending on the magnitude of the upstream flow.
\citet{JSS10} made similar computations at $\alpha$\,=\,$0^\circ$ and \Rey{}\,=\,$10^4$, $M$\,=\,0.2 in addition to those at \Rey{}\,=\,$5 \times 10^4$,  $\alpha$\,=\,$5^\circ$, $M$\,=\,0.4, and found that global instability could be maintained through an acoustic feedback loop emanating from trailing edge tonal noise.  The trailing edge characteristics could thus determine the frequency selection of upstream instabilities through an acoustic feedback loop.
\citet{JS11} performed two-dimensional simulations for \Rey{}\,=\,$10^5$ and $\alpha$\,=\,[0, 0.5, 1, 2$^\circ$], $M$\,=\,0.4 to investigate the possible effect of tonal noise originating at the trailing edge whose frequency selection is most sharp at low $\alpha$. Though the resulting acoustic forcing frequencies were significantly lower than the estimated most unstable boundary-layer modes. the tonal forcing could sustain and drive upstream instabilities. 

Given the cost of true DNS, there have been a number of studies using some flow modeling strategy.  Large-eddy simulations (LES) for the same airfoil were presented by \citet{AJS10}, who determined the spanwise domain size required to resolve LSB bursting and by \citet{LNOF15}, who compared the results of simulations based on different numerical schemes and experiments.  The effect of varying LES modelling approaches has also been investigated by \citet{cadieux:16}, showing that a truncated Navier-Stokes solution with periodic filtering could succeed in greatly decreasing computational cost with reasonable solution accuracy for an LSB over a flat plate. \citet{GV10, visbal11, uranga11, BBFFGHM14} have all presented simulation results of the low-Reynolds number flow over a SD 7003 through implicit LES.  \citet{DLR16} have investigated the flow transition of the same geometry at several \Rey{} through DNS and reported a partial lock-in of the shedding from the LSB and the Karman vortices. 
The studies show how location and size of the LSB changes with \Rey{} and $\alpha$ through various flow instabilities that result in considerable changes of the integrated forces over the airfoil.

Lower fidelity methods, such as Reynolds-Averaged Navier-Stokes (RANS) or integral boundary layer methods (e.g. \textit{XFoil} \citep{xfoil}), often fail to accurately predict the transition and the aerodynamic forces \citep{uranga11}, because the solution is neither time nor space resolved and depends on the choice of turbulent closure models \citep{Durbin18}.  Large-eddy simulations \citep{RM84,sengupta2007large} or detached-eddy simulations (DES) \citep{Spalart09} capture the larger structures of the unsteady flow, but similarly employ user-defined models to account for the the sub-grid scale fluid motions.

\subsection{Airfoil shapes and applications}
\begin{figure}
    \makebox[\textwidth][c]{
    \includegraphics[width=0.312\textwidth]{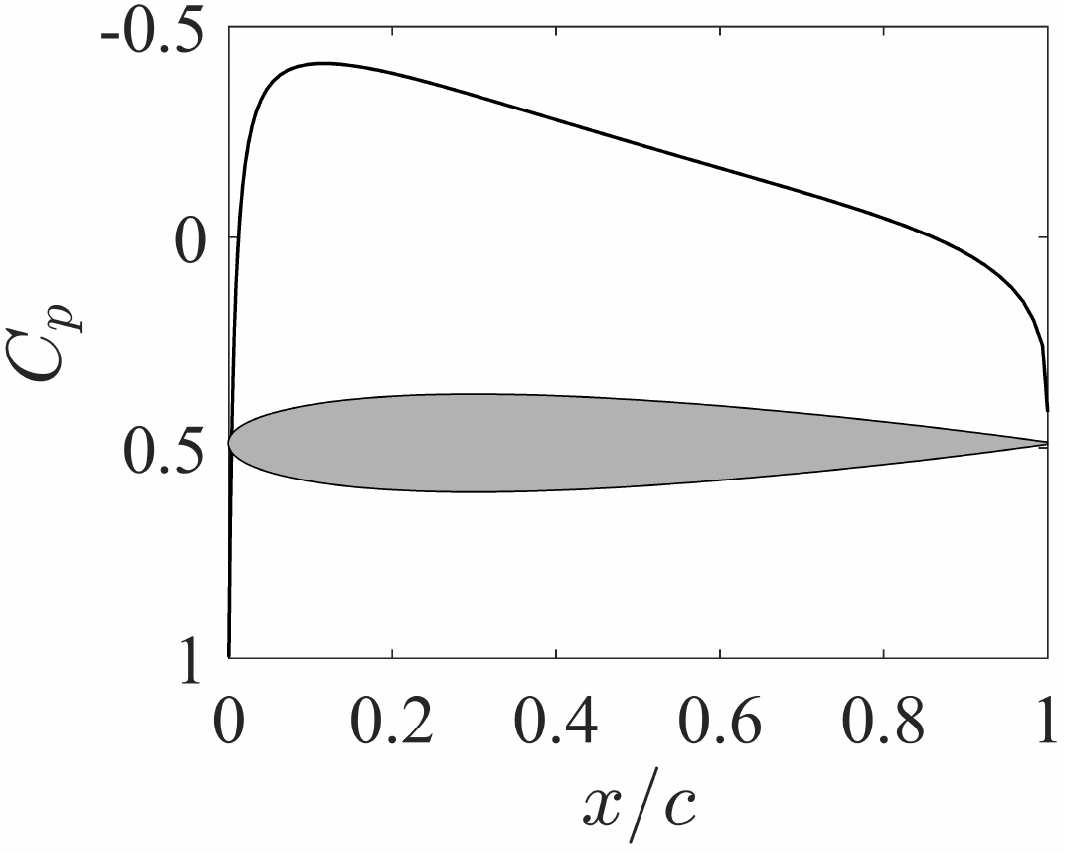}
    \hfill
    \includegraphics[width=0.3\textwidth]{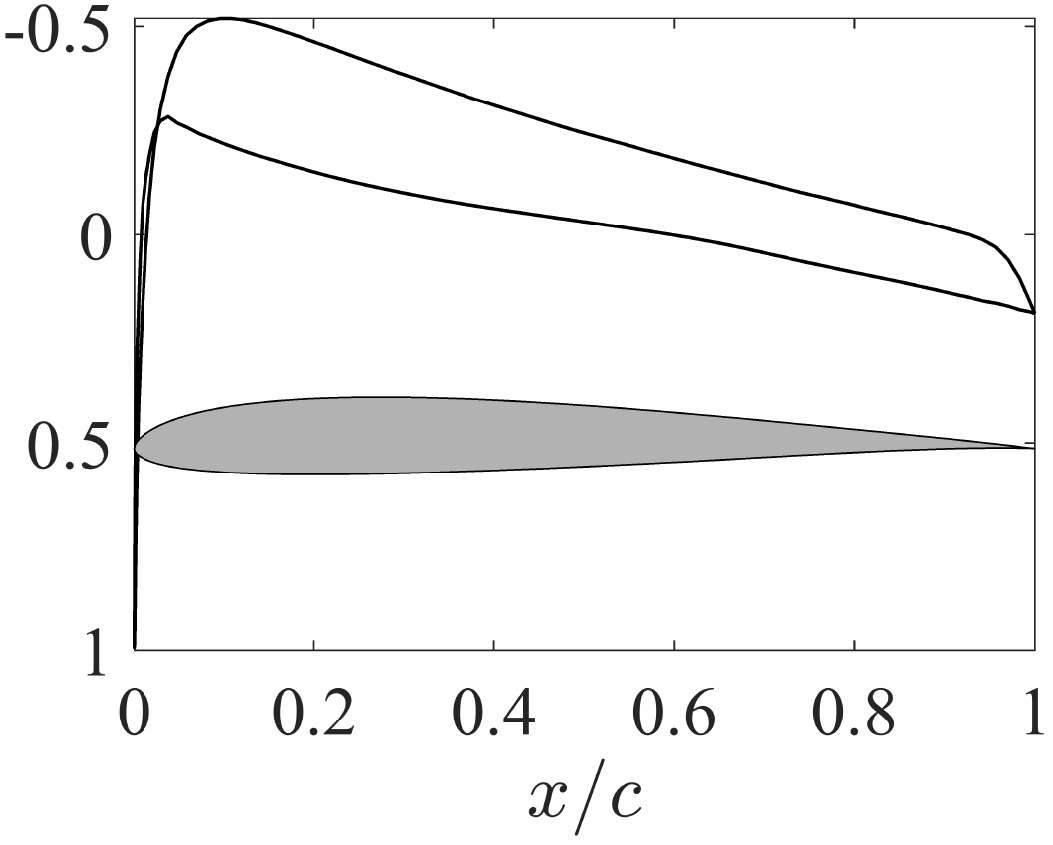}
    \hfill
    \includegraphics[width=0.3\textwidth]{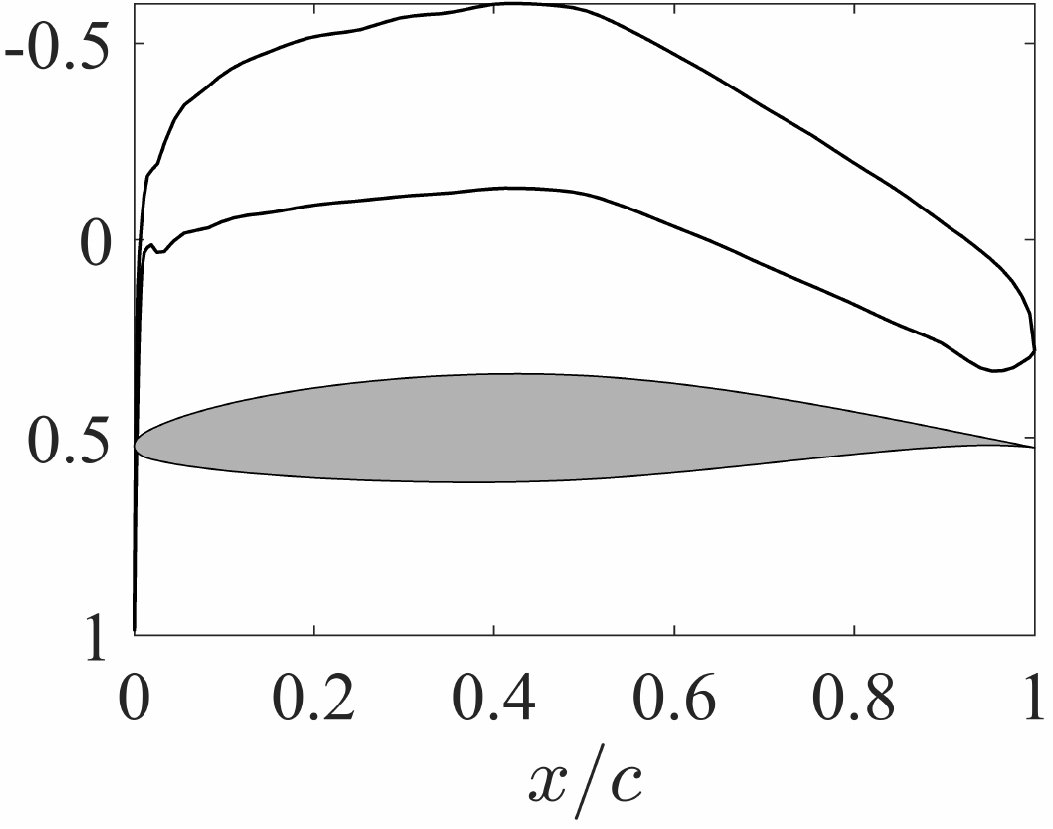}
    }
    \makebox[\textwidth][c]{
    \makebox[0.312\textwidth][c]{(a) NACA 0012}
    \hfill
    \makebox[0.312\textwidth][c]{(b) SD 7003}
    \hfill
    \makebox[0.3\textwidth][c]{(c) NACA 65(1)-412}
    }
	\caption{Pressure coefficients (black) for inviscid flow over airfoils (grey) at $\alpha = 0^\circ$. NACA 0012 (a), SD 7003 (b), and NACA 65(1)-412 (c) \citep{xfoil}.}
	\label{fig:airfcomp}
\end{figure}

Given the known sensitivities to airfoil geometry, an intensive study should select the target with care to make practical application more straight-forward.  The blades used in axial flow compressors are commonly selected from the NACA 65-series \citep{HEE51, Wright74} or Eppler series.  These airfoils and associated aerodynamics are distinct from the NACA 0012 and SD 7003 in that they are designed with the objective of extended laminar flow over the wing by having the suction peak located well downstream of the leading edge, at $x/c$\,=\,0.5, as the nomenclature states.
The NACA 65(1)-412 has its maximum thickness at 40\% chord, and so differs from the NACA 0012 and the SD 7003 where it is at 30\% and 25\% chord, respectively.  The pressure distributions over the example airfoils are summarized in figure \ref{fig:airfcomp}, where the inviscid pressure coefficients for the NACA 0012, SD 7003, and NACA 65(1)-412  are plotted for $\alpha = 0^\circ$. The pressure distribution of the NACA 65(1)-412 differs significantly and features an extended favorable pressure gradient over half of the chord.  At higher Re, this design approach assures attached flow up until this point, but the same may not be found at lower Re, and a substantial portion of the chord may be strongly influenced by the complex dynamics downstream of an LSB.

\subsection{Objectives}
This goal of this paper is to provide a comprehensive and detailed description of the flow topology over a cambered NACA 65(1)-412 airfoil at \Rey{}\,=\,$2\times10^4$ for $\alpha$ from 0$^\circ$ to 10$^\circ$ through highly-resolved direct numerical simulations.
The simulations are conducted with a high-order, compressible, discontinuous, Galerkin spectral element method using a large span (0.5$c$) and large domain length and height (30$c$) to accurately capture the instabilities without altering the separation and transition dynamics through domain blockage.
We set out to identify three-dimensional instabilities and show how they are connected to the formation of three-dimensional tubular structures and very large regions of turbulent coherence (puffs).
For selected $\alpha$, we analyze the interaction of Karman vortices, which originate from the suction-side shear layer at the trailing edge of the airfoil, and the separated pressure-side shear layer with its associated Kelvin-Helmholtz instabilities. 
These flow structures are shown in high detail in the near wake field and are connected to the topology of the far wake several chord-lengths downstream of the airfoil, as well as to time-averaged surface profiles of the skin friction and the pressure coefficients.
We further compare the lift and drag polars of DNS results with wind tunnel experiments, \textit{Xfoil} data, and RANS simulations.

%% file: content/gov_eqn.tex
\subsection{Conservation Laws}
We compute solutions to the compressible Navier-Stokes equations, which can be written in non-dimensional form as
\begin{equation}\label{eq:n-s}
\partial_t\mathbf{U} + \nabla\cdot\mathbf{F} = 0,
\end{equation}
where $\mathbf{U}$ represents the vector of the conserved variables,
\begin{equation}\label{eq:var}
\mathbf{U} = \left[\,\rho \quad \rho u \quad \rho v \quad \rho w \quad \rho e\,\right]^T.
\end{equation}
$\rho$ is the density and $u$, $v$, and $w$ are the velocity components. 
The specific total energy is $\rho e\,=\,p/(\gamma-1)+\frac{1}{2}\rho(u^2+v^2+w^2)$ and the system is closed by the equation of state,
\begin{equation}
p = \frac{\rho T}{\gamma M_f^2},
\end{equation}
where $p$, $T$, and $\gamma$ are the pressure, temperature, and the ratio of specific heats, respectively, and $M_f$ is the reference Mach number.
The flux vector $\mathbf{F}$ comprises an advective (superscript \textit{a}) and a viscous part (superscript \textit{v}),
\begin{equation} \label{eq:flux}
\nabla\cdot\mathbf{F} = \partial_x\mathbf{F}^a + \partial_y\mathbf{G}^a + \partial_z\mathbf{H}^a - \frac{1}{Re_f}\left(\partial_x\mathbf{F}^v + \partial_y\mathbf{G}^v + \partial_z\mathbf{H}^v\right),
\end{equation}
where
\begin{align}
\begin{split}
\mathbf{F}^a &= \left[\,
\rho u \quad p {+} \rho u^2 \quad \rho u v \quad \rho u w \quad u(\rho e {+} p)
\,\right]^T,
\\
\mathbf{G}^a &= \left[\,
\rho v \quad \rho v u \quad p {+} \rho v^2 \quad \rho v w \quad v(\rho e {+} p)
\,\right]^T,
\\
\mathbf{H}^a &= \left[\,
\rho w \quad \rho w u \quad \rho w v \quad p {+} \rho w^2 \quad w(\rho e {+} p)
\,\right]^T,
\end{split}
\end{align}
\begin{align}
\begin{split}
\mathbf{F}^v &= \left[\,
0 \quad \tau_{xx} \quad \tau_{yx} \quad \tau_{zx} \quad u \tau_{xx} {+} v \tau_{yx} {+} w \tau_{zx} {+} \frac{\kappa}{\left(\gamma -1\right)Pr M_f^2} T_x
\,\right]^T,
\\
\mathbf{G}^v &= \left[\,
0 \quad \tau_{xy} \quad \tau_{yy} \quad \tau_{zy} \quad u \tau_{xy} {+} v \tau_{yy} {+} w \tau_{zy} {+} \frac{\kappa}{\left(\gamma -1\right)Pr M_f^2} T_y
\,\right]^T,
\\
\mathbf{H}^v &= \left[\,
0 \quad \tau_{xz} \quad \tau_{yz} \quad \tau_{zz} \quad u \tau_{xz} {+} v \tau_{yz} {+} w \tau_{zz} {+} \frac{\kappa}{\left(\gamma -1\right)Pr M_f^2} T_z
\,\right]^T.
\end{split}
\end{align}
$Re_f$ is the reference Reynolds number, $Pr$ the Prandtl number, and the stress tensor is $\tau_{ij}$\,=\,$2\mu(S_{ij}-S_{mm}\delta_{ij}/3)$
with the strain rate tensor $S_{ij}$. The viscosity $\mu$ is calculated following Sutherland's law,
\begin{equation}
    \mu = \frac{(1+R_T)T^{3/2}}{T+R_T},
\end{equation}
where $R_T$ denotes the ratio of the Sutherland constant $S$ to the reference temperature $T_f$.
All quantities are non-dimensionalized with respect to the airfoil chord length $c$, the free-stream velocity $U_\infty$, density $\rho_\infty$, and temperature $T_\infty$.

\subsection{Boundary layer relations}
The boundary layer velocity profile is extracted from DNS data according to the methods described by \citet{AS00} and \citet{uranga11}, who use a pseudo-velocity profile inside the rotational boundary layer flow based on the spanwise vorticity
\begin{equation} \label{eq:ustar}
    \mathbf{u}^*(s,\eta) = \int_0^\eta \mathbf{\boldsymbol{\omega}}(s,\tilde{\eta})\times\mathbf{n}(s)\,\mathrm{d}\tilde{\eta},
\end{equation}
where $s$ and $\eta$ refer to the wall-tangential and normal coordinates respectively and $\mathbf{n}(s)$ is the wall-normal unit vector.
The boundary layer edge $\eta_e$ is located at a distance where the vorticity magnitude and gradient are below a certain threshold and the flow is assumed to be irrotational \citep{uranga11}.
The displacement thickness $\delta^*$ and momentum thickness $\theta$ are computed by integrating the velocity profile across the boundary layer
\begin{align}
    \delta^*(s) &= \int_0^{\eta_e} \left(1-\frac{u_s(s,\eta)}{u_e(s)}\right)\mathrm{d}\eta,\\
    \theta(s) &= \int_0^{\eta_e} \frac{u_s(s,\eta)}{u_e(s)}\left(1-\frac{u_s(s,\eta)}{u_e(s)}\right)\mathrm{d}\eta.
\end{align}
Here, $u_s$ is the local, tangential velocity component and $u_e$ the velocity magnitude evaluated at the boundary layer edge $\eta_e$.
The shape factor is defined as the ratio of displacement to momentum thickness, $H$\,=\,$\delta^*/\theta$.

The exact locations of flow separation and reattachment are based on the zero-crossings of the time and space-averaged skin friction coefficient according to theory by \citet{Haller04}.
In accordance with \citet{uranga11}, the transition point indicates the location of a local maximum in the shape factor.
We note, however, that the definition of the transition point is not unique; \citet{AS00}, for example, use the point of maximum negative skin friction.

%% file: content/setup.tex
The flow over a cambered NACA 65(1)-412 airfoil is simulated in two and three dimensions at a chord-based Reynolds number of $Re_c$\,=\,$2\times10^4$ and a free-stream Mach number of $M$\,=\,0.3. 
At this Mach number, the compressibility effect in terms of the pressure coefficient deviations are expected to be on the order of 5\% in relation to incompressible flow, according to the Prandtl-Glauert correction $C_{p,M}/C_{p,i}$\,=\,$1/\sqrt{1-M^2}$.
While the Mach number in comparable wind tunnel experiments is usually closer to 0.1, the increased stiffness of the explicit numerical scheme results in time step sizes of the order of $\mathcal{O}(10^{-6})$ that result in an excessive computational cost for three-dimensional simulations.
The Prandtl number is set to $Pr$\,=\,0.72. The Sutherland constant $R_T$\,=\,$S/T_f$\,=\,110/200, and ratio of specific heats $\gamma$\,=\,1.4 are chosen in accordance with \citet{nelson}.

We discretize the Navier-Stokes equations \eqref{eq:n-s} with a discontinuous Galerkin spectral element method (DGSEM). 
The method and code is extensively discussed, tested and used for DNS in previous work (\citet{kopriva,KJK20} and references therein).
The conservative variables \eqref{eq:var} are spatially approximated on a \textit{N\textsuperscript{th}} order polynomial basis and collocated on quadrature nodes of Legendre polynomials. 
A Roe upwind scheme is used for the advective interface fluxes and a Bassi-Rebay formulation is used for the viscous part. 
A fourth-order explicit Runge-Kutta adaptive time-stepping scheme is used with time step sizes ranging between $2.3\times 10^{-5}$ $\leq$ $\Delta t$ $\leq$ $8.4\times 10^{-6}$, depending on the element size and the polynomial order.

Similar to the simulations by \citet{uranga11} and \citet{BBFFGHM14}, Riemannian free-stream conditions are applied at the outer boundaries of the domain.
Spurious oscillations from exiting vortices are decreased through grid coarsening towards the outflow, as well as a damping layer on the energy term to reduce the reflected pressure waves \citep{JKM03}.
The surface of the airfoil is treated as no-slip adiabatic wall and, to account for its curvature, we fit the neighboring boundary elements to a spline representing the profile of the airfoil according to \citet{nelson16}.
For 3D simulations, the mesh is extruded in the spanwise direction and the boundaries are set to be periodic to model an infinite wing. 

The simulations are run until the flow has fully transitioned to a three-dimensional state and the solution has reached quasi-steady state with the lift and drag coefficients fluctuating around a mean. 
Flow statistics are recorded subsequently, with the integration times given in table \ref{tab:sim}.

\subsection{Domain Size}
\begin{figure}
\centering
    \begin{minipage}{0.4\textwidth}
    \includegraphics[width=\textwidth]{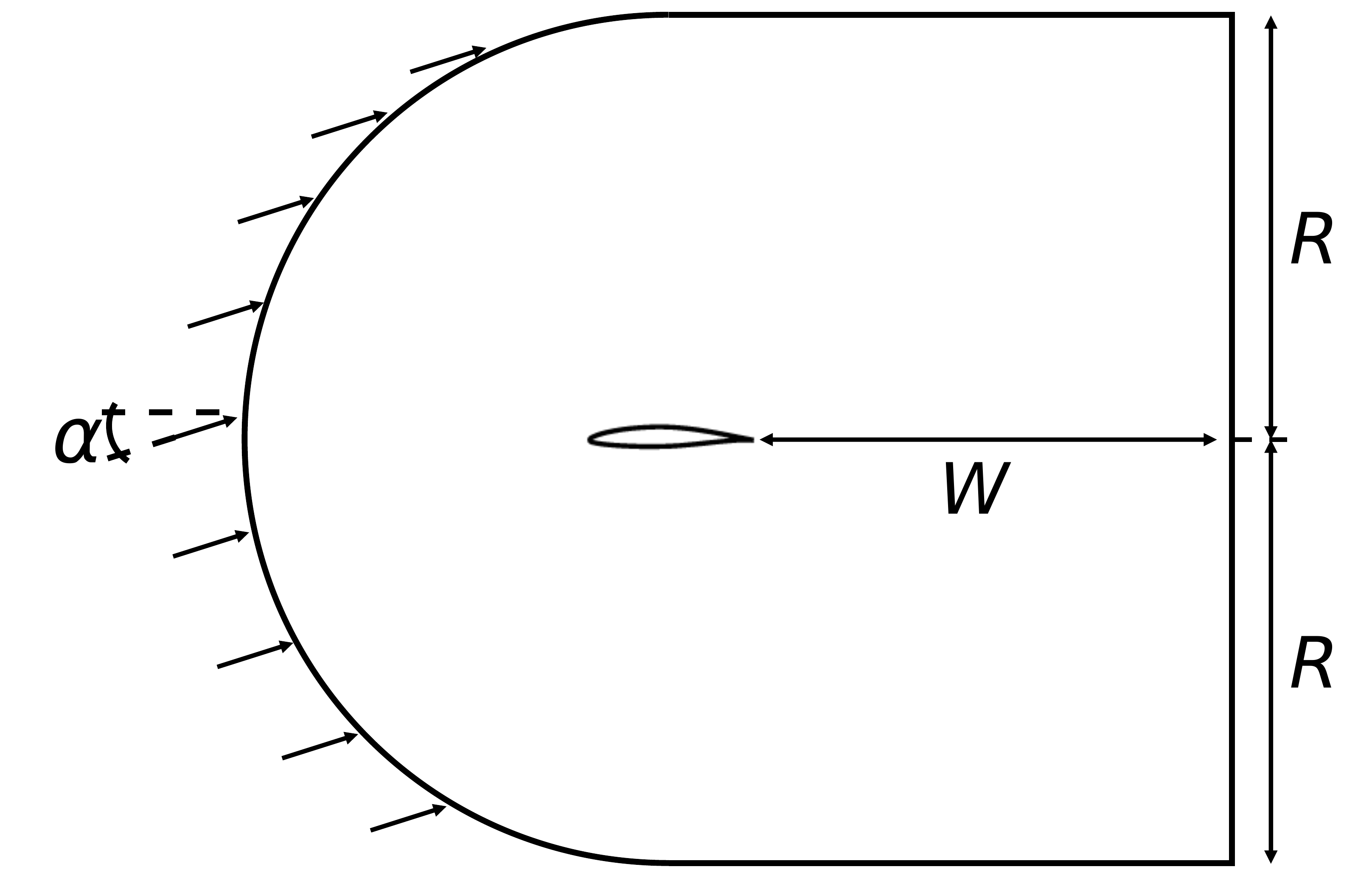}
    \makebox[\textwidth][c]{(a) Domain Parameters}
    \end{minipage}\hfill
    \begin{minipage}{0.45\textwidth}
    \includegraphics[width=0.9\textwidth]{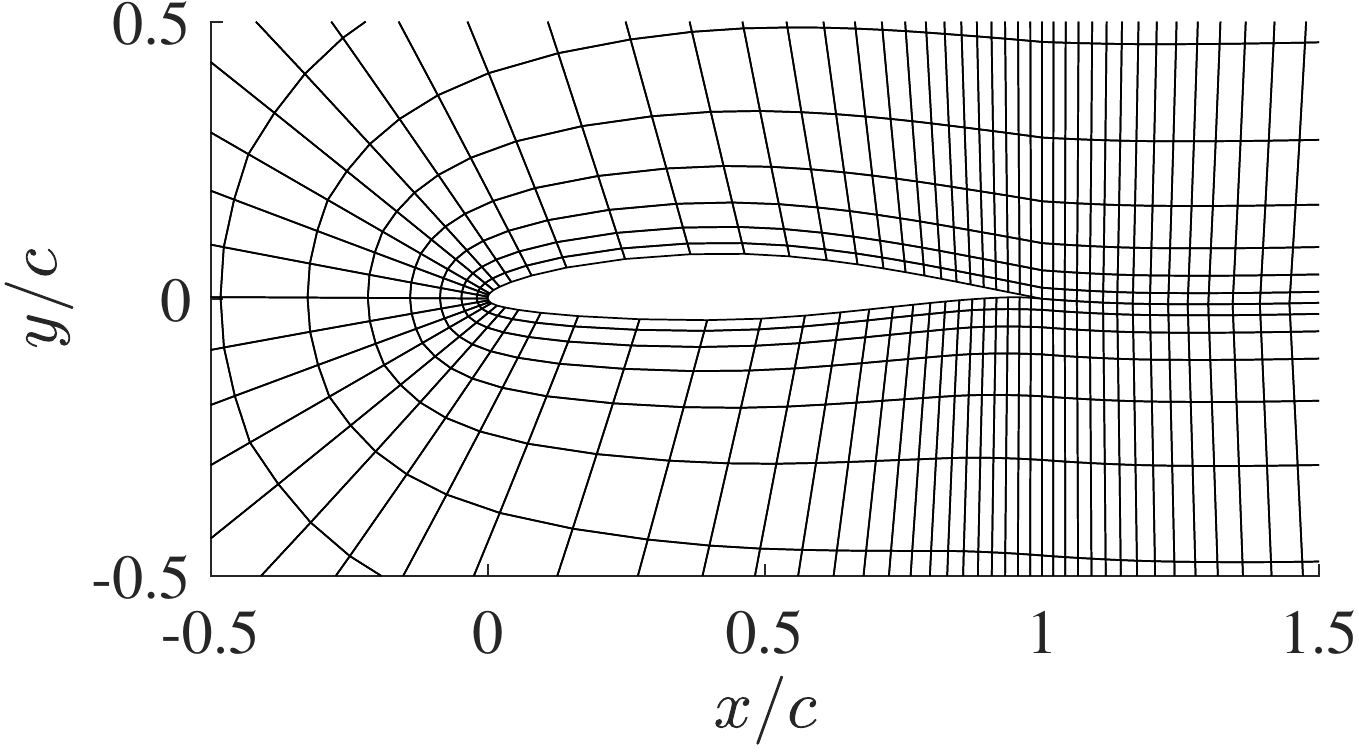}
    \makebox[0.9\textwidth][c]{(b) Grid 1}
    \vskip\baselineskip
    \includegraphics[width=0.9\textwidth]{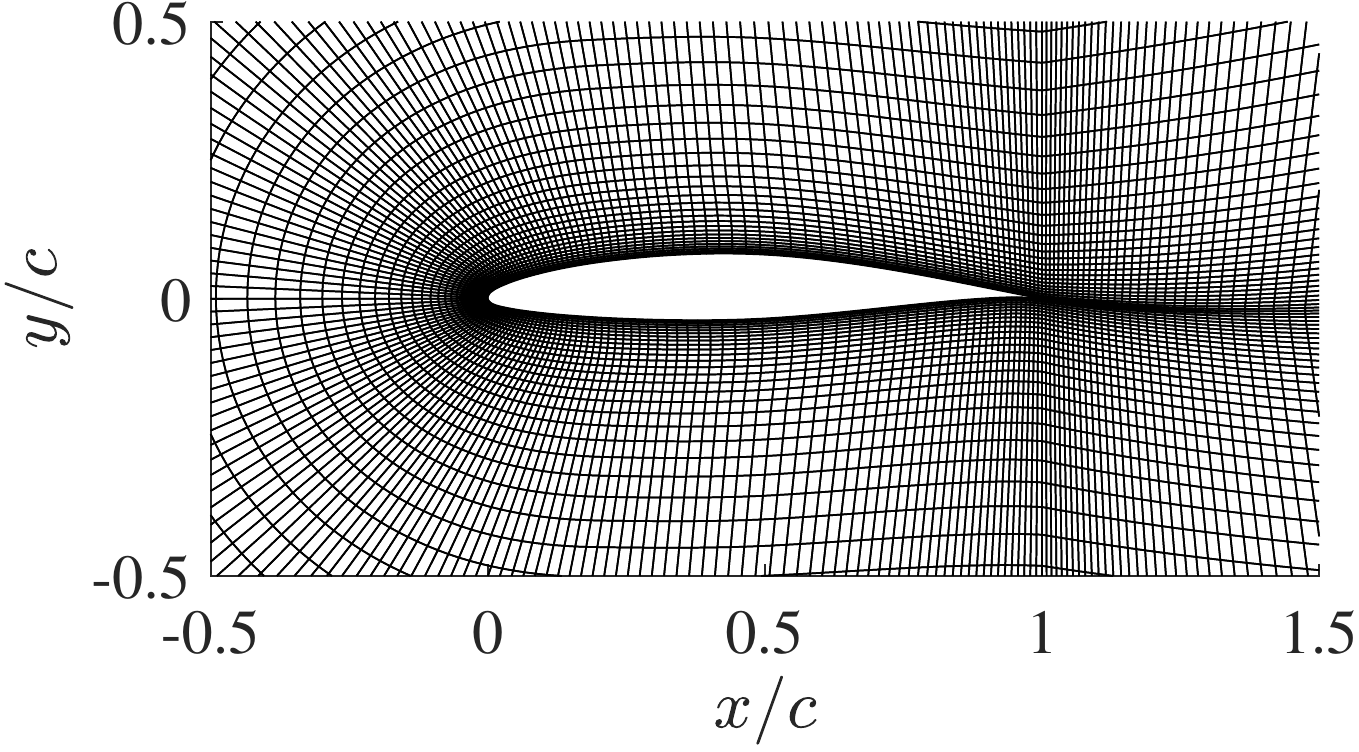}
    \makebox[0.9\textwidth][c]{(c) Grid 2}
    \end{minipage}
	\caption{(a) C-type computational domain with general parameters.
	Elements of 2D computational meshes \textit{Grid 1} (b) and \textit{Grid 2} (c) around the airfoil. Only elements without interior Gauss nodes are shown.}
	\label{fig:domain}
\end{figure}

The size of the computational domain impacts the numerical solution through blockage and spurious reflections from the outer boundaries. 
A compromise has to be made between a domain large enough to minimize such boundary effects and the available computational resources that necessarily limit the number of grid points. 
A C-type computational domain is used in this work and schematically shown in figure \ref{fig:domain}(a), with the radius \textit{R} and the wake length \textit{W} indicated.

Table \ref{tab:domain} summarizes domain sizes used in selected airfoil simulations in literature. 
The domain radius of these airfoil DNS ranges from 4\textit{c} \citep{DJL07} to 100\textit{c} \citep{visbal11,BBFFGHM14}. 
\begin{table}
    \begin{center}
    \begin{tabular}{lccc}
        Source  	        & $R/c$     & $W/c$     & $L_z/c$ \\
        \hline
        \citet{DJL07}       & 4         & 3         & 0.1  \\
        \citet{JSS08}       & 7.3       & 5         & 0.2  \\
        \citet{uranga11}    & 6         & 6.4       & 0.2  \\
        \citet{visbal11}    & 100       & 100       & 0.1-0.8  \\
        \citet{BBFFGHM14}   & 100       & 100       & 0.2  \\
        \citet{LNOF15}      & 25        & 25        & 0.2  \\
        \citet{ZCGQS15}     & 6         & 10        & 0.1-0.8  \\
        \citet{balakumar}	& 15        & 15        & 0.2  \\
        \citet{SMS17}       & 15        & 10        & 0.5-2.0 \\ 
        present	            & 30        & 30        & 0.5 \\
    \end{tabular}
    \caption{Domain sizes of selected airfoil studies.}
    \label{tab:domain}
    \end{center}
\end{table}

In the present study, two different C-type meshes are applied, where each grid has a sharp trailing edge and a domain radius, $R$, and wake length, $W$, of $30c$. 
These values are higher than in most comparable studies (Table \ref{tab:domain}), but a large domain size was found to be necessary to minimize spurious reflections from the outflow boundaries or changes in the separation bubble shape \citep{BBFFGHM14}.
For three-dimensional computations with periodic boundaries, the domain is extruded in z-direction by $L_z$\,=\,0.5$c$, as recommended by \citet{AJS10} in their LES study of the NACA 0012.

\subsection{Resolution - Is it DNS?}
Two C-type meshes are employed in this study: 
\textit{Grid 1} consists of 3,366 quadrilateral elements in the \textit{x}-\textit{y} plane and is extruded by 10 elements along the span for 3D simulations (see figure \ref{fig:domain}b).
\textit{Grid 2} is refined with 23,400 elements per 2D plane and 50 elements in spanwise direction (see figure \ref{fig:domain}c). 
For \Rey{}\,=\,$2\times10^4$ and $\alpha$\,=\,4$^\circ$, \citet{nelson16,KJK20} have reported a grid-converged solution at a polynomial of $N$\,=\,12 for a mesh nearly identical to \textit{Grid 1}, but limited to $R$\,=\,5$c$ and $W$\,=\,15$c$. 
\citet{KJK20}, however, show that numerical instabilities occur at higher angles of attack (10$^\circ$) if the standard DGSEM scheme is applied, but that these can be prevented by stabilization through a kinetic energy conserving formulation of the advective volume fluxes based on the split form by \citet{pirozzoli}. 
While the coarse grid is proven to give a converged solution at $N$\,=\,12 for lower angles of attack ($\alpha$\,=\,0$^\circ$, 4$^\circ$), we use the refined \textit{Grid 2} with a lower order of $N$\,=\,4 and $N$\,=\,6 for computations at higher $\alpha$\,=\,7$^\circ$, 8$^\circ$, respectively.
The higher near-wall resolution of the fine grid adds more resolution to the boundary layer and hence is more suitable for turbulent flow and the accompanying increase in wall shear stress.
For $\alpha$\,=\,8$^\circ$, the maximum wall-normal height of the first element (\textit{Grid 2}) at the leading edge is $\eta_{e}^+$\,=\,8 and places the first Gauss point at a wall coordinate of $\eta_{g}^+$\,=\,0.2. The maximum tangential grid spacing, based on the average spacing per element, is $\zeta^+$\,=\,4 and occurs at the reattachment point of the LSB at $x/c$ $\approx$ 0.4.
These values are well within the limits accepted for DNS \citep{GRF10}.
Grid independent solutions are established in two dimensions first, and presented in appendix \ref{appendix_2D}.
These studies are a reference point for the three-dimensional simulations to establish grid-independence of the primary, two-dimensional vortex shedding.

For $\alpha$\,=\,10$^\circ$, we have a marginal resolution of all scales at best, and we will not claim DNS for that angle but rather refer to it as implicit LES (ILES) computation.
Because the flow is past the critical transition angle (7$^\circ$--8$^\circ$), we employ a computationally more efficient setup and resolve the flow on \textit{Grid 1} with twelfth order polynomials in the near field and reduced order elements in the outer field.
A spectral filter reduces spurious oscillations from the decreasing order approximations away from the airfoil.
The test matrix of three-dimensional simulations is collated in table \ref{tab:sim} for different meshes, polynomial orders, and refinements. 
\begin{table}
    \begin{center}
    \begin{tabular}{lccccccccc}
        $Re$          & $\alpha$         & Mesh        & $R/c$  & Scheme & $N_i$($N_o$)  & $T_{init}$ & $T_{fin}$ & $T_{stat}$ & DOF ($\times 10^6$)\\
        \hline
        $2\times10^4$  & 0$^\circ$        & Grid 1      & 30     & GL-SF  & 12      & 0$^a$           & 43.2 & 10.1 & 74.0 \\
        $2\times10^4$  & 4$^\circ$        & Grid 1      & 30     & GL-SF  & 12      & 0$^a$           & 46.7 & 10.2 &  74.0 \\
        $2\times10^4$  & 6$^\circ$        & Grid 2      & 30     & G      & 4       & 0$^b$           & 5.4  & - & 146.3 \\
        $2\times10^4$  & 7$^\circ$        & Grid 2      & 30     & G      & 4       & 0$^b$           & 23.3 & 14.5 & 146.3 \\
        $2\times10^4$  & 8$^\circ$        & Grid 2      & 30     & G      & 6       & 25.8        & 39.0 & 8.2 & 401.3  \\
        $2\times10^4$  & 10$^\circ$       & Grid 1      & 30     & GL-SF  & 12(1)   & 15.6        & 36.1 & 15.9 & 30.7 \\
    \end{tabular}
    \caption{Overview of 3D simulations. $Re$ = free-stream Reynolds number, $\alpha$ = angle of attack, $R/c$ = domain radius, G = standard Gauss DGSEM (* = with spectral filter), GL-SF = split form DGSEM with Gauss-Lobatto nodes, $T_{init}$/$T_{fin}$ = initial/final convective time of run, $^a$ = initialized with uniform velocity field, $^b$ = initialized with 2D result, $T_{stat}$ = integration time of statistics, (2x) = h-refined, $N_i$($N_o$) = polynomial order inner (outer) region, DOF = degrees of freedom (number of high-order nodes).}
    \label{tab:sim}
    \end{center}
\end{table}

%% file: content/results3D.tex
\begin{figure}
    \centering
    \makebox[\textwidth][c]{
    \includegraphics[width=0.47\textwidth]{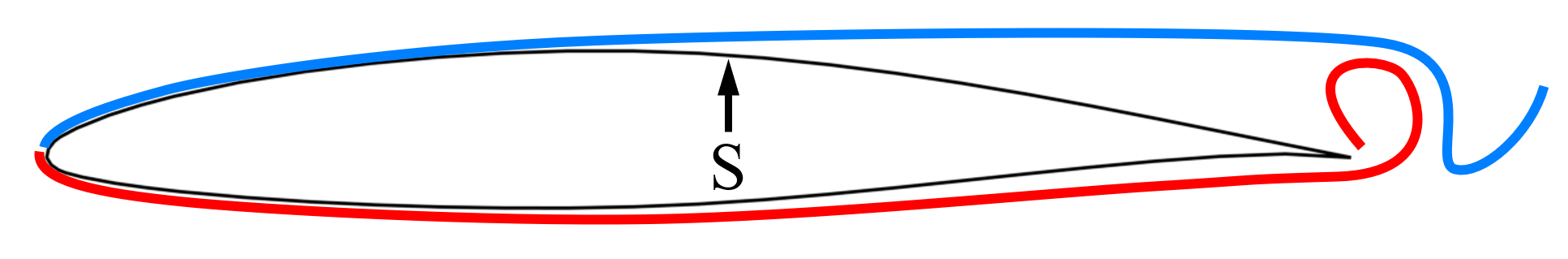}
    \hfill
    \includegraphics[width=0.47\textwidth]{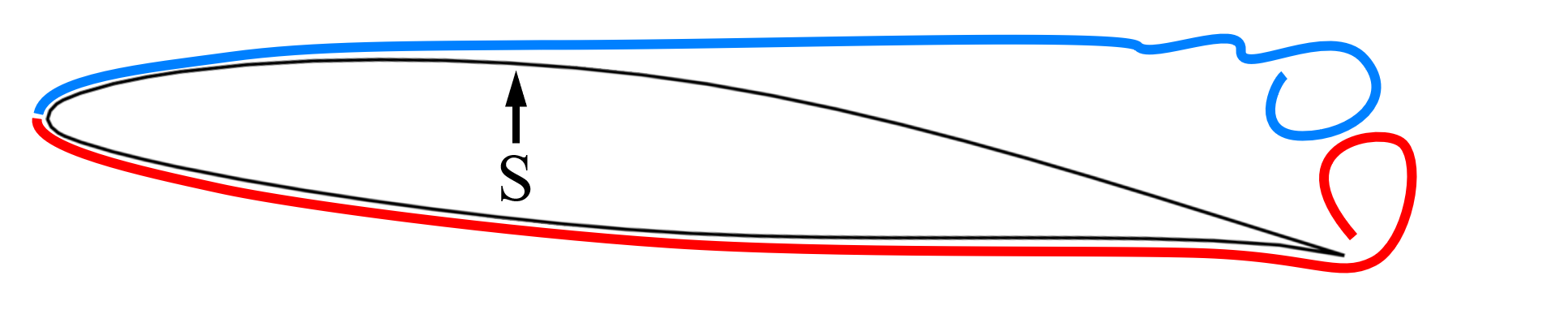}
    }
    \makebox[\textwidth][c]{
	\makebox[0.47\textwidth][c]{(a) $\alpha$\,=\,0$^\circ$}
	\hfill
	\makebox[0.40\textwidth][c]{(b) $\alpha$\,=\,6$^\circ$}
	}
	\vskip\baselineskip
	\makebox[\textwidth][c]{\includegraphics[width=0.47\textwidth]{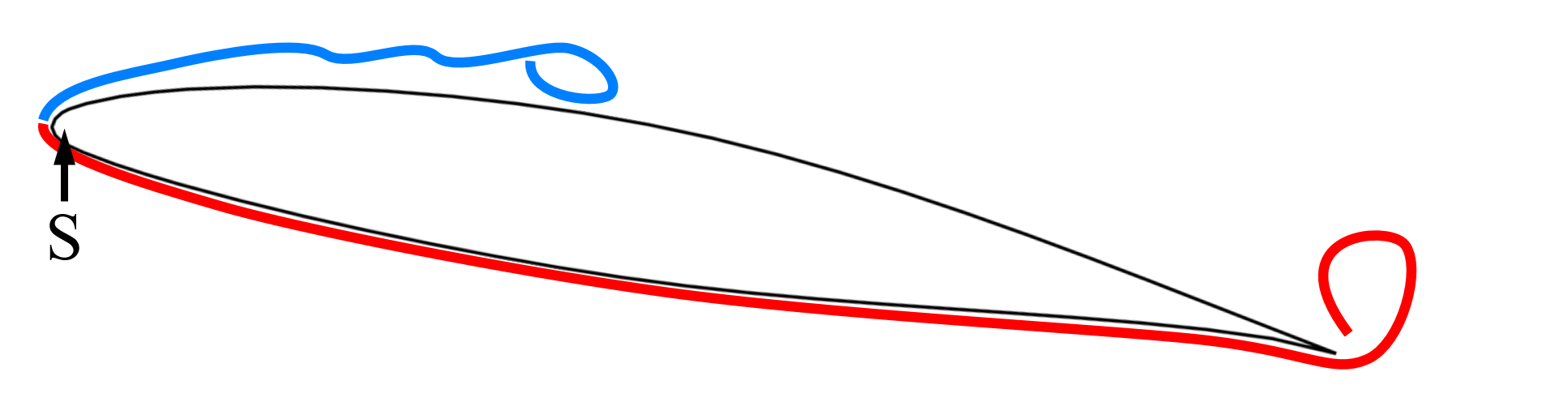}}
	\makebox[\textwidth][c]{\makebox[0.40\textwidth][c]{(c) $\alpha$\,=\,10$^\circ$}}
	\caption{Schematic of the vortex roll-up at the trailing edge and instability within the separated shear layer for $\alpha$\,=\,0$^\circ$, 6$^\circ$, 10$^\circ$.}
	\label{fig:schematic}
\end{figure}

The main flow features are schematically illustrated in figure \ref{fig:schematic} and include the separation and transition of the suction (top) side shear layer and its interaction with the shedding of the pressure (bottom) side boundary layer.
These features can also be recognized in the detailed flow visualizations in figure \ref{fig:vorticity}.
The surface curvature of the profile induces an adverse pressure gradient on both sides of the airfoil, but on the pressure side, the shear layer withstands the adverse force and grows without separating from the wall until it forms a left-turning vortex at the sharp trailing edge. 
On the suction side, the higher surface curvature of the cambered profile results in a stronger adverse pressure force, as compared to the pressure side, that leads to a laminar separation of the boundary layer. 
At low $\alpha$ (see figure \ref{fig:schematic}a), the bottom trailing edge vortex induces a vortex of opposite rotation in the upper, separated shear until the vortex pair sheds off and forms a vortex street in the wake.
Kelvin-Helmholtz instabilities cause the growth of additional modes within the upper, separated shear layer (see figure \ref{fig:schematic}b).
These modes are amplified as they travel downstream and eventually yield vortical flow structures. 
As $\alpha$ increases, the adverse pressure gradient steepens, and the separation point moves upstream. 
At a critical $\alpha$ the transition point of the shear layer has moved upstream far enough for sizable flow structures to form upstream of the trailing edge, and they then transport flow momentum towards the wall, resulting in reattachment of the flow and the formation of a LSB.
Both the LSB and the trailing edge now shed vortices separately (but not necessarily independently, as \citet{DLR16} have shown that LSB and Karman shedding are locked-in for the flow over a SD 7003). 

\subsection{Flow topology -- separation, transition, and wake}
\begin{figure}
    \centering
    \makebox[\textwidth][c]{
    \includegraphics[width=0.32\textwidth]{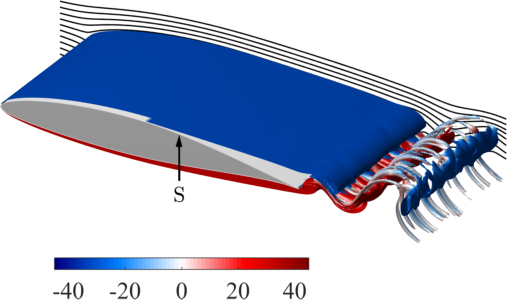}
    \hfill
    \includegraphics[width=0.32\textwidth]{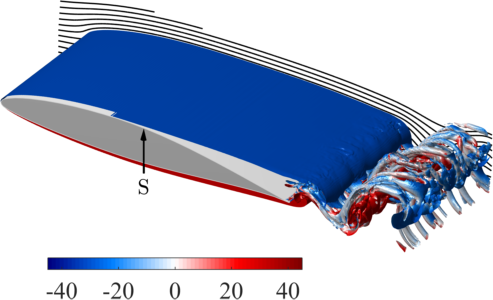}
    \hfill
    \includegraphics[width=0.32\textwidth]{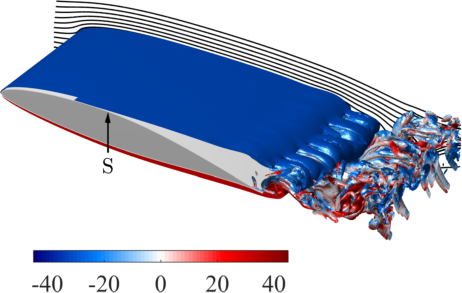}
    }
    \makebox[\textwidth][c]{
	\makebox[0.2\textwidth][c]{(a) $\alpha$\,=\,0$^\circ$}
	\hspace{0.12\textwidth}
	\makebox[0.2\textwidth][c]{(b) $\alpha$\,=\,4$^\circ$}
	\hspace{0.12\textwidth}
	\makebox[0.2\textwidth][c]{(c) $\alpha$\,=\,6$^\circ$}
	\hfill
	}
	\vskip\baselineskip
	\makebox[\textwidth][c]{
    \includegraphics[width=0.32\textwidth]{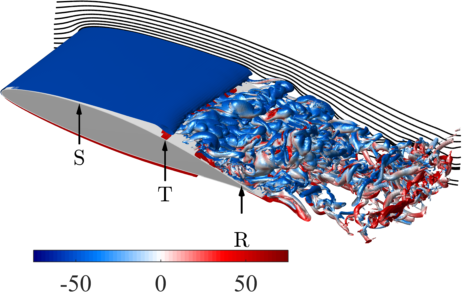}
    \hfill
    \includegraphics[width=0.32\textwidth]{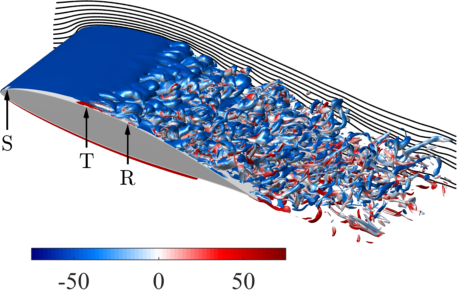}
    \hfill
    \includegraphics[width=0.32\textwidth]{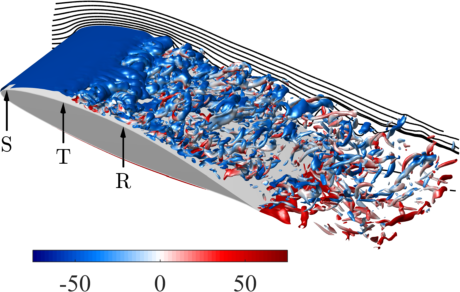}
    }
    \makebox[\textwidth][c]{
    \makebox[0.2\textwidth][c]{(d) $\alpha$\,=\,7$^\circ$}
    \hspace{0.12\textwidth}
	\makebox[0.2\textwidth][c]{(e) $\alpha$\,=\,8$^\circ$}
	\hspace{0.12\textwidth}
	\makebox[0.2\textwidth][c]{(f) $\alpha$\,=\,10$^\circ$}
	\hfill
	}
	\caption{Iso-vorticity surfaces for $\alpha$ from 0$^\circ$ (a) to 10$^\circ$ (f).
	\textit{S}, \textit{T}, and \textit{R} indicate the mean locations of separation, transition, and reattachment.}
	\label{fig:vorticity}
\end{figure}

Figure \ref{fig:vorticity} shows iso-surfaces of the instantaneous vorticity magnitude $|\mathbf{\omega}|$\,=\,$\sqrt{\omega_x^2+\omega_y^2+\omega_z^2}$ colored by the spanwise vorticity component $\omega_z$ to indicate the rotational direction of the vortical structures.
The vorticity is computed as the curl of the velocity vector $\mathbf{\omega}$\,=\,$\nabla\times\mathbf{u}$ and normalised by the advective frequency $U_\infty/c$.
Laminar separation without reattachment is observed at $\alpha$ from 0$^\circ$ to 6$^\circ$.
The separation point \textit{S}, i.e. the time-averaged location of zero skin friction, is located at $x_s/c$\,=\,0.6 (0$^\circ$), 0.49 (4$^\circ$), and 0.4 (6$^\circ$), as shown in figure \ref{fig:vorticity}(a--c). 
For $\alpha$\,=\,0$^\circ$ and $4^\circ$, the flow over the airfoil is quasi-two dimensional and three-dimensional structures develop only at the trailing edge and in the wake.
The vortex system comprised of spanwise Karman and longitudinal braid vortices, resembles the three-dimensional structure found in bluff body wakes at low Reynolds number (such as behind a circular cylinder \citep{Williamson96a}).
A detailed discussion of these near-wake instabilities is given in section \ref{sec:inst_low_aoa}.
At $\alpha$\,=\,6$^\circ$, a Kelvin-Helmholtz (K-H) instability causes the development of spanwise vortices within the separated, upper shear layer  upstream of the trailing edge (figure \ref{fig:vorticity}c).
Although the vortices cause the disruption of the separated shear layer upstream of the pressure side vortex roll-up, the instability does not cause the flow to reattach and the separated shear layer encloses a large recirculation region between the separation point and the trailing edge for $\alpha$ from $0^\circ$ to $6^\circ$.

As $\alpha$ is further increased, the streamlines are increasingly displaced and result in a lower pressure at the airfoil's leading edge and consequently in a stronger adverse pressure gradient than at lower incidence. 
At $\alpha$\,=\,7$^\circ$, the flow separates at $x_s/c$\,=\,0.26 and a K-H instability drives the formation of large, spanwise vortices along the separated shear layer in the region 0.5 $\leq$ $x/c$ $\leq$ 0.6.
Upon their generation, these vortices are quasi-two dimensional but loose their coherence as they roll over the airfoil surface and transition into turbulence at $x_t/c$\,=\,0.62 (marked as \textit{T} in figure \ref{fig:vorticity}). 
Downstream of the transition point, the vortices can combine and burst into large-scale three-dimensional turbulent clouds, which we shall refer to as ``puffs'', as we elaborate in section \ref{sec:inst_high_aoa} in more detail. 
The turbulent fluid motion transports momentum from the mean flow towards the airfoil surface and re-energizes the boundary layer, resulting in the reattachment of the flow at $x_r/c$\,=\,0.93 (see figure \ref{fig:vorticity}d).
The turbulent reattachment at the trailing edge results in a slender LSB that has a maximum wall-normal height of $h_{LSB}$\,=\,3.3\% of the chord length and stretches over 67\% of the airfoil.

At $\alpha$\,=\,7$^\circ$ the LSB abruptly shifts from the trailing edge of the airfoil to the leading edge with the flow separating at $x_s/c$\,=\,0.014 and reattaching at mid-chord ($x_r/c$\,=\,0.48) for an angle of attack of $\alpha$\,=\,8$^\circ$ (see figure \ref{fig:vorticity}e). 
The proportions of the LSB are similar to those at $\alpha$\,=\,7$^\circ$, with the bubble height measuring $h_{LSB}$\,=\,2.5\% of the chord length and the ratios of LSB height to length being 0.054 (8$^\circ$) and 0.052 (7$^\circ$).
These values are lower than the ratios reported by \citet{GV10} for the SD 7003 airfoil, which are typically between 0.08 (\Rey{}\,=\,$4\times10^4$, $\alpha$\,=\,4$^\circ$) and 0.15 (\Rey{}\,=\,$3\times10^4$, $\alpha$\,=\,8$^\circ$).
The more slender bubble over the NACA 65(1)-412 stems from the geometrical differences with the SD 7003, which favours formation of an LSB, whereas the present airfoil is designed to extend laminar flow by having its maximum thickness near mid-chord.
At $\alpha$\,=\,8$^\circ$, laminar flow is present over two thirds of the bubble until the separated shear layer transitions at $x_t/c$\,=\,0.32.
The topology downstream of the transition point is governed by hairpin vortices and the break up of the laminar, spanwise vortices that shed off the LSB shear layer.
Horseshoe, hairpin, or loop vortices all describe vortex tubes that, starting from a cross-stream alignment, bend upwards away from the wall and stretch in the streamwise direction as the upper portion (head) is subjected to the higher velocity in the boundary layer \citep{Robinson91}. 
Because the bubble trailing edge has shifted from $x_r/c$\,=\,0.93 at $\alpha$\,=\,7$^\circ$ to $x_r/c$\,=\,0.48 at $\alpha$\,=\,8$^\circ$, a turbulent boundary layer develops past mid-chord over the downstream section of the airfoil until it sheds off at the trailing edge.
A similar flow topology is found at $\alpha$\,=\,10$^\circ$, where the LSB has slightly changed in size with an earlier transition point ($x_t/c$\,=\,0.23), marginally shorter bubble length ($x_s/c$\,=\,0.012, $x_r/c$\,=\,0.46), and increased height $h_{LSB}/c$\,=\,3.9\%.
The height-to-length ratio consequently increases to 0.087, which is more than 60\%  higher than the values found at $\alpha$\,=\,7$^\circ$ and 8$^\circ$.

\begin{figure}
    \centering
    \makebox[\textwidth][c]{
    \includegraphics[width=0.45\textwidth,trim={0 25pt 0 0},clip]{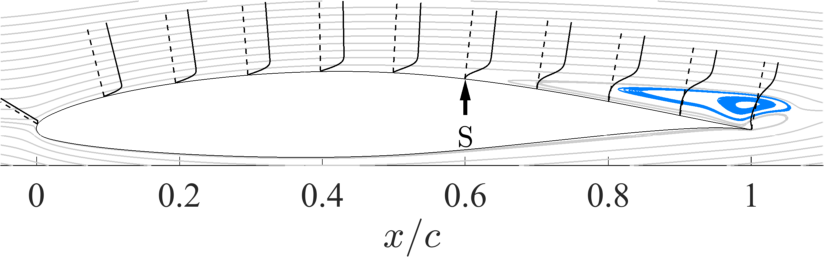}
    \hfill
    \includegraphics[width=0.45\textwidth,trim={0 25pt 0 0},clip]{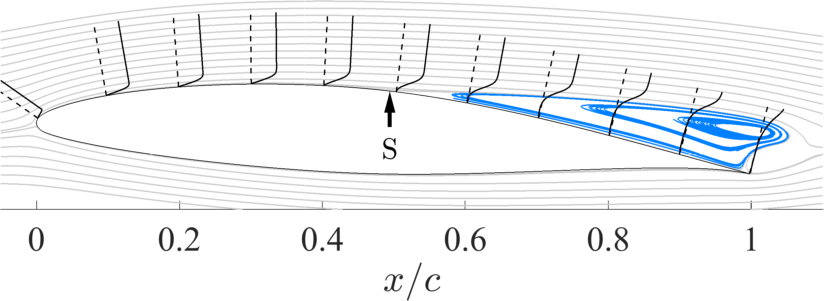}
    }
    \makebox[\textwidth][c]{
	\makebox[0.45\textwidth][c]{(a) $\alpha$\,=\,0$^\circ$}
	\hfill
	\makebox[0.45\textwidth][c]{(b) $\alpha$\,=\,4$^\circ$}
	}
	\vskip\baselineskip
	\makebox[\textwidth][c]{
    \includegraphics[width=0.45\textwidth,trim={0 25pt 0 0},clip]{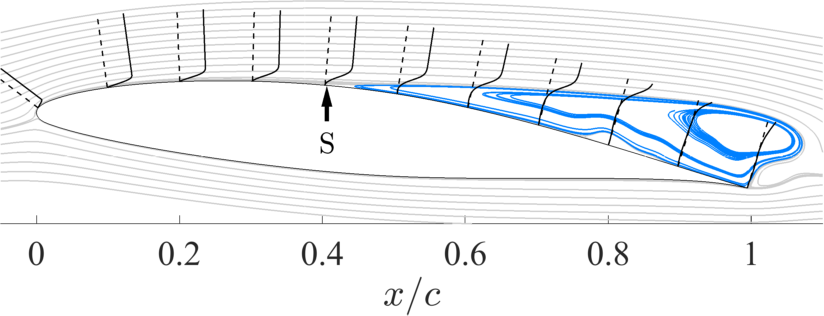}
    \hfill
    \includegraphics[width=0.45\textwidth,trim={0 25pt 0 0},clip]{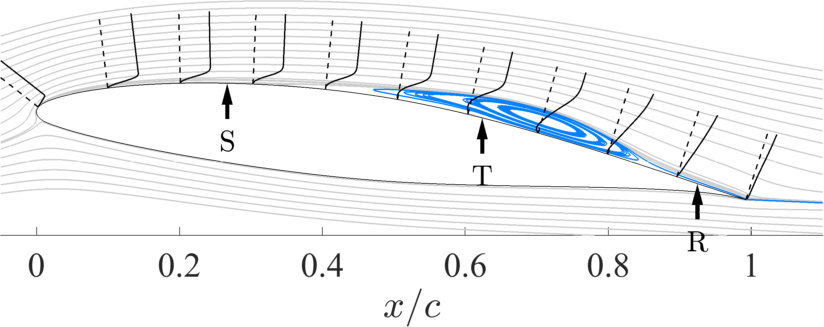}
    }
    \makebox[\textwidth][c]{
	\makebox[0.45\textwidth][c]{(c) $\alpha$\,=\,6$^\circ$}
	\hfill
	\makebox[0.45\textwidth][c]{(d) $\alpha$\,=\,7$^\circ$}
	}
	\vskip\baselineskip
	\makebox[\textwidth][c]{
	\includegraphics[width=0.45\textwidth]{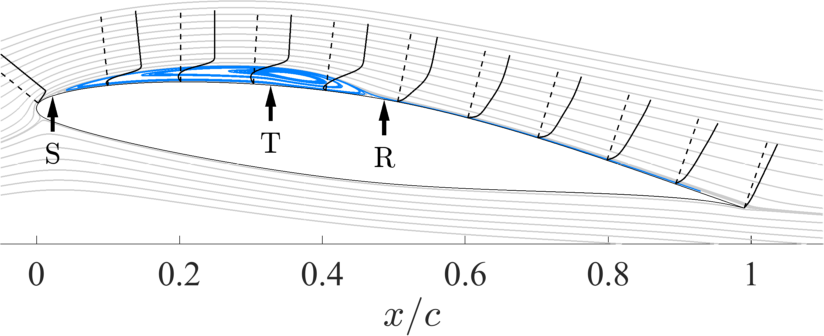}
	\hfill
    \includegraphics[width=0.45\textwidth]{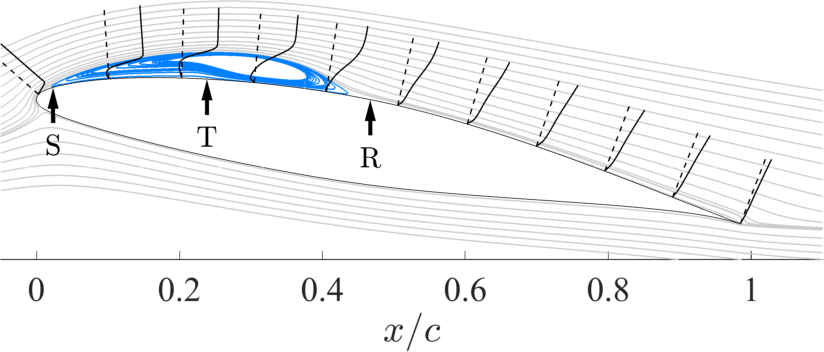}
    }
    \makebox[\textwidth][c]{
    \makebox[0.45\textwidth][c]{(e) $\alpha$\,=\,8$^\circ$}
    \hfill
	\makebox[0.45\textwidth][c]{(f) $\alpha$\,=\,10$^\circ$}
	}
	\caption{Time and space-averaged streamlines for $\alpha$ from 0$^\circ$ (a) to 10$^\circ$ (f).
	Recirculating flow in blue.
	\textit{S}, \textit{T}, and \textit{R} indicate the mean locations of separation, transition, and reattachment.}
	\label{fig:streamline}
\end{figure}

A comparison of the time and space-averaged streamline patterns in figure \ref{fig:streamline} demonstrates the change in flow topology from a region of separated, recirculating flow at the trailing edge into a LSB and its swift shift (within one degree $\alpha$) towards the leading edge.
Because the maximum height of the airfoil is at $x/c$\,=\,0.4, the LSB is either formed upstream or downstream of that point and not at mid-chord. 

\subsubsection{Three-dimensional instability at lower angles} \label{sec:inst_low_aoa}
\begin{figure}
    \centering
    \makebox[\textwidth][c]{
    \includegraphics[width=0.3\textwidth]{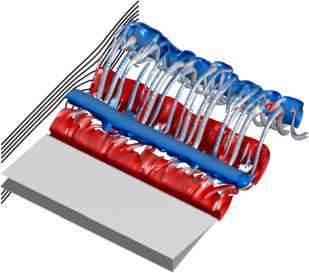}
    \hfill
    \includegraphics[width=0.3\textwidth]{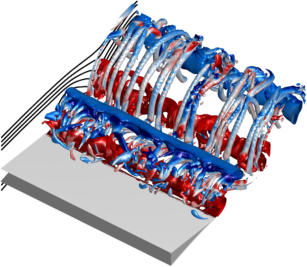}
    \hfill
    \includegraphics[width=0.3\textwidth]{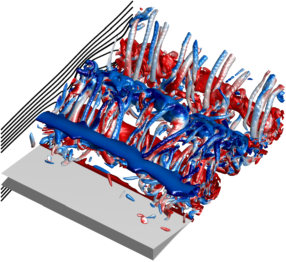}
    }
    \vskip\baselineskip
    \makebox[\textwidth][c]{
    \includegraphics[width=0.25\textwidth,trim={60pt 0 0 0},clip]{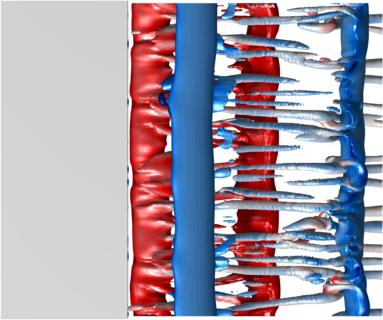}
    \hfill
    \includegraphics[width=0.25\textwidth,trim={60pt 0 0 0},clip]{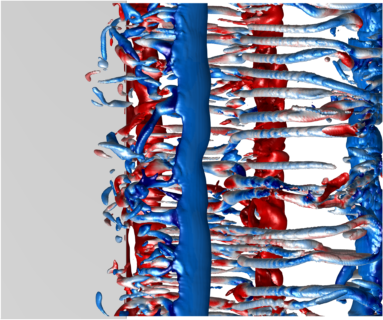}
    \hfill
    \includegraphics[width=0.275\textwidth,trim={60pt 0 0 0},clip]{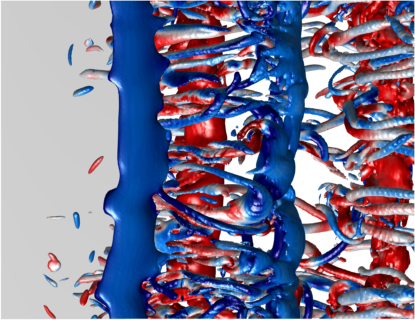}
    }
    \makebox[\textwidth][c]{
	\makebox[0.32\textwidth][c]{(a) $\alpha$\,=\,0$^\circ$}
	\hfill
	\makebox[0.32\textwidth][c]{(b) $\alpha$\,=\,4$^\circ$}
	\hfill
	\makebox[0.32\textwidth][c]{(c) $\alpha$\,=\,6$^\circ$}
	}
	\caption{Trailing edge view of iso-$Q$ surfaces for $\alpha$\,=\,0$^\circ$ (a), 4$^\circ$ (b), and 6$^\circ$ (c).
	Coloring by spanwise vorticity $\omega_z$.}
	\label{fig:q_detail}
\end{figure}

Figure \ref{fig:q_detail} visualizes vortical structures at the trailing edge for $\alpha$\,=\,0$^\circ, 4^\circ, \;\textrm{and} \; 6^\circ$ through iso-surface plots of $Q$ $\equiv$ $(u_{i,i}^2 - u_{i,j}u_{j,i})/2$ \citep{JH95} and coloring according to the spanwise vorticity component $\omega_z$. 
The flow topology is characterized by the shedding of rollers, also called `Strouhal' or `Karman' vortices, that are the primary structures associated with the vortex shedding behind bluff bodies \citep{WS86,Williamson96}. 
Longitudinal braids envelop and connect these spanwise vortices and are common flow structures in wakes \citep{Williamson96a} and shear layers \citep{LC88}. 
K-H instabilities also form within the upper separated shear layer and can drive the formation of additional quasi-two-dimensional vortices as shown in figure \ref{fig:vorticity}(c).

Instabilities in vortices and the resulting three-dimensional flow structures can be driven by different mechanisms.
Given an initial perturbation, a Crow instability \citep{Crow70} can result in the displacement of local vortex segments that in turn leads to self and mutually-induced rotation and straining \citep{LLW16}. 
The axial wavelength of this instability varies and can be large (5-10 times the spacing between the vortex cores) for vortices of equal strength or be as low as of the order of the spacing between the cores for unequal pairs of vortices \citep{CCBRAK08}.
Here, the weaker of the vortices is stretched into so-called $\Omega$ loops that envelop the stronger vortex \citep{LLW16}.

A perturbation of the vortex can also be caused by Kelvin waves of the elliptic flow within the vortex core and is therefore labeled an elliptic instability \citep{Kerswell02}. 
The axial wavelength of the elliptic instability scales with the diameter of the vortex core where the most unstable wavelength has been found to be $\lambda_{z}$\,=\,2$D_{inv}$ \citep{LW98} to $\lambda_{z}$\,=\,3$D_{inv}$ \citep{Williamson96}, with $D_{inv}$ being the invariant streamline of the core (i.e. the streamline with zero radial velocity).

Crow and elliptic instabilities can occur together and develop combined vortical structures that result in the rapid breakdown into turbulent flow \citep{LW98,LLW16}.
The elliptic instability has also been identified by \citet{Williamson96} to be one of two instability modes responsible for the transition to three-dimensional structures in the wake behind a circular cylinder, called \textit{mode A}.
The other instability (called \textit{mode B}) is of shorter wavelength and scales with the hyperbolic flow along the braid shear layer \citep{Williamson96,Williamson96a,LW98a}.
Both instability modes, \textit{mode A} and \textit{mode B}, result in the formation of streamwise vortex loops:
The larger \textit{mode A} deformations of the primary vortices are stretched in the braid shear layer and roll up into a pair of streamwise loops.
The braid region between the Karman vortices, on the other hand, is a hyperbolic shear flow and a three-dimensional instability along the braid shear layer stretches vortex filaments into streamwise pairs of loops at a wavelength that scales with the braid shear layer thickness (\textit{mode B}) \citep{Williamson96a,LW98a}.
These vortex pairs induce sinusoidal velocity perturbations within the upstream braid region that, in a self-sustaining manner, result in the continued formation of the streamwise vortices.
For the wake behind a circular cylinder, \citet{Williamson96} shows that the \textit{mode A} has a wavelength $\lambda_z$ $\approx$ 3$D$ -- 4$D$ and \textit{mode B} a wavelength of $\lambda_z$ $\approx$ 1$D$, where $D$ is the diameter of the cylinder but approximately equals the diameter of the near-wake vortex \citep{Williamson96}.

Because the spanwise wavelength of the three-dimensional instability is known to scale with the characteristic length of the aerodynamic body, the underlying instability mechanisms in the wake of the NACA 65(1)-412 can also be analyzed by the instantaneous vortex topologies. 
\citet{JSS08} compare the spanwise wavelength of streamwise vorticity streaks in the vortex shedding of a LSB over a NACA 0012 airfoil to the diameter of the shed vortex and find that the wavelength of the streamwise vortices corresponds to a \textit{mode B} instability that is observed in the wake of bluff bodies. 

To determine the origin of the three-dimensional instability observed in the wake of the NACA 65(1)-412 (see figure \ref{fig:q_detail}), we compare the diameter of the Karman vortices to the spanwise distribution of the braid vortices.
The average diameter of the near-wake Karman vortices, as measured by the region of positive $Q$, is approximately $D_{v}$\,=\,0.04$c$--0.06$c$.
We note that this may not be an objective measure, but that the value also closely matches the vortex diameter of 0.05$c$ reported for the LSB shedding over the NACA 0012 at a Reynolds number of $5\times10^4$ \citep{JSS08}.

The elliptic \textit{mode A} instability is expected to induce modes with wavelengths of 3$D_{v}$ -- 4$D_{v}$ that corresponds to the range 0.12$c$ $\leq$ $\lambda_{ell}$ $\leq$ 0.24$c$, or 2--4 waves per span.
Accordingly, the hyperbolic \textit{mode B} instability is expected to induce 8--12 waves per span.
Figure \ref{fig:q_detail} shows the emergence of a series of streamwise loop vortices that originate from the vortex roll-up at the trailing edge and, as the flow angle increases from 0$^\circ$ to 4$^\circ$ and 6$^\circ$, combine into clusters of longitudinal coherent vortex structures with a tubular shape (see figure \ref{fig:q_detail}c). 
\citet{Williamson96a} notes that the process of loop generation is self-sustaining as the downstream loops induce a corresponding velocity mode in the upstream vortex.

As first proposed by \citet{JSS08} and later by \citet{MLR13}, the self-sustained turbulence in a LSB is driven by a combination of elliptic instability within the K-H vortices and hyperbolic shear instabilities between them. 
The occurrence of both three-dimensional instability mechanisms can also be inferred in the present airfoil flow.
The lower $\alpha$ (0$^\circ$ -- 6$^\circ$) are distinctly characterized by the small-scale braids with approximately 16 braid loops per span (0.5$c$).  
The loops are generated as pairs so there are two vortices per wavelength, matching the number for a (hyperbolic) \textit{mode B} instability.

While the braids are rather evenly distributed at 0$^\circ$, they start to cluster at 4$^\circ$ and most notably at 6$^\circ$, while the primary Karman vortices increasingly deform compared to the topology at 0$^\circ$ (see figure \ref{fig:q_detail}). 
The clustering of braids and deformation of the spanwise vortex tubes implies the presence of an additional low-frequency elliptic instability mode that is superposed onto the hyperbolic shear instability. 
At 6$^\circ$, the upper (blue) Karman vortex appears to have four waves over the span with the braids also clustering into four groups.
With the expected wavelength of the elliptic instability in the range 0.12$c$ $\leq$ $\lambda$ $\leq$ 0.24$c$, the four waves shown in figure \ref{fig:q_detail}(c) match the expected wavelength and the combination of both, an elliptic \textit{mode A} and a hyperbolic \textit{mode B} instability drives the transition of the near-wake vortex shedding into three-dimensional turbulent structures.

\subsubsection{Three-dimensional instability at higher angles} \label{sec:inst_high_aoa}
\begin{figure}
    \centering
    \makebox[\textwidth][c]{
    \includegraphics[width=0.45\textwidth,trim={0 35pt 0 0},clip]{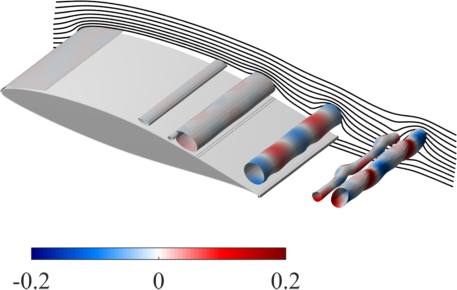}
    \hfill
    \includegraphics[width=0.45\textwidth,trim={0 35pt 0 0},clip]{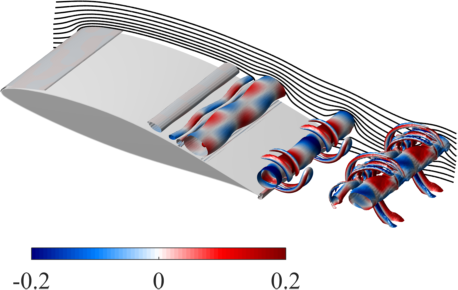}
    }
    \makebox[\textwidth][c]{
	\makebox[0.46\textwidth][c]{(a) $t$\,=\,7.8}
	\hfill
	\makebox[0.48\textwidth][c]{(b) $t$\,=\,8.8}
	}
	\vskip\baselineskip
	\makebox[\textwidth][c]{
    \includegraphics[width=0.45\textwidth]{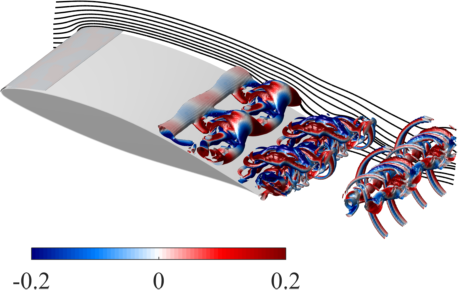}
    \hfill
    \includegraphics[width=0.45\textwidth]{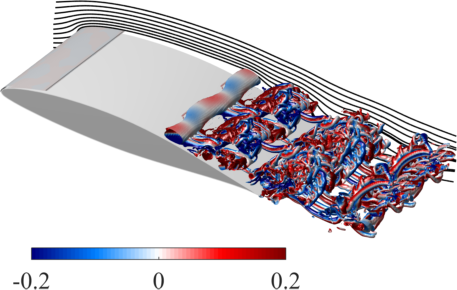}
    }
    \makebox[\textwidth][c]{
	\makebox[0.46\textwidth][c]{(c) $t$\,=\,9.8}
	\hfill
	\makebox[0.48\textwidth][c]{(d) $t$\,=\,10.8}
	}
	\caption{Iso-surfaces of $Q$-criterion (level: 100) colored by transverse velocity from $t$\,=\,7.8 (a) to $t$\,=\,10.8 (d).
	Instantaneous streamlines in black.
	$\alpha$\,=\,7$^\circ$}
	\label{fig:temp_transition_7deg}
\end{figure}

The flow at higher $\alpha$ is characterized by the formation of a closed and thin LSB and subsequent transition to turbulence. 
The evolution of the instabilities at $\alpha$\,=\,7$^\circ$ are particularly insightful in the mechanism by which the transition takes place.
A series of snapshots from $t$\,=\,7.8 to $t$\,=\,10.8 in figure \ref{fig:temp_transition_7deg} shows that the flow transitions from a nominally two-dimensional state to three-dimensional turbulent structures.
In the figure, iso-surfaces of $Q$ colored by the spanwise velocity component $w$ show the emergence of three-dimensional modes within the vortices.
Note that no perturbation or forcing is added to the flow and the transition occurs naturally. 
The flow is initially quasi-two-dimensional and characterized by the formation of spanwise Kelvin-Helmholtz vortices from the separating shear layer at mid-chord (see figure \ref{fig:temp_transition_7deg}a).
A well-defined low-frequency perturbation mode along the vortex at the trailing edge, as well as within the advected vortex pair downstream in the wake, is made visible by the $w$-velocity coloring at $t$\,=\,7.8.
The smaller of the two downstream vortices has attained noticeable bends whereas the larger tube is only weakly bulging along the span. 
One convective time unit later, the snapshot at $t$\,=\,8.8 (figure \ref{fig:temp_transition_7deg}b) shows that the three-dimensional instability grows over time and result in an increasing spanwise velocity component and stronger bending of the vortex tubes over the airfoil.
Downstream, the continued deformation of the vortex pair has resulted in the smaller vortex being stretched into two pairs of streamwise loops that resemble hairpin-like structures and envelop the larger spanwise vortex. 

The stretching and wrapping of initially spanwise vortices into pairs of rollers and streamwise $\Omega$-shaped loops is driven by a Crow instability for counter-rotating vortex pairs of unequal strength, as described by \citet{LLW16} and visualized by \citet{CCBRAK08}. 
The Crow instability scales on the separation $b$ between the vortex cores, where the axial wavelength is between 6$b$ $\leq$ $\lambda_z$ $\leq$ 10$b$ for vortices of equal strength but can be as low as $\lambda_z$\,$\approx$\,$b$ for unequal pairs \citep{LLW16}. 
At $t$ = 7.8 (figure \ref{fig:temp_transition_7deg}a), the downstream vortex pair has a separation of $b/c$\,=\,0.09, and the axial wavelength of $\lambda_z$\,=\,0.25 is well within the limits of the Crow instability for unequal vortex pairs, as $\lambda_z$\,=\,2.8$b$.
Because the Crow instability acts on pairs of counter-rotating vortices, the near-wake structures are initially not connected through braids for $t$ $\leq$ 9.8, but increase the instability of the upstream vortex pair through induction of the perturbed velocity components.
Only at later times ($t$ $\geq$ 10.8), when the flow becomes more turbulent and the perturbation has moved further upstream, a set of braid vortices link consecutive Kelvin-Helmholtz rollers and establish a continuous wake of three-dimensional, turbulent motion (see figure \ref{fig:temp_transition_7deg}c--d).

While the Crow instability describes the $\Omega$-loop formation of the downstream vortex pairs, the initial perturbation is caused by different instability mechanisms.
The diameter of the K-H vortices at $t$\,=\,7.8, measured by the region of positive $Q$, is in the range $D_{KH}$\,=\,0.06$c$--0.09$c$.
An elliptic instability is therefore expected to occur at wavelengths between 0.18$c$ $\leq$ $\lambda_z$ $\leq$ 0.36$c$ and hyperbolic instabilities between 0.06$c$ $\leq$ $\lambda_z$ $\leq$ 0.09$c$ if we apply the same scaling arguments as above and employed by \citet{JSS08}.
Because the observed spanwise wavelength of $\lambda_z$\,=\,0.25$c$ lies within the range 0.18$c$ $\leq$ $\lambda_z$ $\leq$ 0.36$c$, it is likely that the initial perturbation of the spanwise vortices is caused by an elliptic instability within the vortex cores, with the bending mode of the upstream vortices over the airfoil matching the topology presented by \citet{LLW16}.
The Crow and the elliptic instability then cause the formation of the $\Omega$-loops in the downstream vortex pairs.

\begin{figure}
    \makebox[\textwidth][c]{
    \includegraphics[width=0.3\textwidth]{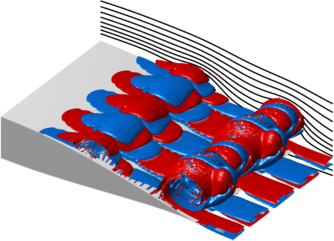}
    \hfill
    \includegraphics[width=0.3\textwidth]{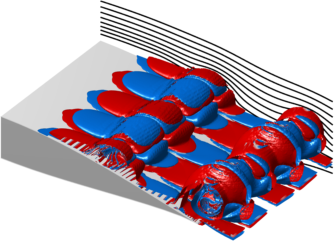}
    \hfill
    \includegraphics[width=0.3\textwidth]{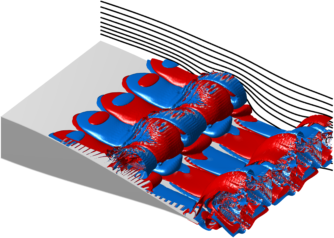}
    }
    \makebox[\textwidth][c]{
    \makebox[0.3\textwidth][c]{(a) $t$\,=\,8.7}
	\hfill
	\makebox[0.3\textwidth][c]{(b) $t$\,=\,8.8}
    \hfill
	\makebox[0.3\textwidth][c]{(c) $t$\,=\,8.9}
	}
	\vskip\baselineskip
	\makebox[\textwidth][c]{
    \includegraphics[width=0.3\textwidth]{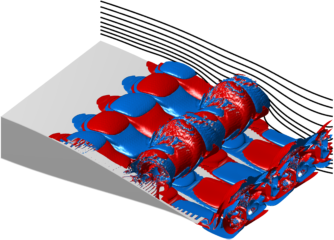}
    \hfill
    \includegraphics[width=0.3\textwidth]{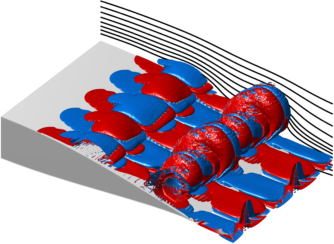}
    \hfill
    \includegraphics[width=0.3\textwidth]{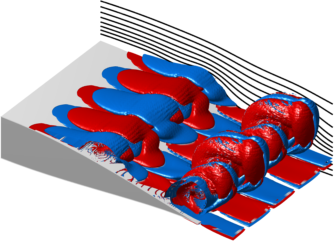}
    }
    \makebox[\textwidth][c]{
    \makebox[0.3\textwidth][c]{(d) $t$\,=\,9.0}
	\hfill
	\makebox[0.3\textwidth][c]{(e) $t$\,=\,9.1}
    \hfill
	\makebox[0.3\textwidth][c]{(f) $t$\,=\,9.2}
	}
	\caption{Iso-surfaces of the streamwise vorticity $+\omega_x$ (red) and $-\omega_x$ (blue) for a level $|\omega_x|$\,=\,1.
	Rear section of the airfoil shown between $x/c$\,=\,0.4 and $x/c$\,=\,1.1 for $t$\,=\,8.7 (a) to $t$\,=\,9.2 (f).
	Instantaneous streamlines in black.
	$\alpha$\,=\,7$^\circ$}
	\label{fig:omega_x_7deg}
\end{figure}

To monitor the development of the three-dimensional instability over the airfoil, we consider a time series of streamwise vorticity iso-surfaces $|\omega_x|$\,=\,1 in figure \ref{fig:omega_x_7deg}.
The vortical structures identified in this way only relate to rotating fluid along the streamwise axis and hence detect three-dimensional flow patterns without being obscured by the dominating two-dimensional topology.
The $\omega_x$ surfaces in figure \ref{fig:omega_x_7deg} are flat layers that are stacked on the airfoil surface and lifted off by passing spanwise vortices.
A similar topology of streamwise vorticity surfaces has been reported by \citet{JSS08} for the LSB shedding over a NACA 0012 and by \citet{SJD20} for the instability of a laminar separation bubble under a solitary wave. 

The time series in figure \ref{fig:omega_x_7deg} illustrates that streamwise vorticity is present within a thin layer at the trailing edge in a slender region of recirculating fluid (bubble height = 0.026$c$).
As the next vortex forms, it is bent by the existing rotating flow near the airfoil surface and, as the vortex line tilts, induces streamwise vorticity itself, thereby amplifying the spanwise velocity component. 
The cycle repeats until the bending of the spanwise vortices towards the trailing edge at a location where patches of positive and negative $\omega_x$ induce a wall-normal upwelling fluid movement (see figure \ref{fig:omega_x_7deg}f).
The iso-surfaces of $Q$ shown in figure \ref{fig:temp_transition_7deg}(b-c) indicate that this asymmetric bending is associated with the generation of loop vortices that eventually grow into the enveloping braids at later times.

According to \citet{AS00} and \citet{JSS08}, laminar separation bubbles require a reverse flow velocity $U_R$ of 15\% to 20\% of the local boundary layer edge velocity to develop an absolute stability, while \citet{Theofilis11} found that lower levels of reverse flow of $\mathcal{O}$(10\%) are sufficient to sustain a three-dimensional instability mode, which is also confirmed by \citet{MLR13}.
For the present airfoil flow at $\alpha$\,=\,7$^\circ$, the maximum level of the reverse velocity component is $U_R$\,=\,10.9\% (10.3\% at 8$^\circ$ and 11.5\% at 10$^\circ$), relative to the local velocity magnitude at the boundary layer edge. 
These values therefore seem to not qualify for an absolute instability mode, but are sufficient for a three-dimensional mode.
The temporal and spatial growth of three-dimensional perturbations (figure \ref{fig:temp_transition_7deg} and \ref{fig:omega_x_7deg}) distinctly shows that turbulence is first induced through the amplification of the three-dimensional flow that originates from an elliptic instability.
Three-dimensional flow is then amplified within the braid shear layer through smaller-scale loop vortices.

\vskip\baselineskip
\paragraph{\centerline{\textit{Trailing-edge LSB and the formation of turbulent puffs}}}
\par
\begin{figure}
    \centering
    \makebox[\textwidth][c]{
    \includegraphics[width=0.51\textwidth,trim={0 60pt 0 0},clip]{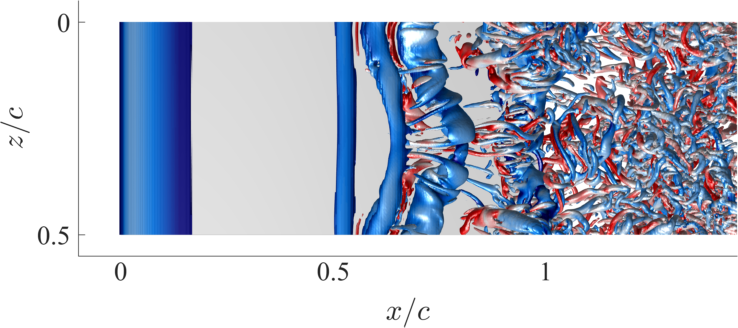}
    \hfill
    \includegraphics[width=0.45\textwidth,trim={70pt 60pt 0 0},clip]{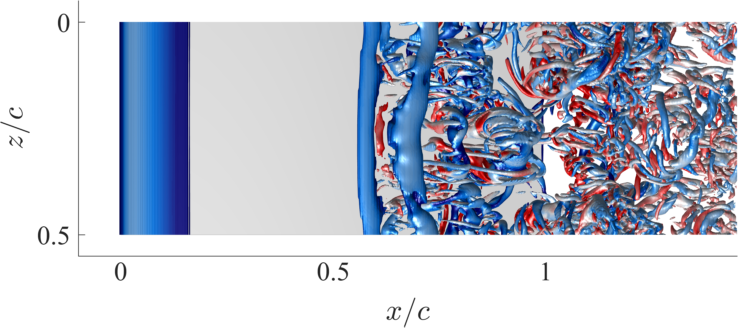}
    }
    \makebox[\textwidth][c]{
	\makebox[0.51\textwidth][c]{(a) $t$\,=\,16.1}
	\hfill
	\makebox[0.45\textwidth][c]{(b) $t$\,=\,16.4}
	}
	\vskip\baselineskip
	\makebox[\textwidth][c]{
    \includegraphics[width=0.51\textwidth,trim={0 0 0 0},clip]{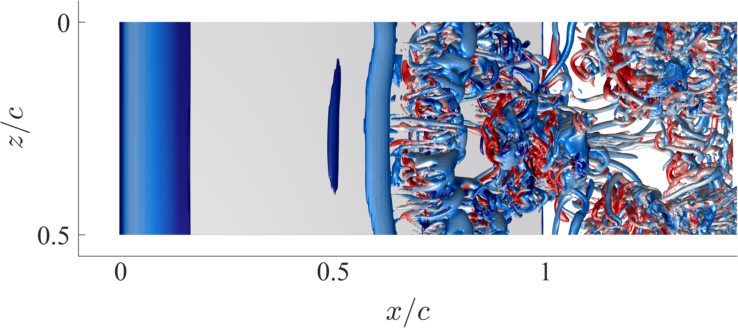}
    \hfill
    \includegraphics[width=0.45\textwidth,trim={70pt 0 0 0},clip]{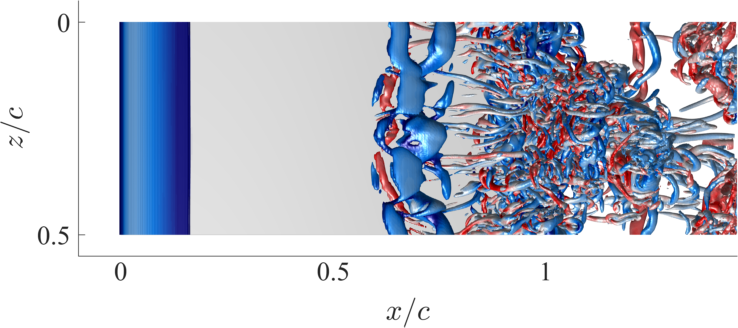}
    }
    \makebox[\textwidth][c]{
	\makebox[0.51\textwidth][c]{(c) $t$\,=\,16.7}
	\hfill
	\makebox[0.45\textwidth][c]{(d) $t$\,=\,17.0}
	}
	\caption{Iso-$Q$ surfaces at $t$\,=\,16.3 (a), 16.5 (b), 16.7 (c), and 16.9 (d).
	Coloring by spanwise vorticity $\omega_z$.}
	\label{fig:puffs}
\end{figure}

Low-frequency spanwise modes within the Kelvin-Helmholtz vortices also are also present at later times and drive the formation of large-scale turbulent structures.
At $t$ $\approx$ 16, the bending of spanwise vortices results in a horseshoe-type vortex structure that extends over the span of the airfoil and bursts into a turbulent cloud or ``puff''.
The process is outlined in figure \ref{fig:puffs} and starts with the bending of the K-H vortex along the span in streamwise direction (figure \ref{fig:puffs}a).
While the nominally two-dimensional flow structure breaks down and forms smaller-scale loop vortices (figure \ref{fig:puffs}b), a large-scale coherent horseshoe-shaped vortex system develops (figure \ref{fig:puffs}c), which then bursts and sheds off at the trailing edge (figure \ref{fig:puffs}d).
Because the wavelength of this mode ($\lambda_z$\,=\,0.5$c$) is twice the expected wavelength of an elliptic instability for the given vortex diameter ($D_{v}$\,$\approx$\,0.06), the large-scale deformation of the K-H vortex in figure \ref{fig:puffs}(a) is at least partly the result of the interaction with downstream vortices through induction of velocity in the upstream vortices (Crow-type instability).
The series in figure \ref{fig:puffs}(a--d) shows how the interplay of three-dimensional large scale instability modes and the small-scale loop vortices results in the formation and bursting of such large coherent turbulent structures.

Given that these structures are of the same size as the spanwise extent of the computational domain, an open question remains whether even larger vortex systems exist if a larger span is chosen. 
The study on the effect of the spanwise domain length is, however, not subject of this paper and we will therefore leave the answer to this question to future research. 

\vskip\baselineskip
\paragraph{\centerline{\textit{Leading-edge LSB}}}
\par
\begin{figure}
    \centering
    \makebox[\textwidth][c]{
    \includegraphics[width=0.51\textwidth,trim={0 0 0 0},clip]{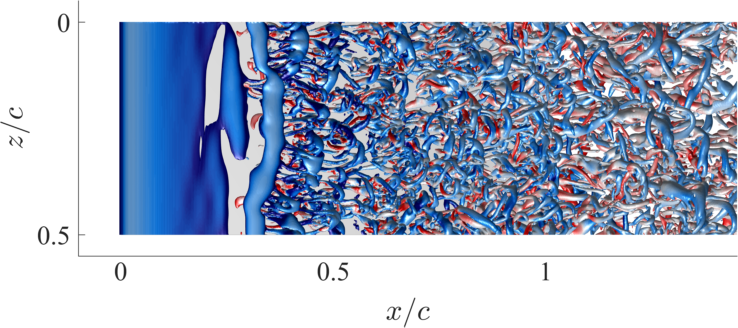}
    \hfill
    \includegraphics[width=0.45\textwidth,trim={70pt 0 0 0},clip]{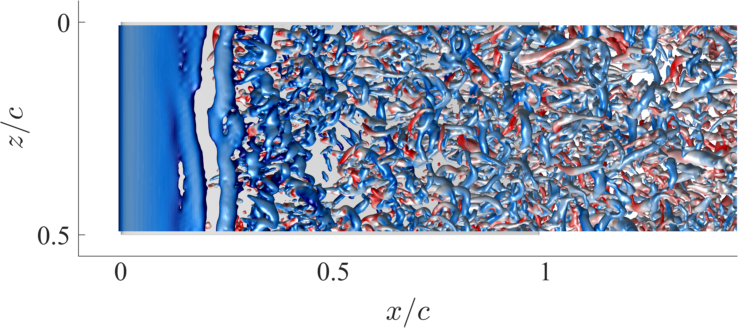}
    }
    \makebox[\textwidth][c]{
	\makebox[0.51\textwidth][c]{(a) $\alpha$\,=\,8$^\circ$}
	\hfill
	\makebox[0.45\textwidth][c]{(b) $\alpha$\,=\,10$^\circ$}
	}
	\caption{Iso-$Q$ surfaces for $\alpha$\,=\,8$^\circ$ (a) and 10$^\circ$.
	Coloring by spanwise vorticity $\omega_z$.}
	\label{fig:le_lsb}
\end{figure}

As the impact of the elliptic instability grows with increasing $\alpha$ from 0$^\circ$ to 7$^\circ$, the flow structures are less distinct in the the leading-edge LSB at $\alpha$\,=\,8$^\circ$. 
Figure \ref{fig:le_lsb} shows the vortices over the suction side of the airfoil at $\alpha$\,=\,8$^\circ$ and 10$^\circ$.
The separated laminar shear layer sheds K-H vortices which break down before mid-chord and loose their spatial coherence as the flow transitions to a turbulent boundary layer. 
Non-zero spanwise velocity and three-dimensional vortex structures are also present throughout the upstream, laminar section of the bubble and result in the non-uniform generation of the K-H vortices with vortex displacements and re-connections visible.
Low-frequency deformations of the K-H vortices and the generation of hairpin loops within the shear region point to the same instability mechanisms observed at lower $\alpha$, but their occurrence is less pronounced and masked by the rapid transition to turbulence.

\subsubsection{Effect on the wake topology}
\begin{figure}
    \centering
    \makebox[\textwidth][c]{
    \makebox[0.03\textwidth][l]{}
    \hfill
    \makebox[0.45\textwidth][c]{\includegraphics[width=0.25\textwidth]{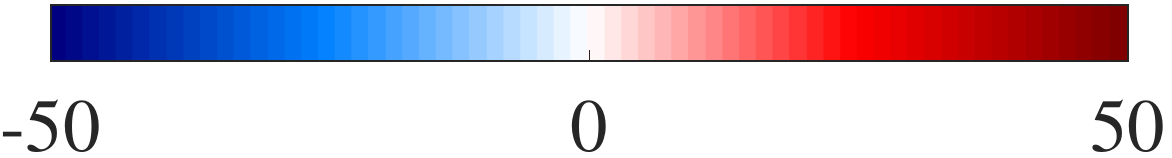}}
    \hfill
    \makebox[0.45\textwidth][c]{\includegraphics[width=0.25\textwidth]{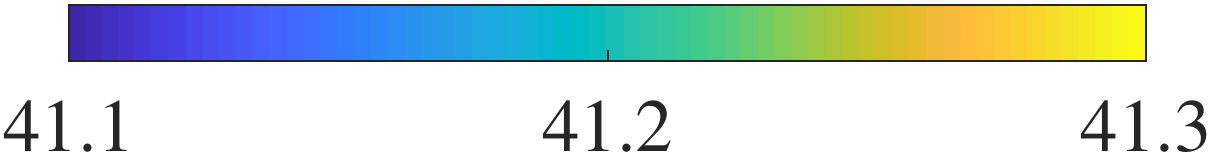}}
    }
    \makebox[\textwidth][c]{
    \makebox[0.03\textwidth][l]{\rotatebox{90}{\hspace{8pt} $\alpha$\,=\,0$^\circ$}}
    \hfill
    \includegraphics[width=0.45\textwidth,trim={0 25pt 55pt 0},clip]{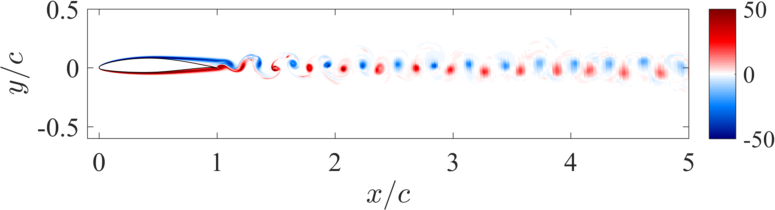}
    \hfill
    \includegraphics[width=0.45\textwidth,trim={0 25pt 60pt 0},clip]{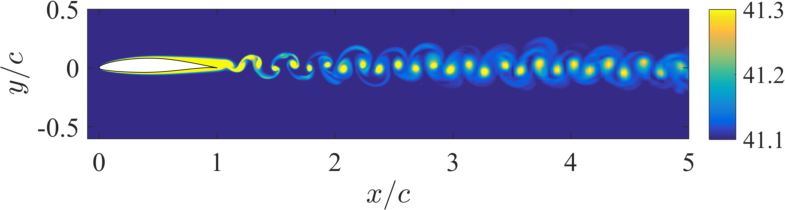}
    }
	\vskip\baselineskip
	\makebox[\textwidth][c]{
	\makebox[0.03\textwidth][l]{\rotatebox{90}{\hspace{8pt} $\alpha$\,=\,4$^\circ$}}
    \hfill
    \includegraphics[width=0.45\textwidth,trim={0 25pt 55pt 0},clip]{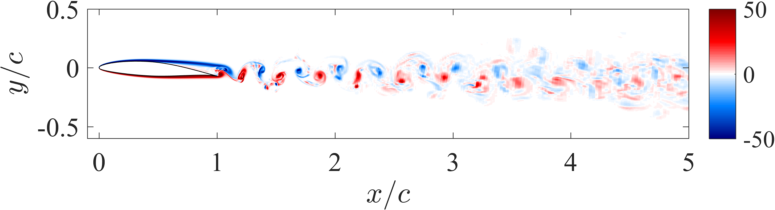}
    \hfill
    \includegraphics[width=0.45\textwidth,trim={0 25pt 60pt 0},clip]{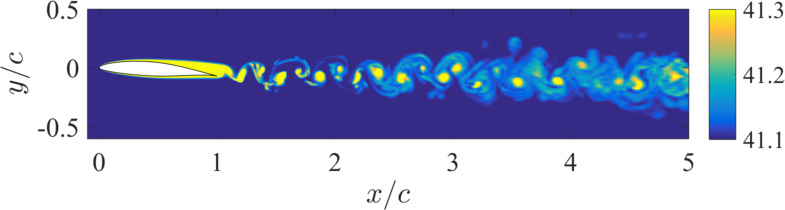}
    }
	\vskip\baselineskip
	\makebox[\textwidth][c]{
	\makebox[0.03\textwidth][l]{\rotatebox{90}{\hspace{8pt} $\alpha$\,=\,7$^\circ$}}
    \hfill
    \includegraphics[width=0.45\textwidth,trim={0 25pt 55pt 0},clip]{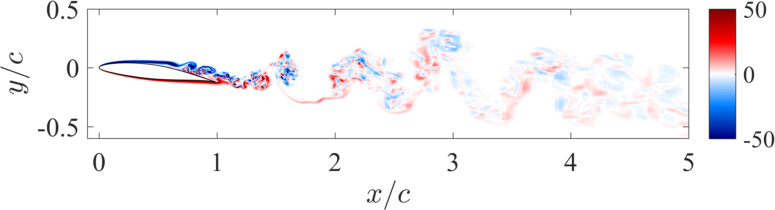}
    \hfill
    \includegraphics[width=0.45\textwidth,trim={0 25pt 60pt 0},clip]{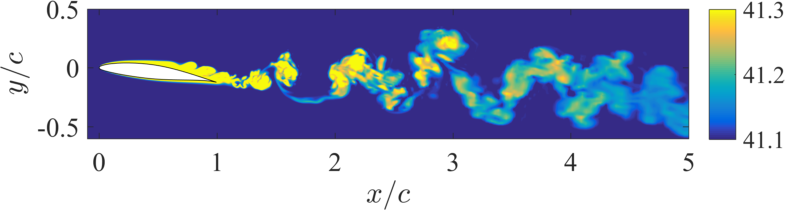}
    }
    \vskip\baselineskip
	\makebox[\textwidth][c]{
	\makebox[0.03\textwidth][l]{\rotatebox{90}{\hspace{8pt} $\alpha$\,=\,8$^\circ$}}
    \hfill
    \includegraphics[width=0.45\textwidth,trim={0 0 55pt 0},clip]{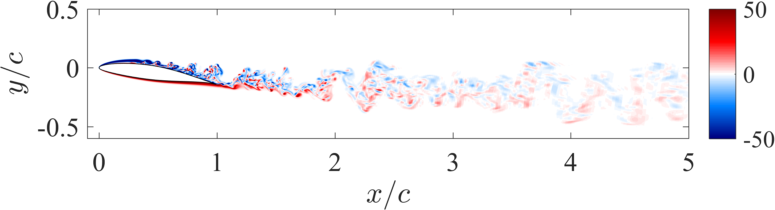}
    \hfill
    \includegraphics[width=0.45\textwidth,trim={0 0 60pt 0},clip]{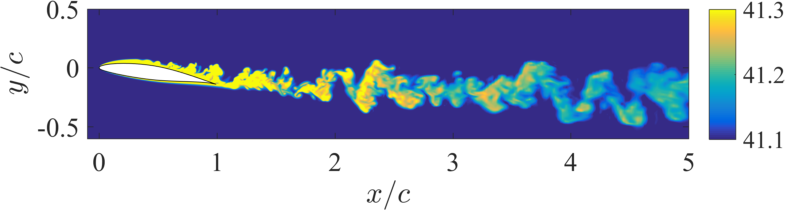}
    }
    \makebox[\textwidth][c]{
    \makebox[0.03\textwidth][l]{}
    \hfill
	\makebox[0.45\textwidth][c]{(a) Vorticity $\omega_z$}
	\hfill
	\makebox[0.45\textwidth][c]{(b) Specific entropy $s$}
	}
	\caption{Instantaneous snapshots of the vorticity $\omega_z$ (a) and the specific entropy $s$\,=\,$\ln(p/\rho^\gamma)/(\gamma(\gamma-1)M_f^2)$ (b) along a slice at $z/c$\,=\,0.025 for $\alpha$\,=\,0$^\circ$, 4$^\circ$, 7$^\circ$, and 8$^\circ$ (top to bottom).}
	\label{fig:wake}
\end{figure}

The spatial development of the vortex street in the airfoil wake is visualized in figure \ref{fig:wake} for $\alpha$\,=\,0$^\circ$, 4$^\circ$, 7$^\circ$, and 8$^\circ$, where contours of the instantaneous, spanwise vorticity component $\omega_z$ are plotted in (a) and contours of the specific entropy $s$ in (b).
Vorticity is generated at the airfoil wall and transported downstream,
where stretching of vortex filaments, mixing, and diffusion results in the decay of vorticity and the spreading of the street.
We also visualize this transport of fluid from the airfoil into the wake through contours of specific entropy,  $s$\,=\,$\ln(p/\rho^\gamma)/(\gamma(\gamma-1)M_f^2)$, which follows a scalar transport equation with a production term for irreversible processes \citep{SA08,CJDAM16}.
Entropy is generated by viscous dissipation in the wall boundary layer \citep{CJDAM16}, and then transported into the wake and consequently highlights the associated flow topology. 

At $\alpha$\,=\,0$^\circ$, the Karman vortices remain aligned in a narrow vortex street throughout the wake (figure \ref{fig:wake}a).
The mixing rate with the surrounding fluid is low as the higher levels of entropy generated in the shear layer around the airfoil remain confined to an area of $\pm$0.25$c$ until at least 5 chord lengths behind the airfoil (figure \ref{fig:wake}b). 
The low mixing and spreading rates of the vortex street at $\alpha$\,=\,0$^\circ$ show how the flow topology in the far wake is defined by the organized near-wake structures at the airfoil trailing edge shown in figure \ref{fig:q_detail}(a).

At $\alpha$\,=\,4$^\circ$, the vortex street in the airfoil wake spreads at a higher rate than at 0$^\circ$ and the vorticity and entropy contours appear diffused two chord lengths downstream from the trailing edge, marking  the transition of the flow to turbulence and the accompanying entrainment of surrounding fluid.
As is the case at $\alpha$\,=\,0$^\circ$, the baseline topology in the far wake is defined by the near-wake coherent structures (figure \ref{fig:q_detail}b), but
the transition and break up of the laminar vortex structures in the wake at 4$^\circ$ confirm the existence of additional flow instabilities that are not strongly influential at 0$^\circ$.
As noted before, these instabilities include a three-dimensional mode within the Karman vortices and K-H instabilities within the separated shear layer on the suction side. 
The transition of the wake flow can therefore be attributed to a combination of these modes and their interaction with the pressure-side Karman vortices.

With $\alpha$ increasing to 7$^\circ$ and 8$^\circ$, the wake structures become more irregular as they are governed by turbulent motion following the flow transition upstream of the trailing edge and the development of wall-bounded turbulence. 
In case the LSB forms at the rear side of the airfoil ($\alpha$\,=\,7$^\circ$), the interaction of suction side (K-H shedding) and pressure side (Karman shedding) instabilities is most pronounced and results in a low-frequency vortex street with the vortices forming large-scale turbulent puffs as they shed downstream into the wake.
The vertical momentum induced by the turbulent puffs increases the wake spread and leads to regions of high vorticity followed by quiescent fluid in contrast to the more uniform wake topology at lower $\alpha$ (0$^\circ$ and 4$^\circ$).
While the Karman vortices can still be distinguished several chord lengths downstream from the trailing edge by local maxima in the vorticity and entropy contours at 0$^\circ$ and 4$^\circ$, the structures at 7$^\circ$ appear more diffused and point to a fully turbulent wake. 

At $\alpha$\,=\,8$^\circ$, the LSB is located at the leading edge and the K-H vortices have transitioned at mid-chord into a turbulent boundary layer (figure \ref{fig:le_lsb}a).
The turbulent breakdown of the larger vortices into smaller structures results in a more isotropic flow over the suction side than at 7$^\circ$ and leads to a narrower wake because the continuous shedding of vortices disrupts the roll-up of the pressure-side shear layer into a large trailing-edge vortex.
Consequently, the wake topology is no longer governed by the large-scale puffs that exist at $\alpha$\,=\,7$^\circ$, but shows a turbulent vortex street, that,
despite the shift in flow regimes, still shows the footprint of the interaction of Karman and K-H shedding but on a smaller scale. 


\bigskip

The wide range and abrupt changes of flow topology driven by the combination of different two and three-dimensional instabilities in the flow over a NACA 65(1)-412 airfoil at a single Reynolds number shows the sensitive nature of the low-Reynolds number aerodynamic response to small changes in free-stream conditions.


\subsection{Aerodynamic forces}
\begin{figure}
	\makebox[\textwidth][c]{
    \includegraphics[width=0.35\textwidth]{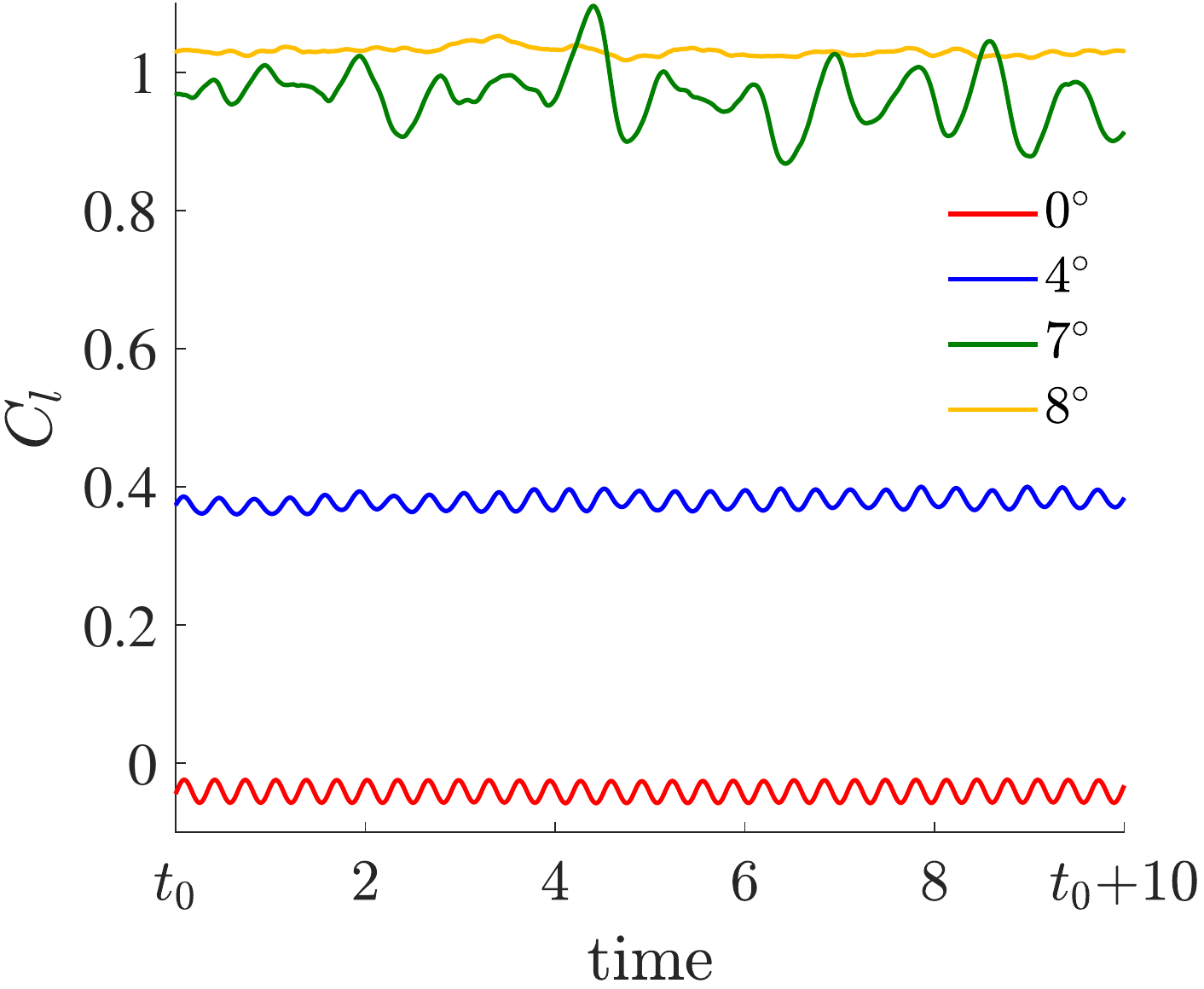}
    \hfill
    \includegraphics[width=0.35\textwidth]{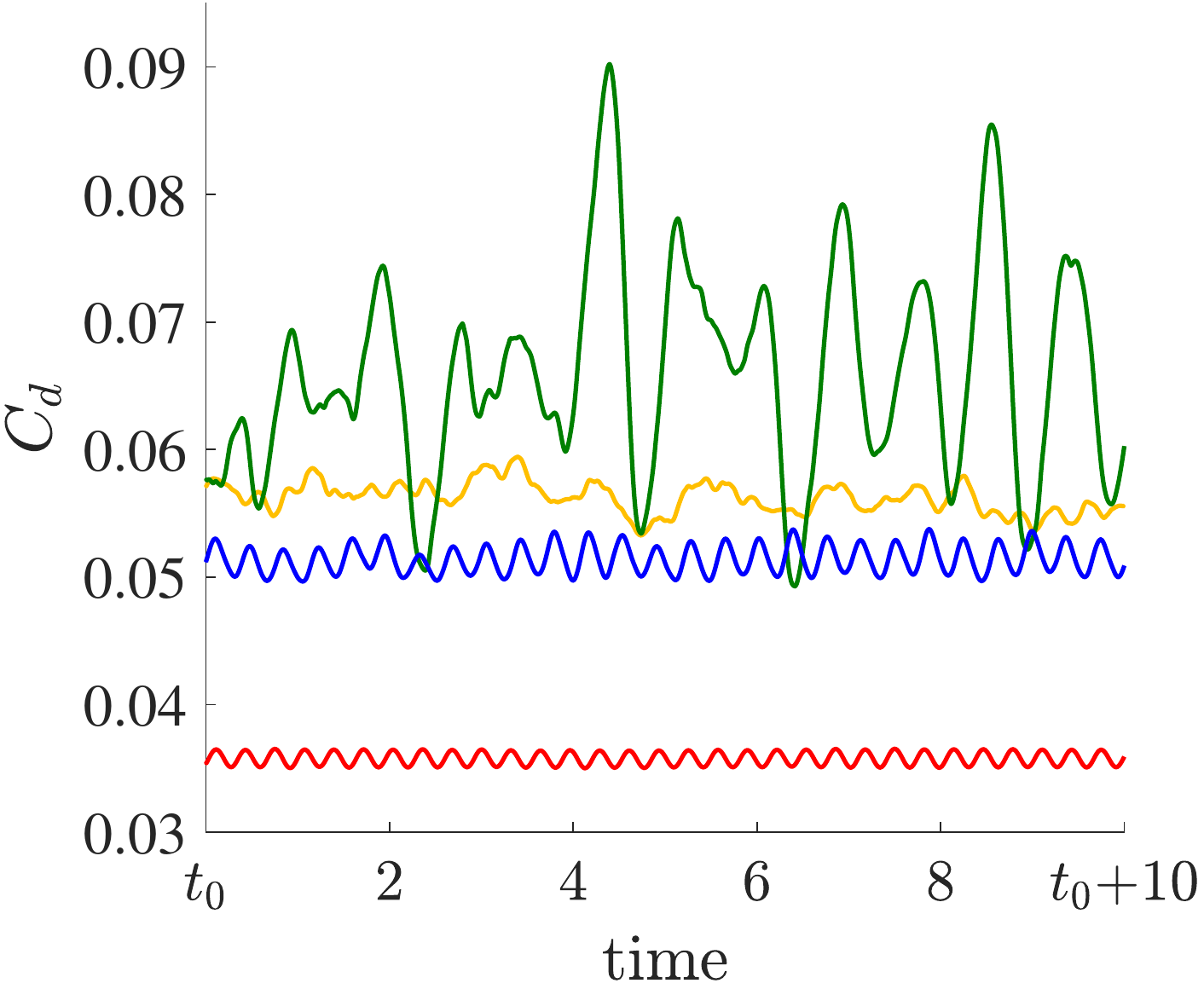}
    \hfill
    \includegraphics[width=0.25\textwidth]{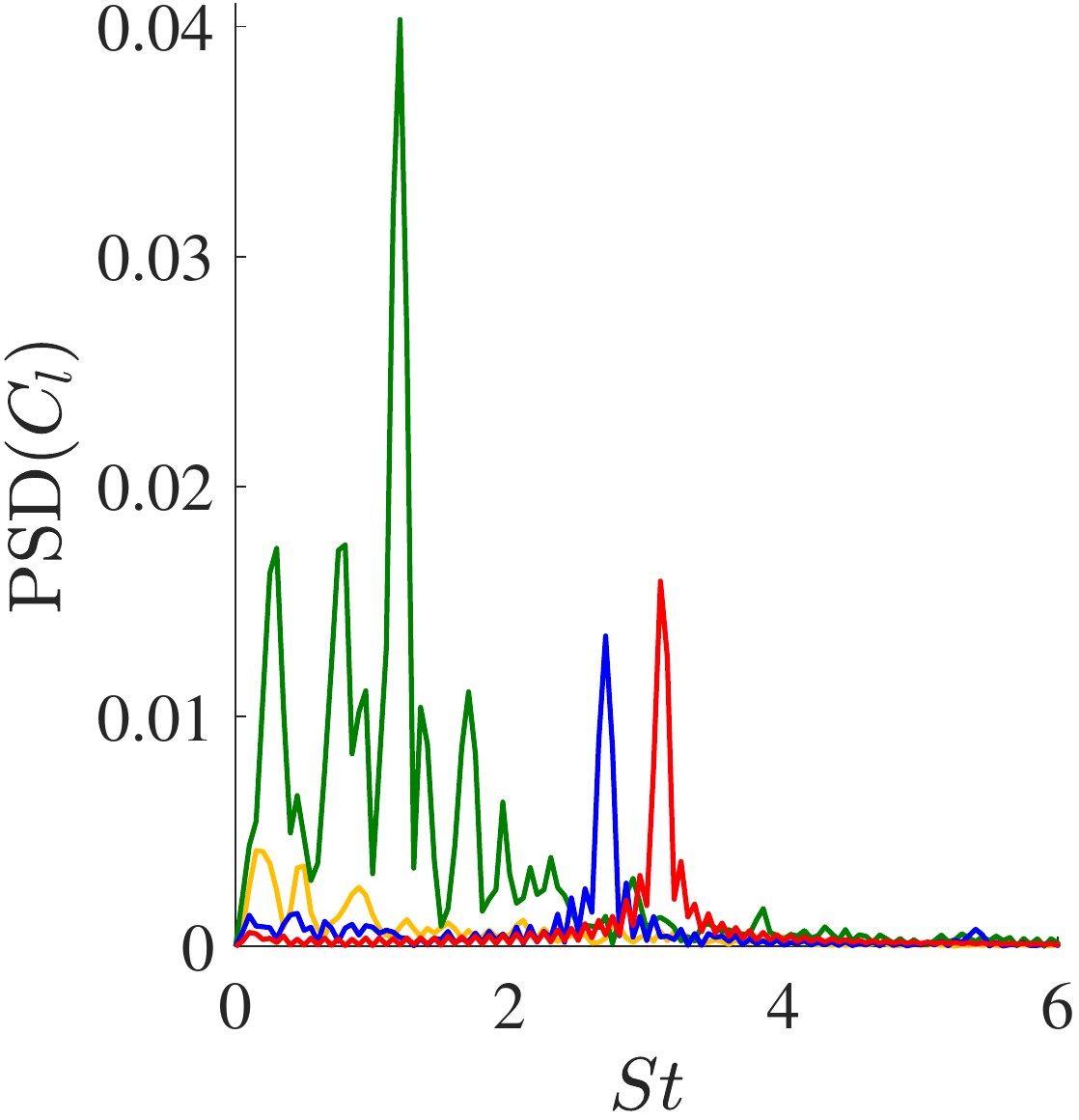}
    }
    \makebox[\textwidth][c]{
    \makebox[0.35\textwidth][c]{(a) Lift coefficient}
    \hfill
    \makebox[0.35\textwidth][c]{(b) Drag coefficient}
    \hfill
    \makebox[0.25\textwidth][c]{(c) Lift spectrum}
    }
	\caption{Lift (a) and drag (b) coefficients over time. 
	(c) Frequency spectrum of the lift coefficient. For $\alpha$\,=\,0$^\circ$, 4$^\circ$, 7$^\circ$, and 8$^\circ$.}
	\label{fig:clcd}
\end{figure}

The changes in flow patterns at different angles of attack are accompanied by changes in the integrated forces on the airfoil.
Figure \ref{fig:clcd} shows the time history of the lift and drag coefficients, as well as the corresponding frequency spectrum for the flow at $\alpha$\,=\,0$^\circ$, 4$^\circ$, 7$^\circ$, and 8$^\circ$. 
The 10$^\circ$ data is omitted here to not overload the plots.
At the lower $\alpha$ (0$^\circ$ and 4$^\circ$), the oscillations of the forces are regular and driven by the shedding of the Karman vortices from the laminar shear layers at the trailing edge that result in the narrow wakes presented in figure \ref{fig:wake}.
Note that the time unit is scaled with the free-stream velocity $U_\infty$ and the chord-length of the airfoil $c$, so the corresponding frequency is a Strouhal number \textit{St}\,=\,$fc/U_\infty$.
At $\alpha$\,=\,0$^\circ$, the lift coefficient yields a single peak at a Strouhal number of \textit{St}\,=\,3.1, indicating that no instability other than the Karman shedding is predominantly driving the flow.
The increasingly unstable separated shear layer at higher angles results in additional low-frequency content of the lift spectrum at $\alpha$\,=\,4$^\circ$, which still shows a dominant peak (at \textit{St}\,=\,2.7), but at a reduced height (by 15\%) caused by the energy transfer to other frequencies.
The dominant mode is therefore still driven by the Karman shedding and the periodic formation of a strong trailing-edge vortex, but the K-H instability within the top shear layer induces waves at different frequencies that lead to additional peaks in the spectrum.

At higher $\alpha$\,=\,7$^\circ$ and 8$^\circ$, the force oscillations are irregular and strongly dependent on the location of the LSB.
When the LSB forms at the trailing edge of the airfoil ($\alpha$\,=\,7$^\circ$), the interaction of suction side (K-H shedding) and pressure side (Karman shedding) instabilities results in large-scale turbulent bursts leading to large-amplitude oscillations of the aerodynamic forces (figure \ref{fig:clcd}a--b).
The corresponding lift spectrum shows a dominant peak at \textit{St}\,=\,1.2 of more than twice the amplitude as the peaks at $\alpha$\,=\,0$^\circ$ or 4$^\circ$, as well as low-frequency peaks at \textit{St}\,=\,0.3 and 0.8 that have the same or higher amplitudes as the peak at $\alpha$\,=\,4$^\circ$ (figure \ref{fig:clcd}c).
The high energy content of these fluctuations results from the interaction of both, Karman (pressure side) and K-H (suction side) instabilities, whereas the shedding at 0$^\circ$ and 4$^\circ$ is mainly driven by the Karman instability induced by the roll-up of the pressure side shear layer.
In case of a leading-edge LSB ($\alpha$\,=\,8$^\circ$), the time-averaged lift force ($\bar{C}_l$\,=\,1.03) is higher than at the other $\alpha$, but the amplitude of the oscillations is reduced (figure \ref{fig:clcd}a).
The lift spectrum (figure \ref{fig:clcd}c) does not show a dominant shedding frequency, but has several low-frequency peaks at only a third of the amplitudes computed for the other cases.
The low-amplitude oscillations are caused by the break up of the K-H vortices and the transition to a turbulent boundary layer at mid-chord, which increases the isotropy of the flow structures on the suction side and interrupts the pressure-side shear layer from rolling into a large vortex.

\begin{figure}
	\makebox[\textwidth][c]{
    \includegraphics[width=0.45\textwidth]{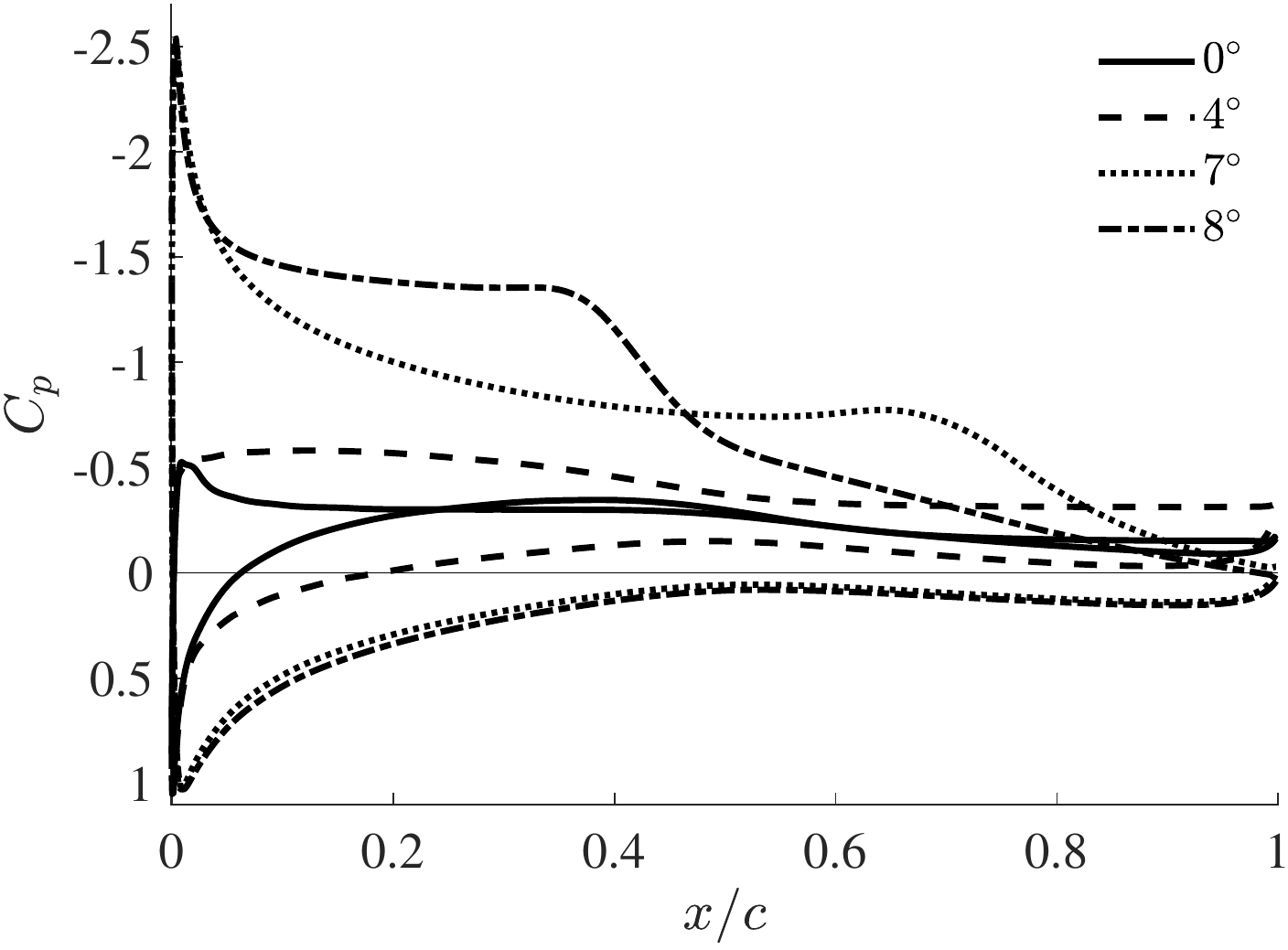}
    \hfill
    \includegraphics[width=0.45\textwidth]{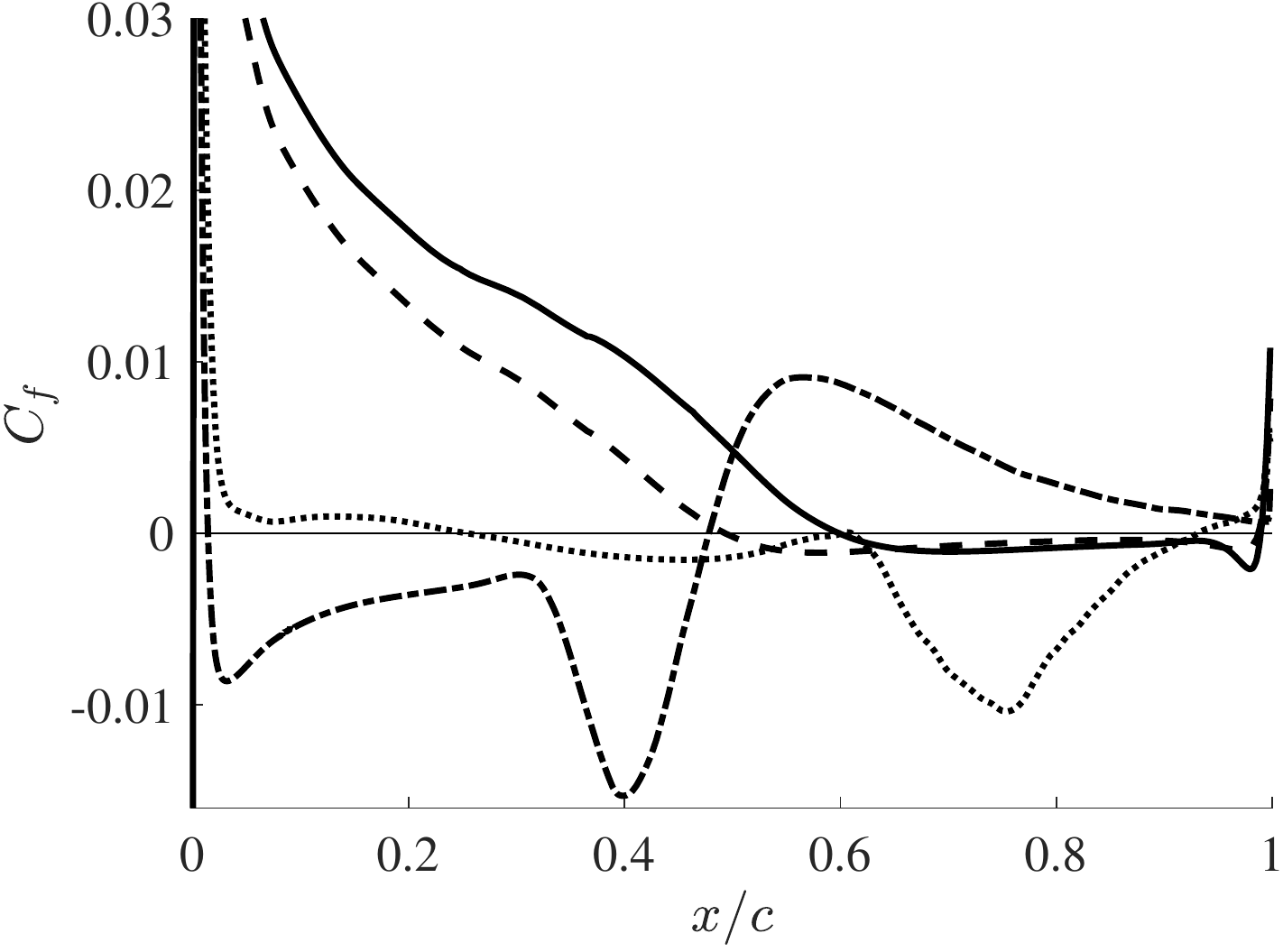}
    }
    \makebox[\textwidth][c]{
    \makebox[0.45\textwidth][c]{(a) Pressure coefficient}
    \hfill
    \makebox[0.45\textwidth][c]{(b) Skin friction coefficient}
    }
	\caption{Time and spanwise averaged pressure (upper and lower side) and skin friction (upper side) coefficients for $\alpha$\,=\,0$^\circ$, 4$^\circ$, 7$^\circ$, and 8$^\circ$.}
	\label{fig:cpcf}
\end{figure}

The time-averaged profiles of the pressure and skin friction coefficients for the flows at $\alpha$\,=\,0$^\circ$, 4$^\circ$, 7$^\circ$, and 8$^\circ$ are plotted over the chord length of the airfoil in figure \ref{fig:cpcf}.
At lower angles ($\alpha$\,$\leq$\,6$^\circ$), the suction peak and the resulting adverse pressure gradient are small and the skin friction coefficient gradually decreases until it becomes negative at the fixed separation point, as identified at the time-averaged zero-skin friction point in \citet{Haller04}, at $x_{s,0^\circ}$\,=\,0.6 and $x_{s,4^\circ}$\,=\,0.49 (figure \ref{fig:cpcf}b). 
Downstream of the separation location, the surface pressure remains constant and does not recover the free-stream value at the trailing edge. 

At higher $\alpha$\,$\geq$\,7$^\circ$, the suction peak increases from $C_p$\,=\,-0.6 at $\alpha$\,=\,4$^\circ$ to  $C_p$\,=\,-2.5 at 7$^\circ$ and 8$^\circ$ and steepens the adverse pressure gradient and promotes flow separation further upstream at $x_{s,7^\circ}$\,=\,0.26 and $x_{s,8^\circ}$\,=\,0.02. 
The skin friction coefficient at 8$^\circ$ shows a shape typical for LSBs (cf. \citet{JSS08}) with a pronounced negative peak around the transition point.
At $\alpha$\,=\,7$^\circ$, the LSB is located at the trailing edge and the skin friction profile indicates that the wall shear stress remains near zero over two thirds of the airfoil until the flow transitions at $x_t/c$\,=\,0.62 and transports momentum to the surface that results in a peak of the shear stress at $x/c$\,=\,0.75.

The skin friction profile at 7$^\circ$ is similar to the data shown by \citet{uranga11} for a trailing edge LSB on the SD 7003 airfoil at \Rey{}\,=\,$2.2\times10^4$ and 4$^\circ$ incidence. 
Instantaneous skin friction data of the NACA 65(1)-412 at 7$^\circ$ (not shown) also illustrates that the shear stress periodically becomes negative at the leading edge and so provides a favourable velocity gradient for the development of the shear instabilities that later result in the shedding of K-H vortices.
Both cases with LSBs at $\alpha$\,=\,7$^\circ$ and 8$^\circ$ recover the free-stream pressure at the trailing edge and thereby reduce the form drag which results in the high-lift low-drag state at the critical angle of attack, $\alpha_{\textrm{crit}}$ between these two regimes.

\vskip\baselineskip
\paragraph{\centerline{\textit{Comparison of numerical and experimental results}}}
\par
There is a rich set of flow phenomena that successively appear as $\alpha$ varies from $0^\circ$ to $10^\circ$.  At low $\alpha$ the airfoil is sheathed in a mainly laminar flow and though the separation point is just aft of mid-chord, the separated region itself is also mostly laminar, up until $\alpha$\,$\approx$\,7$^\circ$ when the LSB instability triggers transition to turbulence on the airfoil itself (figure \ref{fig:vorticity}d).  In a time-averaged sense, the flow now re-attaches.  With further increases in $\alpha$ the separation, transition and reattachment points all move forward (figure \ref{fig:streamline}d-f).  The effect of this change of flow state on the integrated aerodynamic force coefficients is quite large, and the sensitivity of the global flow to small changes that cause significant movement of the separation point is responsible for certain difficulties in comparing experiments and computations, even under nominally similar conditions.

\begin{figure}
    \centering
    \makebox[\textwidth][c]{
    \includegraphics[width=0.48\textwidth]{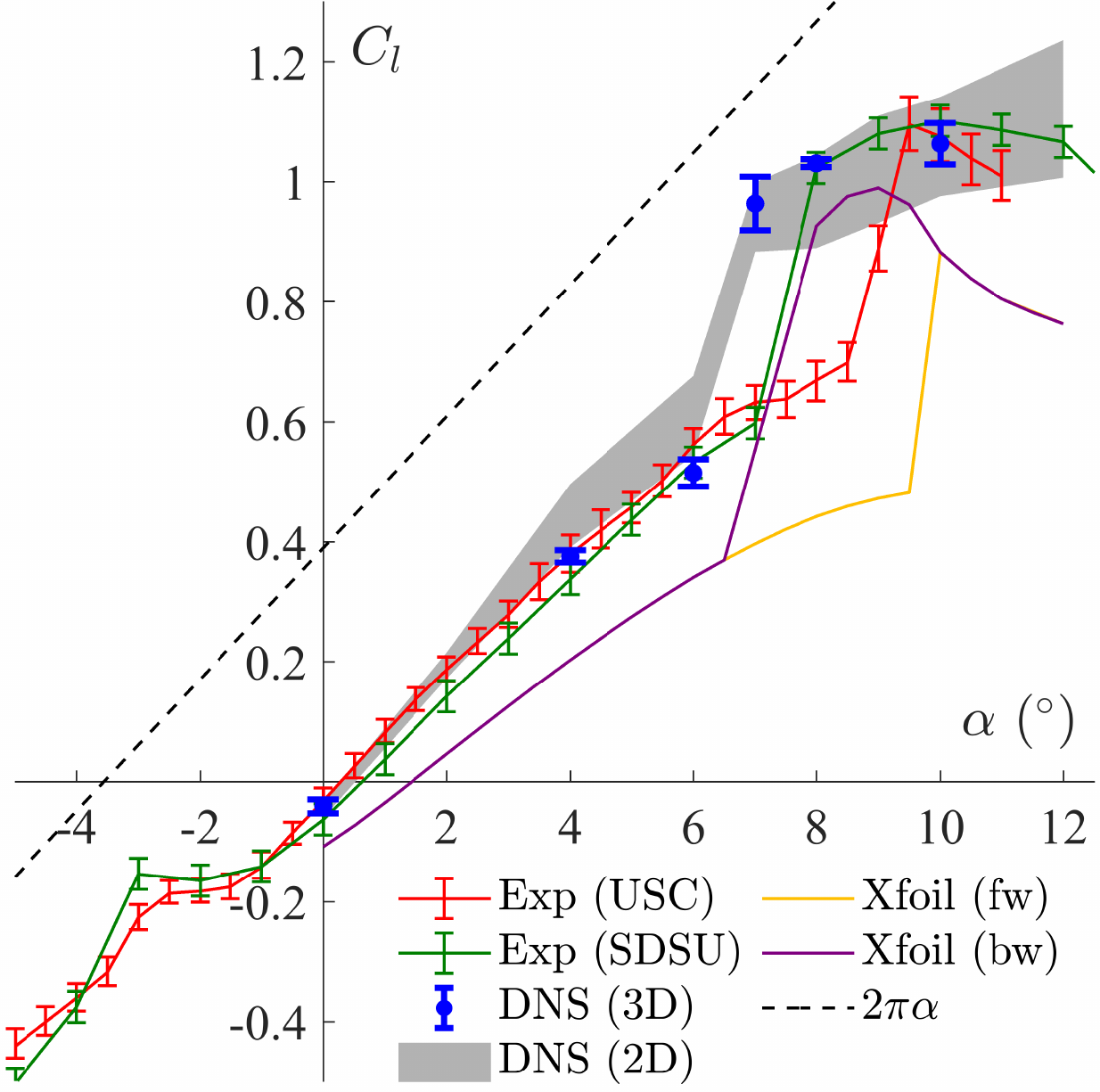}
	\hfill
	\includegraphics[width=0.452\textwidth]{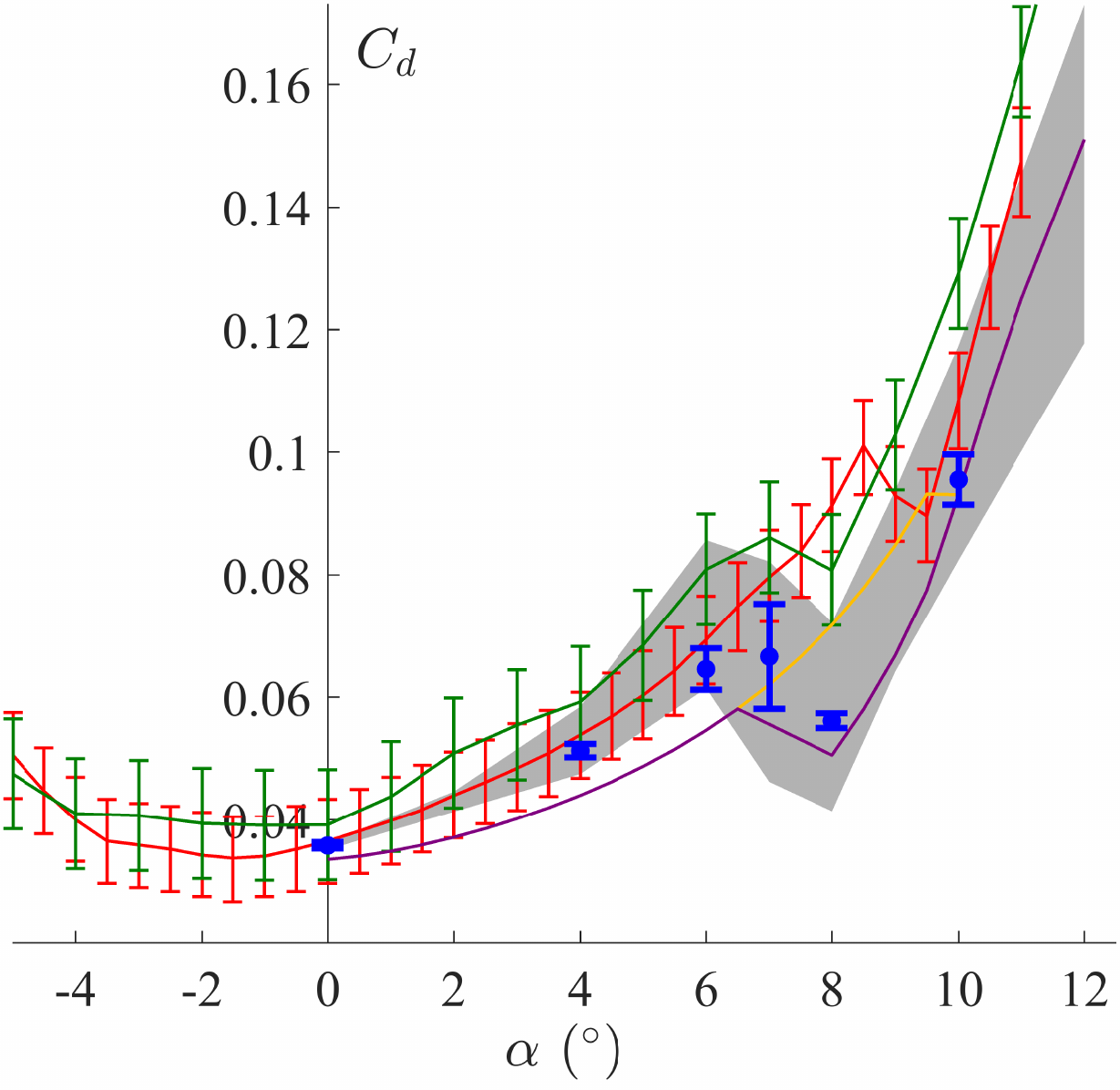}
	}
	\makebox[\textwidth][c]{
	\makebox[0.48\textwidth][c]{(a) Lift coefficient}
	\hfill
	\makebox[0.452\textwidth][c]{(b) Drag coefficient}
	}
	\caption{Lift (a) and drag (b) coefficients obtained from wind tunnel experiments at USC and SDSU, DNS data (2D \& 3D), and \textit{Xfoil} data (forward and backward sweep, $N_{crit}=9$) for a NACA 65(1)-412 at $\Rey{}_c$\,=\,2$\times10^4$. Error bars come from standard deviation of DNS time series and the gray area identifies the total lift and drag range of the parametric 2D study given by the averaged coefficient +/- standard deviation.  The error bars in the experiments come from the variation between separate, repeated experiments.}
	\label{fig:liftdrag}
\end{figure}

Figure \ref{fig:liftdrag} makes such a comparison for DNS in two and three dimensions with measurements from two different wind tunnels \citep{TKJS19a, Choi20}.  Also included are calculations from the panel code \textit{Xfoil} \citep{xfoil} which uses a boundary integral method to estimate separation locations. There is a tunable parameter, $N_{\textrm{crit}}$ for the growth rate of disturbances that depends on the tunnel environment and here $N_{\textrm{crit}}$ is set to 9.

\begin{figure}
    \begin{minipage}{.52\textwidth}
    \centering
    \includegraphics[width=\textwidth,trim={0 20pt 50pt 0},clip]{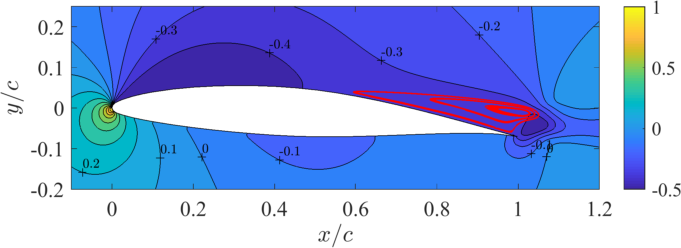}
    \makebox[\textwidth][c]{(a) 3D DNS}
    \vskip\baselineskip
    \includegraphics[width=\textwidth,trim={0 20pt 50pt 0},clip]{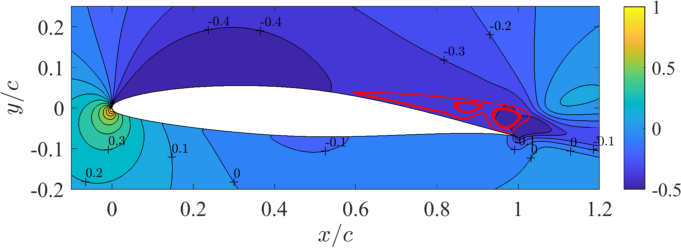}
    \makebox[\textwidth][c]{(b) 2D DNS}
    \vskip\baselineskip
    \includegraphics[width=\textwidth,trim={0 0 50pt 0},clip]{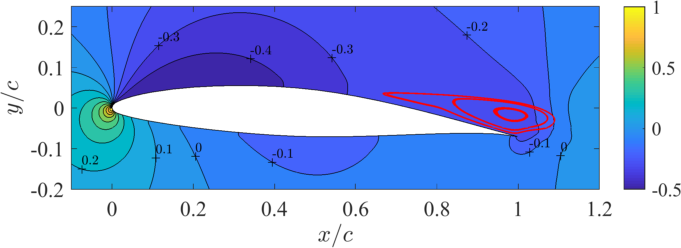}
    \makebox[\textwidth][c]{(c) RANS-SST}
    \end{minipage}
    \hfill
    \begin{minipage}{.45\textwidth}
    \centering
    \includegraphics[width=\textwidth]{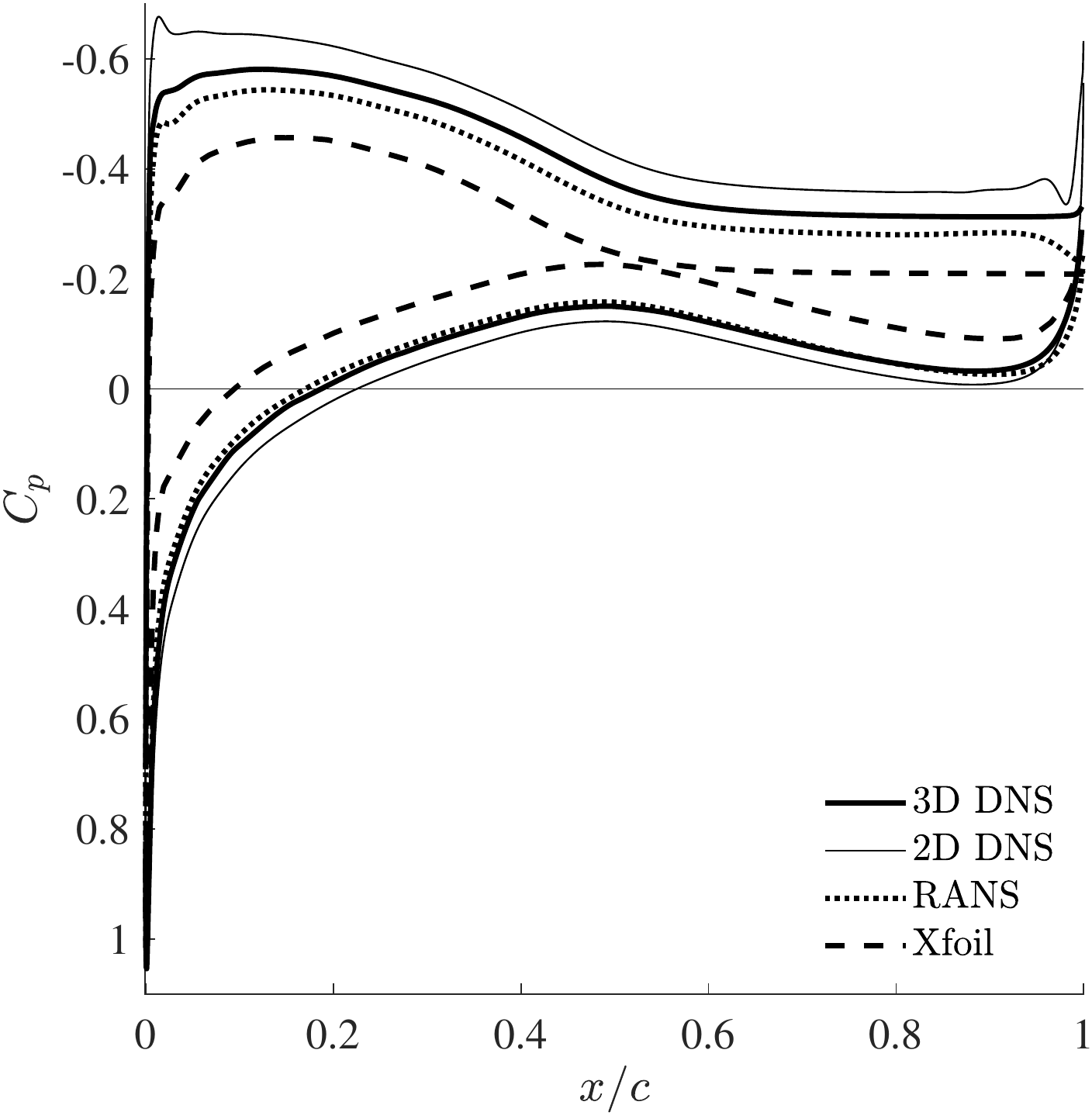}
    \makebox[\textwidth][c]{(d) Surface pressure coefficient}
    \end{minipage}
	\caption{
	Pressure coefficients of two and three-dimensional DNS, RANS-SST, and \textit{Xfoil}.
	}
	\label{fig:cpxfoilrans}
\end{figure}

DNS and experiments all show an abrupt increase in $C_l$ and decrease in $C_d$ at some critical incidence angle, $\alpha_{\textrm{crit}}$.  In both 2D and 3D DNS, $\alpha_{\textrm{crit}}$\,=\,6$^\circ$, which corresponds to the turbulent reattachment of the flow to the aft part of the suction surface.  It is reasonable to assume that the same physical mechanisms occur in experiment, but $\alpha_{\textrm{crit}}$\,=\,9$^\circ$ in the USC data and $\alpha_{\textrm{crit}}$\,=\,7$^\circ$ in the SDSU data.  \textit{XFoil} also predicts a rapid increase in $C_l$ at $9.5^\circ$ in forward sweep and $6.5^\circ$ in backwards sweep.  \textit{Xfoil} underpredicts the lift at almost all $\alpha$.  Before $\alpha_{\textrm{crit}}$, the 3D DNS and experiments lie within uncertainties of each other.  In this same range of $\alpha$, the 2D DNS estimates are all above the 3D data from numerical and laboratory experiment.  All $C_l(\alpha)$ below $\alpha_{\textrm{crit}}$, fall markedly below the inviscid $2\pi\alpha$ line ($\alpha$ at design $C_l$ for this airfoil section at high and infinite \Rey{} is zero degrees; at this \Rey{}, $C_l$\,<\,0 at $\alpha$\,=\,0$^\circ$).  The discrepancies arise from the time-dependent vortex shedding at the trailing edge, where bottom and top shear layers roll up into eddies and induce a local low-pressure region in the time-averaged flow field (see figure \ref{fig:cpxfoilrans}a).
The results therefore vary with the numerical approximation and how accurately the dynamics of the vortex shedding are represented.

The similarities and differences between the various estimates in figure \ref{fig:liftdrag} demonstrate how a correct accounting for the unsteady vorticity field is important at all $\alpha$.  Neither \textit{Xfoil} nor the 2D DNS do this (albeit for very different reasons).  The 2D simulations do yield a correct $\alpha_{\textrm{crit}}$, and since full 3D simulations are expensive, alternative simplified computations may be useful.

We assess the pressure distribution of time and space resolved two and three-dimensional DNS computed with the DGSEM, 2D RANS simulation conducted with \textit{FLUENT} using the transitional shear-stress transport (SST) model \citep{Fluent19}, and \textit{Xfoil}.
A compressible, density-based solver and a second-order upwinding scheme for the derivatives is used for the RANS simulation, where pressure far-field conditions are prescribed at the boundaries of the domain, which is of the same size as the DNS mesh.
All computations use a Mach number of $M$\,=\,0.3.

The time-averaged pressure coefficient and streamlines of the DNS and RANS results are presented in figure \ref{fig:cpxfoilrans} (a--c) and the surface pressure is plotted in (d), together with the \textit{Xfoil} result.
Differences in the streamline pattern of the time-resolved two and three-dimensional DNS are caused by the absence of the three-dimensional vortex dynamics (turning and stretching) in the 2D case, resulting in a more rigid system of vortices within the bubble.
As a result, the surface pressure on the suction side is lower in the two-dimensional DNS and leads to a consistently higher lift force for angles $\leq$ 6$^\circ$ (see figure \ref{fig:clcd}a).
The only RANS model that predicts the recirculation bubble is the SST model, but the result is unsteady and shows vortex shedding.
The data presented in figure \ref{fig:cpxfoilrans} is therefore an averaged solution and shows that the recirculation region is overly diffused and does not include the distinct low pressure region at the trailing edge. 
Accordingly, the surface pressure on the suction side is higher than in the two and three-dimensional DNS.
Because \textit{Xfoil} does not directly model the vortex shedding, the low-pressure region caused by the formation of Karman vortices at the trailing edge is entirely disregarded.
As a result, the surface pressure is misrepresented (figure \ref{fig:cpxfoilrans}b) and the aerodynamic forces are predicted too low, leading to the deviations in the polars shown in figure \ref{fig:clcd}.



%% file: content/conclusion.tex
A comprehensive and detailed overview of the flow topology over a cambered NACA 65(1)-412 airfoil at \Rey{}\,=\,$2\times10^4$ is given for angles of attack from 0$^\circ$ to 10$^\circ$ using direct numerical simulations.
It is shown that the flow is very sensitive to changes in $\alpha$ and multiple flow states emerge within 0$^\circ$\,$\leq$\,$\alpha$\,$\leq$\,10$^\circ$.
The flow regime changes at a critical angle of attack of 7$^\circ$ from laminar separation without reattachment at $\alpha$\,$\leq$\,6$^\circ$ over a LSB at the trailing edge to a LSB at the leading edge for $\alpha$\,$\geq$\,8$^\circ$.
The transition of the flow regimes is governed by the interaction of several instabilities that result in complex three-dimensional structures:
Karman vortices, that are driven by the roll-up of the pressure side boundary layer at the trailing edge, and Kelvin-Helmholtz instabilities within the separated shear layer on the suction side interact with three-dimensional instabilities within the vortex cores and in the braid region and result in three-dimensional tubular structures for $\alpha$\,$\leq$\,6$^\circ$ and large-scale turbulent puffs at $\alpha$\,=\,7$^\circ$.

The topology of the far-wake structures several chord lengths behind the airfoil is governed by the near-wake and the instabilities that transition the flow. 
While a narrow vortex street governs the wake at $\alpha$\,$\leq$\,4$^\circ$, the formation of the LSB at $\alpha$\,=\,7$^\circ$ and the accompanying interaction of pressure and suction side instabilities result in a low-frequency street with large-scale turbulent structures.
The shifting of the LSB to the leading edge at $\alpha$\,=\,8$^\circ$ incidence narrows the wake again, as the wall-bounded turbulence over the airfoil results in a more uniform shedding at the trailing edge compared to $\alpha$\,=\,7$^\circ$.

The flow bifurcation is accompanied by a sudden increase of the lift force and decrease in the drag, as shown by polars from DNS, wind tunnel experiments, and \textit{Xfoil}, and underscores the sensitive nature of low-Reynolds number airfoil aerodynamics.
By comparing DNS, RANS simulation, and \textit{Xfoil} data, it is shown that under-prediction of the lift coefficient in \textit{Xfoil} is related to a low-pressure region at the trailing edge that is caused by vortex formation inside the LSB. These elaborate flow structures and their interactions are more influential at lower Re, and further effort could usefully be put into appropriate modelling strategies when and if simpler models are used, for example in design.

It is the sensitivity and complexity of this flow that makes the comparison of computations and experiments particularly challenging and cause the continued mismatch in the critical angle. 
Although we extensively studied the effect of simulation parameters and reached cross-solver convergence, there are other factors that play a role in comparing with experiments, such as end-wall effects \citep{PMJA01} or acoustics \citep{KSJ21}.
We plan to report on the effect of these parameters in future work.

%% file: content/results2D.tex
In this section we present results of two-dimensional Navier-Stokes simulations of the NACA 65(1)-412, which serve to assess on the effect of resolution, domain size, and Mach number.
Although the physical meaning of these results is limited because vortex stretching is absent in two-dimensional approximation, they are relevant for assessing first-order trends in parametric studies.

\subsection{Effect of Mach number}
Although low-Reynolds number flows also typically operate at low Mach numbers, some applications (e.g. UAV at high altitude) may encounter compressibility effects \citep{Lissaman83}.
The Prandtl-Glauert correction rule to estimate the compressibility effects of the flow is $C_{p,M}/C_{p,i}$\,=\,$1/\sqrt{1-M^2}$. 
For Mach numbers $M$\,=\,0.1 and $M$\,=\,0.3, the correction factors are $C_{p,M=0.1}/C_{p,i}$\,=\,1.005 and $C_{p,M=0.3}/C_{p,i}$\,=\,1.048 respectively and hence we expect deviations of around 4\% -- 5\%.

\begin{figure}
	\makebox[\textwidth][c]{
    \includegraphics[width=0.45\textwidth]{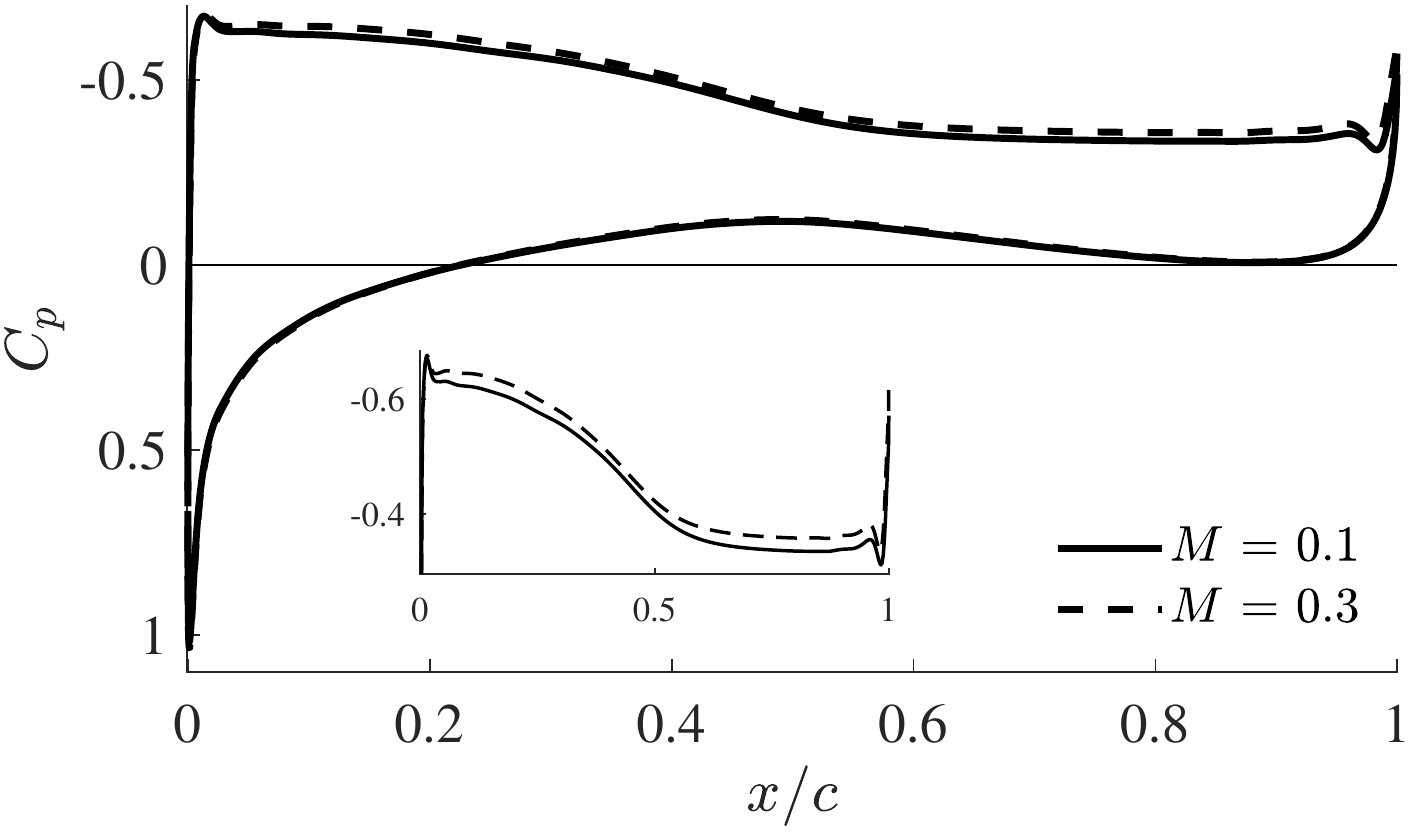}
    \hfill
    \includegraphics[width=0.45\textwidth]{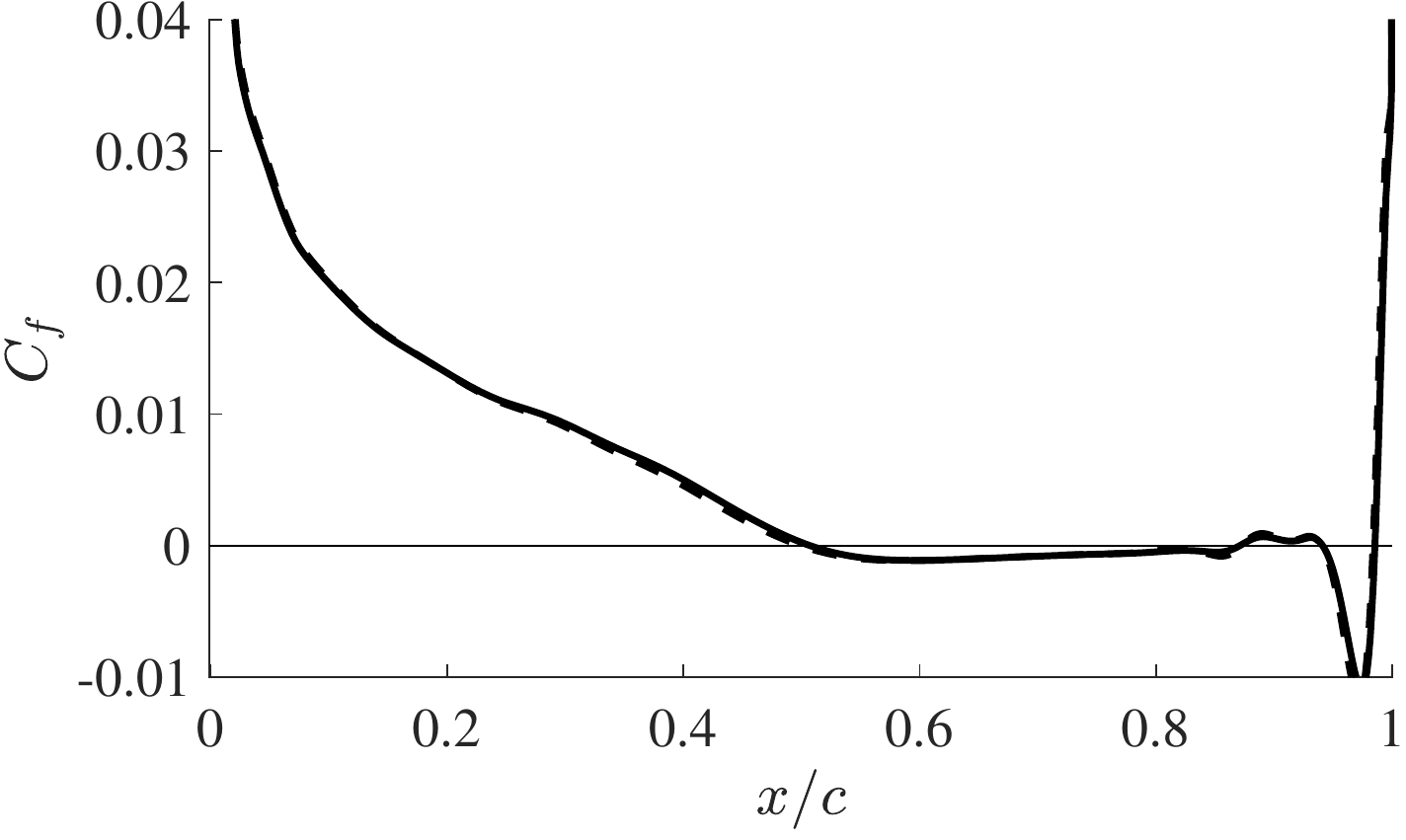}
    }
    \makebox[\textwidth][c]{
    \makebox[0.45\textwidth][c]{(a) Pressure coefficient}
    \hfill
    \makebox[0.45\textwidth][c]{(b) Skin friction coefficient}
    }
	\caption{Time-averaged pressure and skin friction coefficients for $M$\,=\,0.1 and $M$\,=\,0.3 at $\alpha$\,=\,4$^\circ$, $R$\,=\,30$c$.}
	\label{fig:cpcf_mach_4deg}
\end{figure}

At 4$^\circ$ angle of attack, the lower compressibility at $M$\,=\,0.1 results in a larger amplitude of the lift and drag force oscillations, as well as an offset of the time-averaged values by 4\% and 6\% respectively (see table \ref{tab:clcd_mach_4deg}).
These values are in very good agreement with the predicted deviations based on the Prandtl-Glauert correction.
Time-averaged profiles of the the pressure and skin friction coefficients in figure \ref{fig:cpcf_mach_4deg} show that the differences in compressibility effect mainly the pressure distribution on the suction side of the airfoil and have a negligible impact on the skin friction distribution. 

\begin{figure}
	\makebox[\textwidth][c]{
    \includegraphics[width=0.45\textwidth]{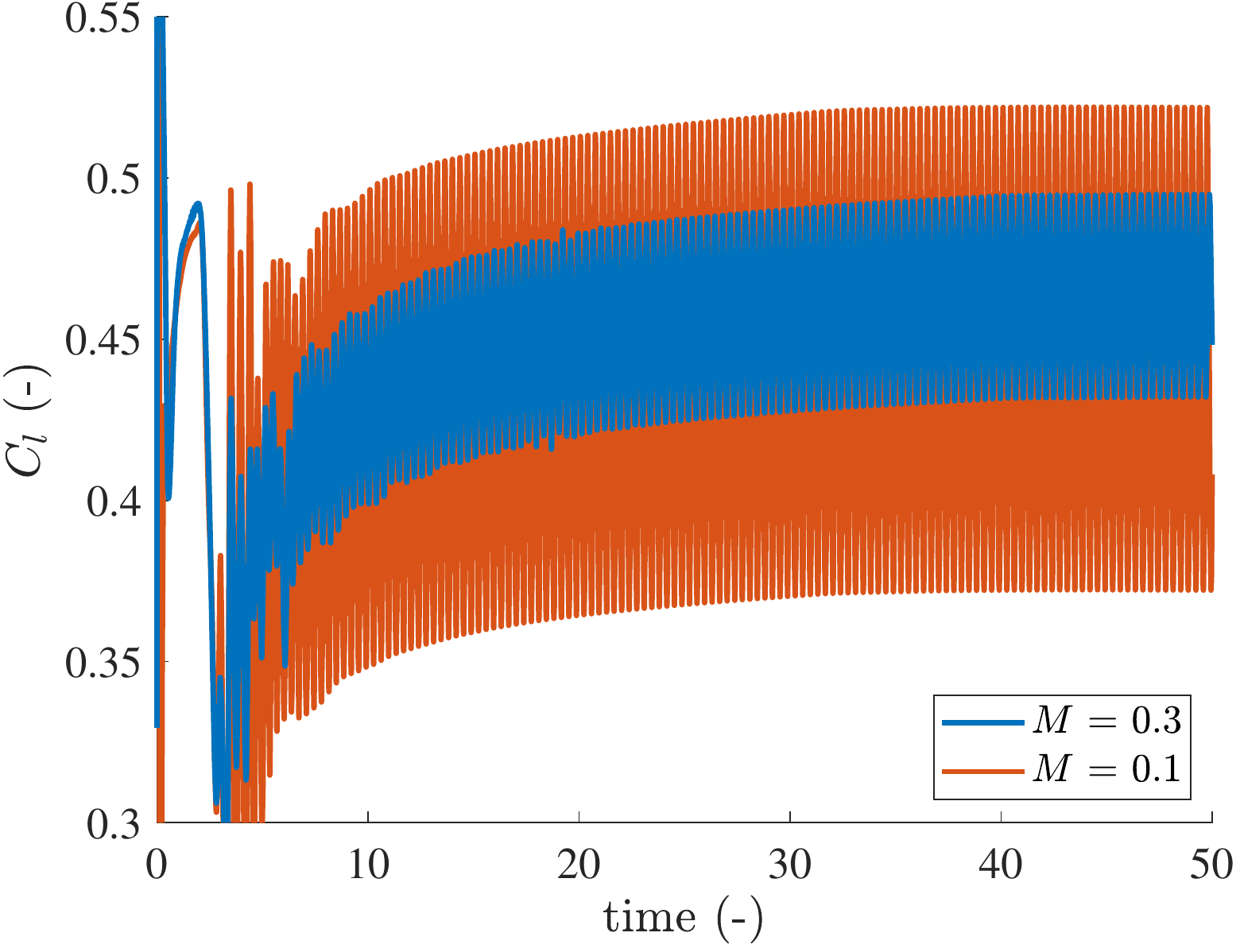}
    \hfill
    \includegraphics[width=0.45\textwidth]{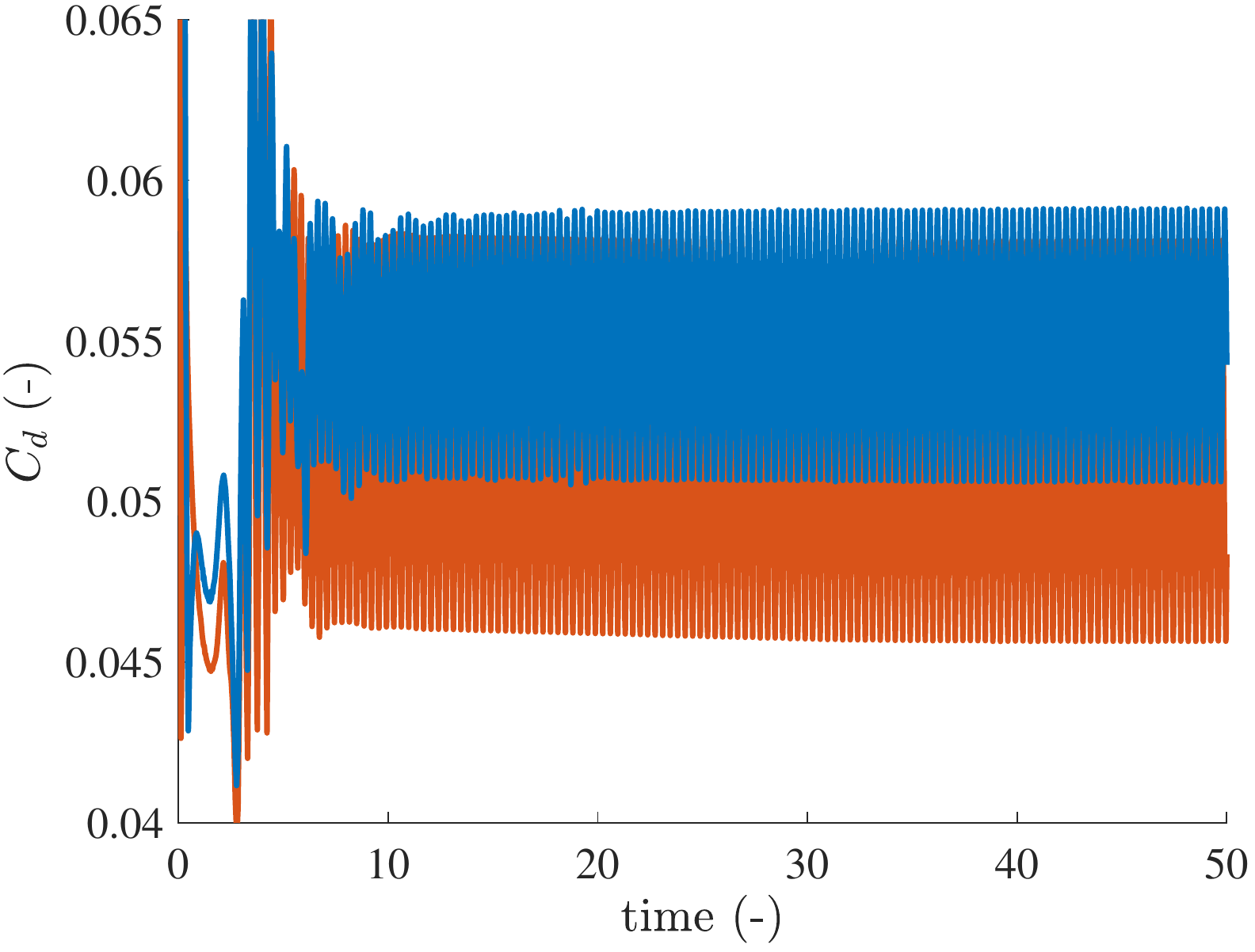}
    }
    \makebox[\textwidth][c]{
    \makebox[0.45\textwidth][c]{(a) Lift coefficient}
    \hfill
    \makebox[0.45\textwidth][c]{(b) Drag coefficient}
    }
	\caption{Lift and drag coefficients at $\alpha$\,=\,4$^\circ$ over time for different Mach numbers. Domain radius $R$\,=\,30$c$.}
	\label{fig:clcd_mach_4deg}
\end{figure}

\begin{table}
    \begin{center}
    \begin{tabular}{lcccccc}
        $M$   & $C_l$     & $C_{l,p}$     & $C_{l,f}$     & $C_d$    & $C_{d,p}$    & $C_{d,f}$ \\
        \hline
        0.1   & 0.444     & 0.443         & 0.001         & 0.051    & 0.033        & 0.019 \\
        0.3   & 0.463     & 0.462         & 0.001         & 0.054    & 0.036        & 0.019 \\
    \end{tabular}
    \caption{Lift and drag forces for $\alpha$\,=\,4$^\circ$ and different Mach numbers.}
    \label{tab:clcd_mach_4deg}
    \end{center}
\end{table}

\begin{figure}
	\makebox[\textwidth][c]{
    \includegraphics[width=0.45\textwidth]{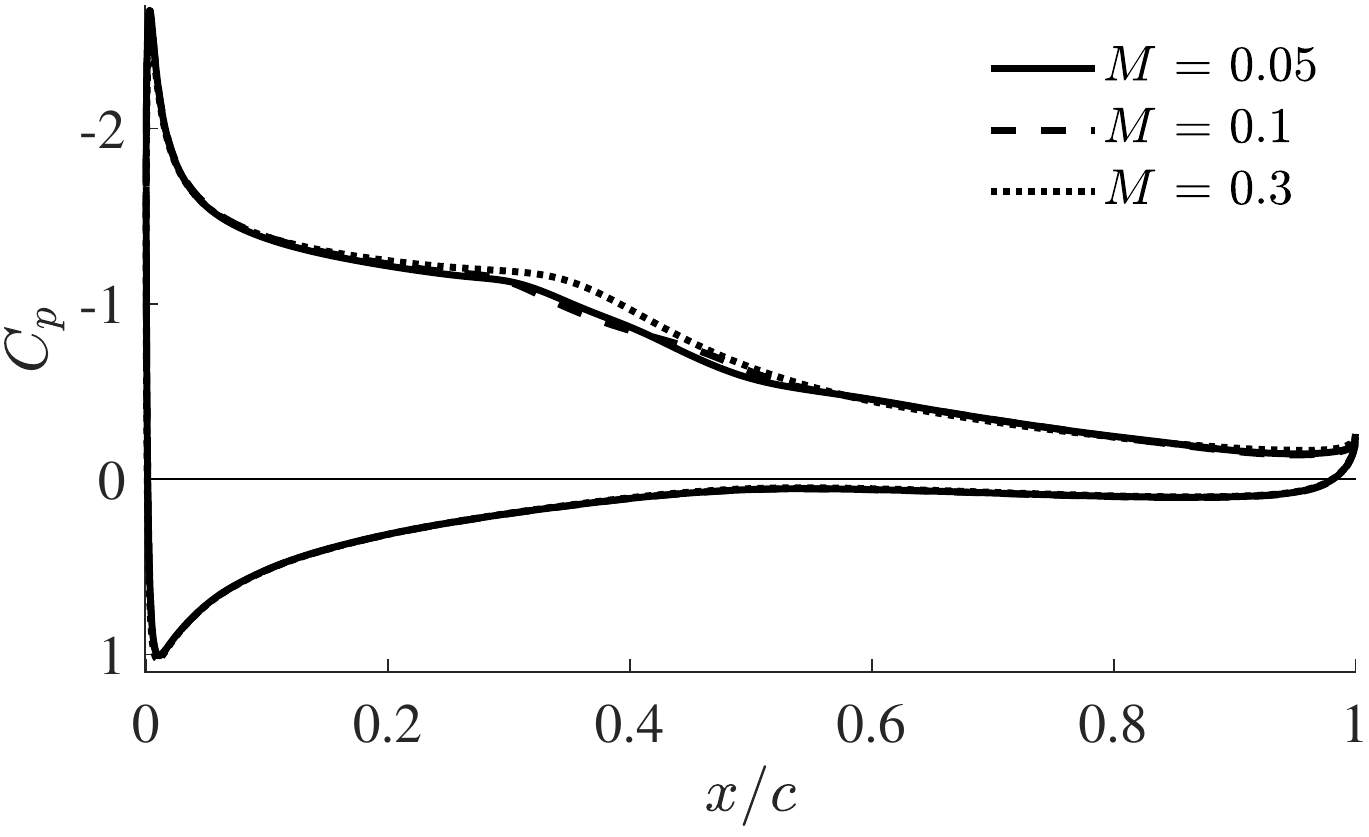}
    \hfill
    \includegraphics[width=0.45\textwidth]{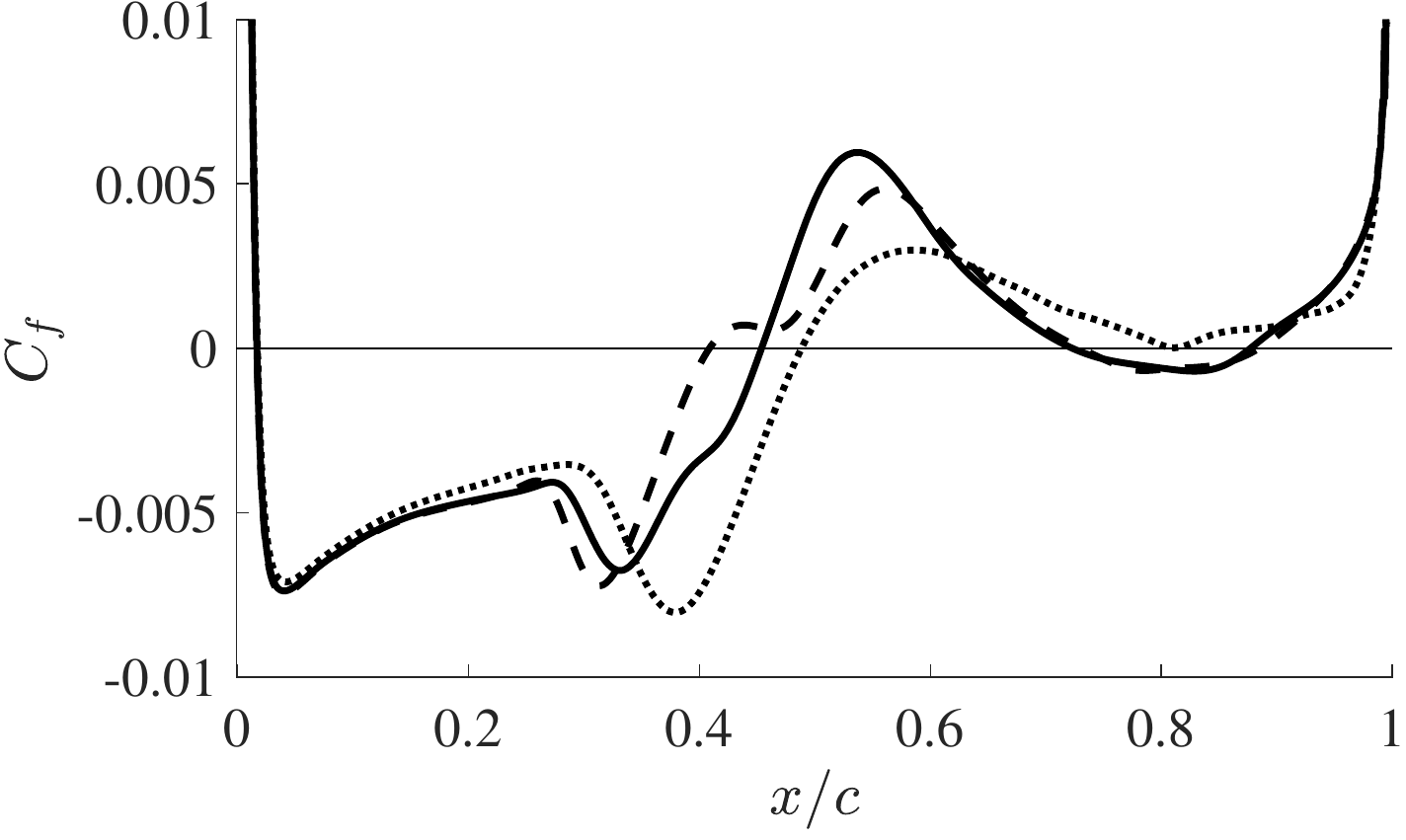}
    }
    \makebox[\textwidth][c]{
    \makebox[0.45\textwidth][c]{(a) Pressure coefficient}
    \hfill
    \makebox[0.45\textwidth][c]{(b) Skin friction coefficient}
    }
	\caption{Time-averaged pressure and skin friction coefficients for different Mach numbers at $\alpha$\,=\,8$^\circ$, $R$\,=\,30$c$.}
	\label{fig:cpcf_mach_8deg}
\end{figure}

\begin{figure}
    \centerline{\includegraphics[width=0.65\textwidth]{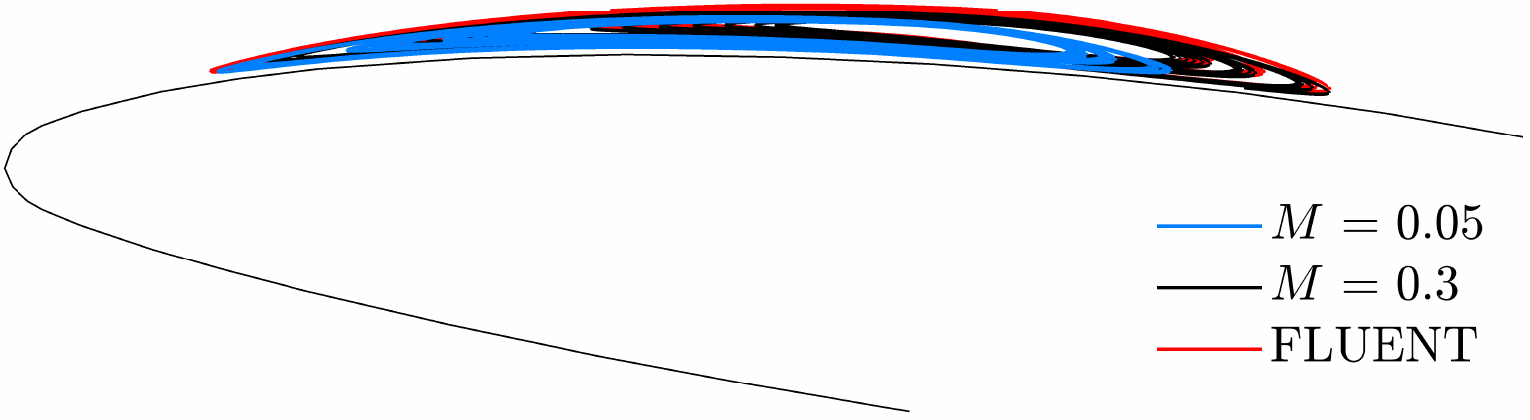}}
	\caption{Laminar separation bubble for $M$\,=\,0.05 and $M$\,=\,0.3 (DGSEM) and incompressible (\textit{FLUENT}) at $\alpha$\,=\,8$^\circ$, $R$\,=\,30$c$.}
	\label{fig:lsb_fluent_8deg}
\end{figure}

At 8$^\circ$ incidence, a slender LSB stretches from the leading until $x_r/c$\,=\,0.45, 0.41, and 0.49 for $M$\,=\,0.05, 0.1, and 0.3 respectively.
The profiles of the pressure and skin friction coefficients are given in figure \ref{fig:cpcf_mach_8deg} and show that the higher compressibility in case of $M$\,=\,0.3 results in a more distinct pressure plateau and elongated separation bubble with downstream reattachment compared to the lower-Mach number cases. 
Streamlines of the time-averaged recirculating flow within the LSB are plotted in figure \ref{fig:lsb_fluent_8deg} and illustrate the difference in bubble sizes.
The lift and drag coefficient averages differ by 2\% and 12\% respectively (see table \ref{tab:clcd_mach_8deg}) and can be attributed to the modified pressure distribution caused by the different LSB sizes.

\begin{table}
    \begin{center}
    \begin{tabular}{lcccccc}
        $M$   & $C_l$     & $C_{l,p}$     & $C_{l,f}$     & $C_d$    & $C_{d,p}$    & $C_{d,f}$ \\
        \hline
        0.05  & 0.941     & 0.940         & 0.002         & 0.052    & 0.041        & 0.010 \\
        0.1   & 0.946     & 0.944         & 0.002         & 0.052    & 0.042        & 0.011 \\
        0.3   & 0.965     & 0.964         & 0.002         & 0.058    & 0.048        & 0.010 \\
    \end{tabular}
    \caption{Lift and drag forces for $\alpha$\,=\,8$^\circ$ and different Mach numbers.}
    \label{tab:clcd_mach_8deg}
    \end{center}
\end{table}

In addition to assessing compressibility effects by computing the flow at different Mach numbers with the compressible DGSEM solver, we also compare our results with incompressible flow simulations performed with \textit{FLUENT}. 
Transient, incompressible computations are conducted with a pressure-based solver, second-order upwind for the spatial discretization, and second-order implicit time-stepping. 
No turbulence model is applied such that only source for artificial viscosity is through numerical dissipation from the upwinding scheme.
A C-type domain with radius and wake length of 30 chords and consisting of 802,300 quadrilateral elements is used. 
The outer boundaries treated as velocity inflow (left, lower, and upper) and pressure outflow conditions (right) and a no-slip condition is applied at airfoil surface.

\begin{figure}
	\makebox[\textwidth][c]{
    \includegraphics[width=0.45\textwidth]{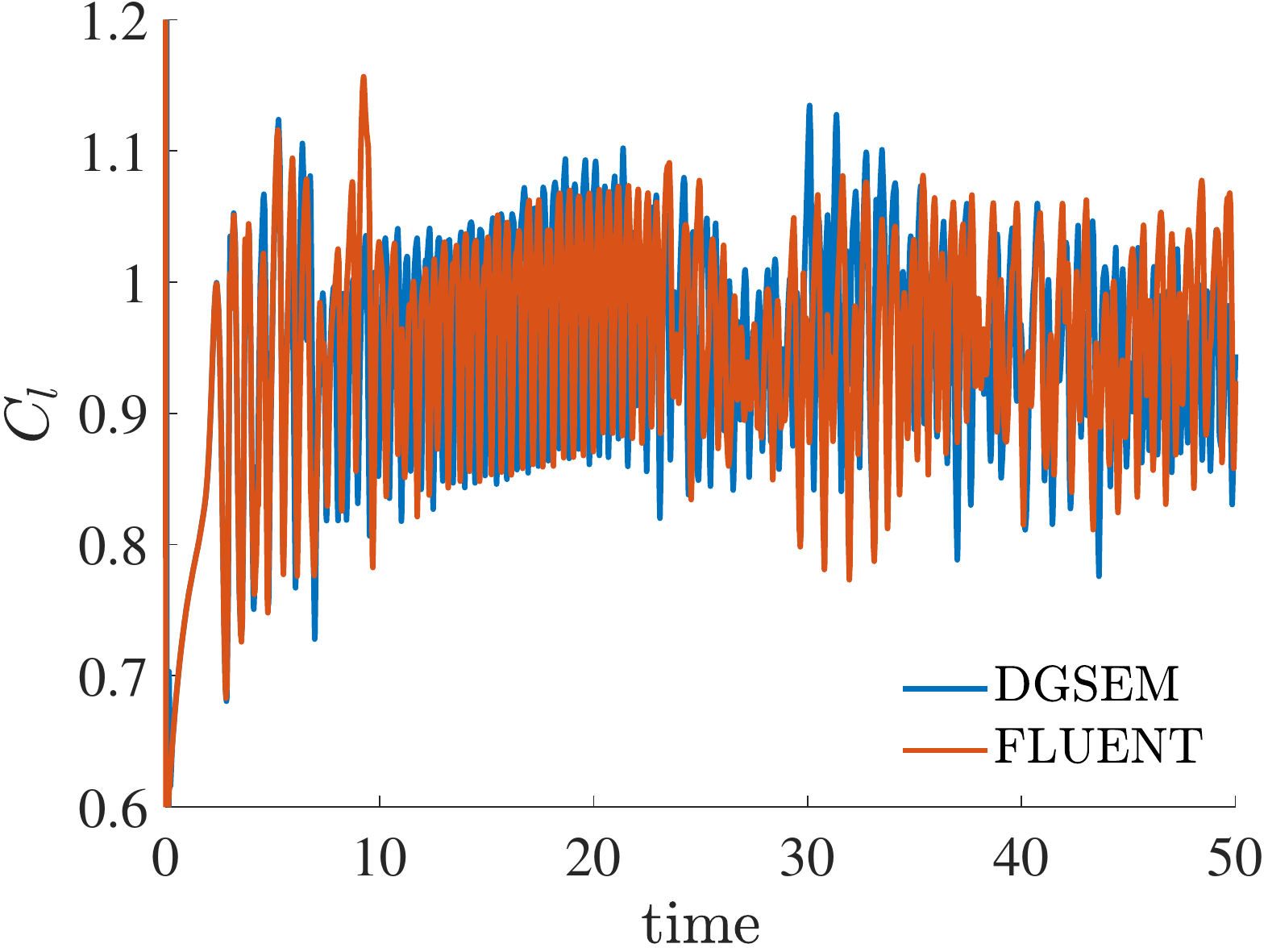}
    \hfill
    \includegraphics[width=0.45\textwidth]{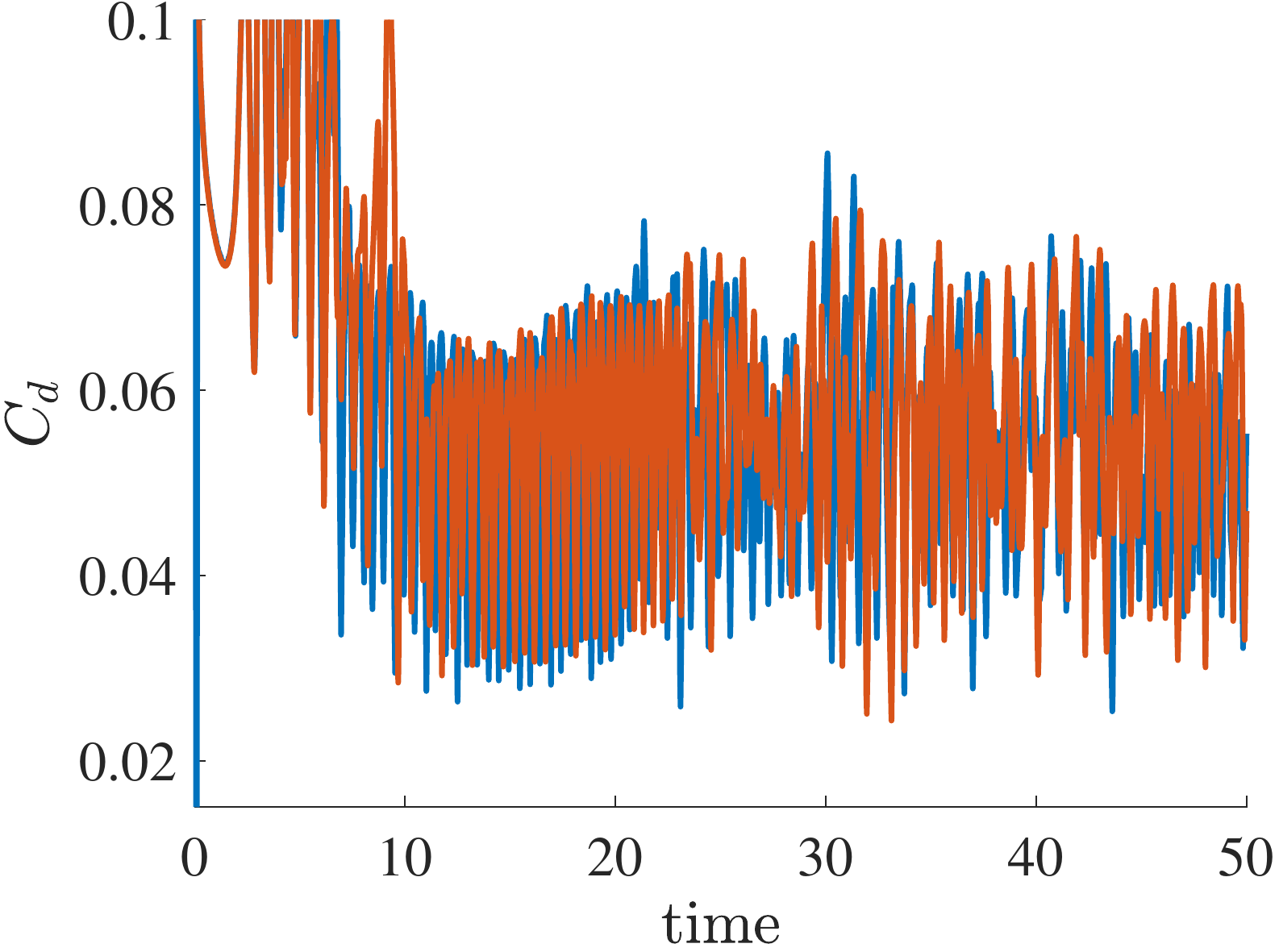}
    }
    \makebox[\textwidth][c]{
    \makebox[0.45\textwidth][c]{(a) Lift coefficient}
    \hfill
    \makebox[0.45\textwidth][c]{(b) Drag coefficient}
    }
	\caption{Lift and drag coefficients at $\alpha$\,=\,8$^\circ$ over time DGSEM ($M$\,=\,0.05) and \textit{FLUENT} (incompressible) computations.}
	\label{fig:clcd_fluent_8deg}
\end{figure}

Figure \ref{fig:clcd_fluent_8deg} shows the history of the lift and drag coefficients obtained from compressible DGSEM computations at a Mach number of $M$\,=\,0.05 and sixth order polynomial representation and incompressible simulations with \textit{FLUENT}. 
The results are in good agreement and confirm that the solution to this particular flow has converged across different numerical solvers. 
The case also shows that compresibility effects are not the cause for the disagreement with the USC wind tunnel experiments at $\alpha$\,=\,8$^\circ$ as all simulations show the transitioned state regardless of the Mach number.

A comparison of the streamlines inside the LSB (see figure \ref{fig:lsb_fluent_8deg}) shows that the bubble size in the \textit{FLUENT} computations is nearly identical with the DGSEM results at $M$\,=\,0.3, but deviates from the topology found at $M$\,=\,0.05. 
While the differences between the DGSEM results are related to the compressibility effects, results from the \textit{FLUENT} simulation are also affected by the lower order accuracy of the spatial and temporal discretization and the increased numerical dissipation of the upwind scheme.

\subsection{Resolution -- is it DNS?}
The resolution in spectral element methods can be adjusted either through mesh refinement (\textit{h}) or by increasing the polynomial order per element (\textit{p}). 
The two meshes employed in this paper, \textit{Grid 1} and \textit{Grid 2}, are based on very different element sizes and polynomial orders (cf. \ref{fig:domain}). 
For 4$^\circ$ incidence, \citet{nelson16} reports a grid-converged solution for a polynomial order of $P$\,=\,12 on \textit{Grid 1}. 
Because the flow transitions to turbulence at higher angles of attack, we compare time-averaged results of the coarser \textit{Grid 1} and the refined \textit{Grid 2} at different polynomial orders for 8$^\circ$ incidence.
At this angle, the wind tunnel experiments at USC and the computations deviate considerably as the experiment is still in the laminar regime below $\alpha_{crit}$ while the DNS simulations has already become turbulent.

\begin{figure}
	\makebox[\textwidth][c]{
    \includegraphics[width=0.45\textwidth]{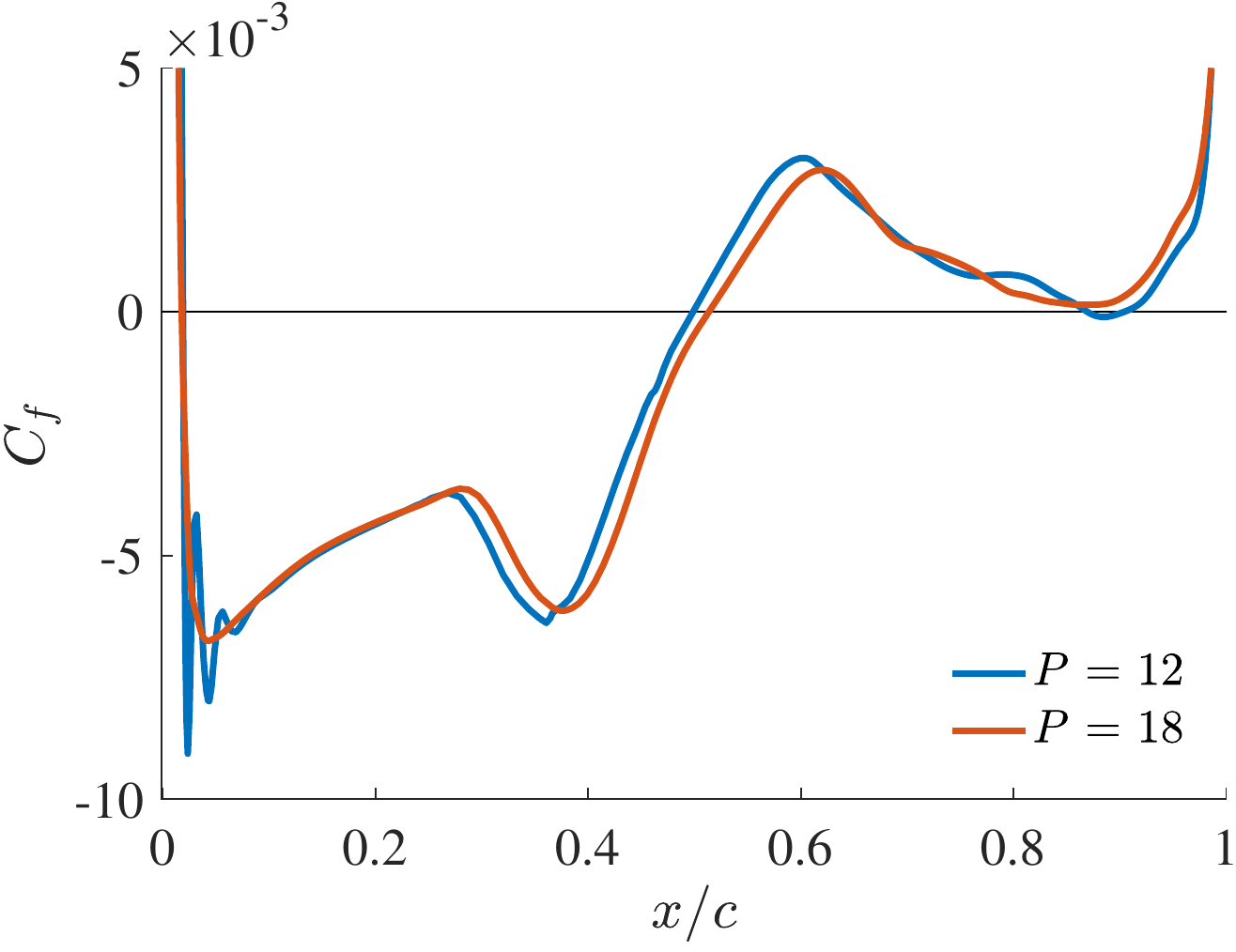}
    \hfill
    \includegraphics[width=0.45\textwidth]{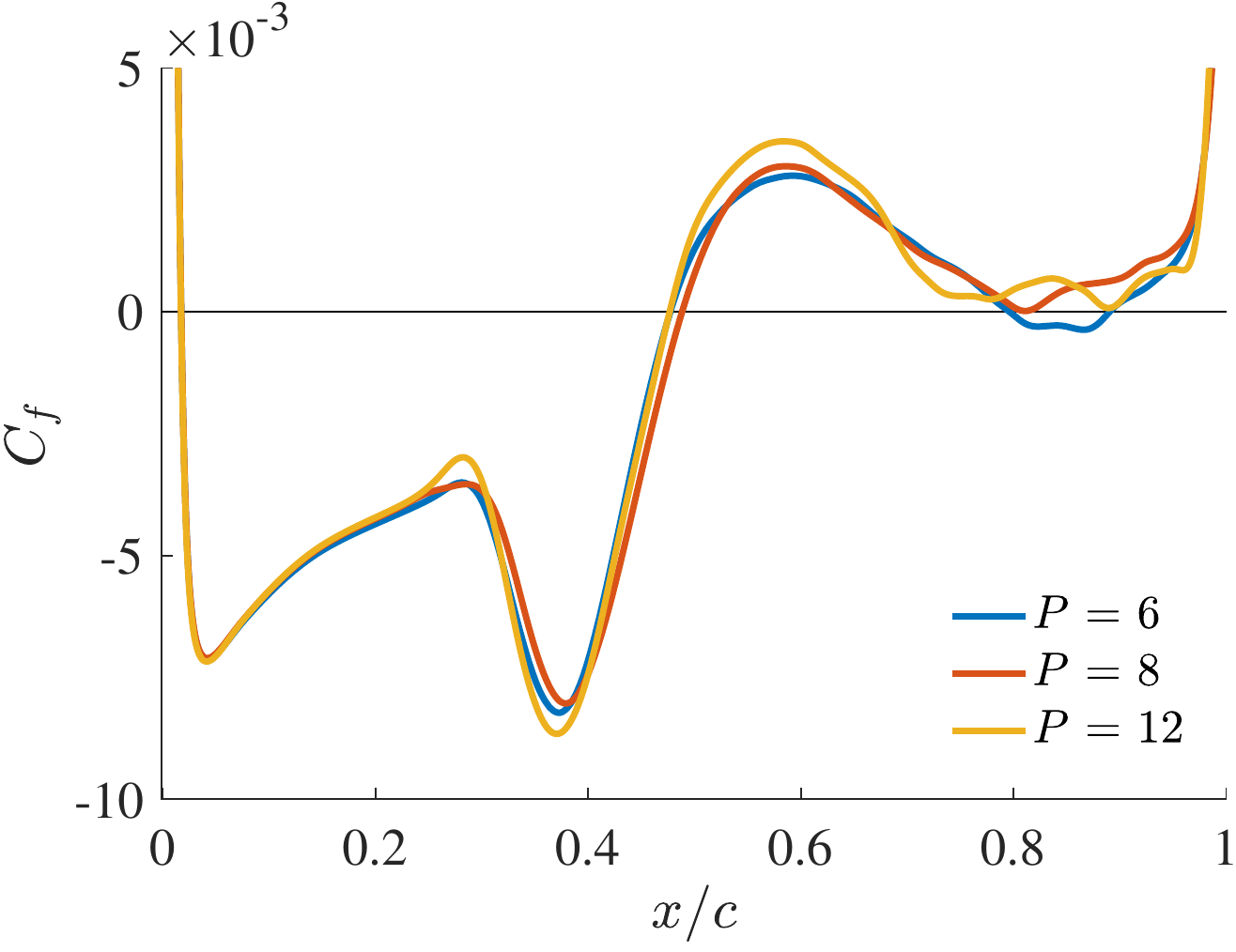}
    }
    \makebox[\textwidth][c]{
    \makebox[0.45\textwidth][c]{(a) \textit{Grid 1}}
    \hfill
    \makebox[0.45\textwidth][c]{(b) \textit{Grid 2}}
    }
	\caption{Time-averaged leading-edge skin friction for \textit{Grid 1} and \textit{Grid 2} at different polynomial orders. $Re_c$\,=\,20,000, $\alpha$\,=\,8$^\circ$.}
	\label{fig:cf_8deg_2D}
\end{figure}

We assess the fidelity of the numerical results by comparison of the skin friction coefficient for different resolutions. 
Figure \ref{fig:cf_8deg_2D}(a) shows that at a polynomial order of $N$\,=\,12, resolution at the airfoil's leading edge is insufficient and causes spurious, numerical oscillations in the solution. 
Although of considerable amplitude, these oscillations remain a local artifact and do not affect the results significantly in comparison to the converged solution at $N$\,=\,18.
The higher mesh refinement of \textit{Grid 2} requires lower polynomial orders to reach a converged solution and the skin friction coefficients plotted in figure \ref{fig:cf_8deg_2D}(b) show good agreement for polynomial orders $N$\,=\,4 -- $N$\,=\,8. The minor deviations can be attributed to the finite number of samples collected for temporal statistics.
Note that differences between figure \ref{fig:cf_8deg_2D}(a) and (b) stem from a larger buffer layer region in \textit{Grid 1} that results in reduced feedback of waves from the wake.

Given that no filter is employed in either of the simulations and the solution shows convergence, we consider the results presented in this paper DNS with the exception of the case at 10$^\circ$, which show some under-resolution at the leading edge and should therefore be considered implicit LES.

\subsection{Domain size}
We assess the effect of domain size, blockage and spurious boundary reflections on the solution by comparing the aerodynamic forces, pressure and skin friction coefficients for different sizes of the computational domain for several angles of attack. 

\begin{figure}
	\makebox[\textwidth][c]{
    \includegraphics[width=0.45\textwidth]{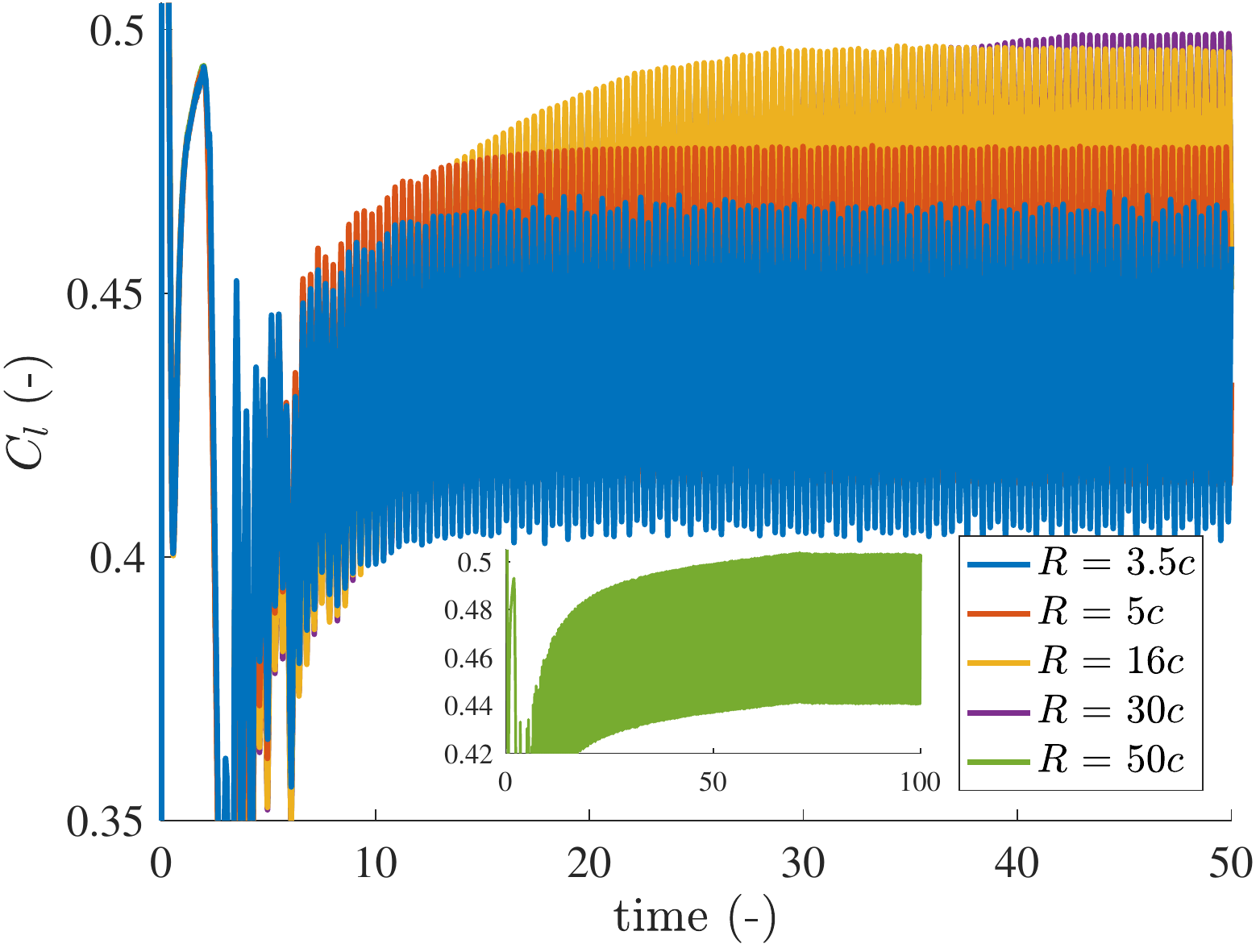}
    \hfill
    \includegraphics[width=0.45\textwidth]{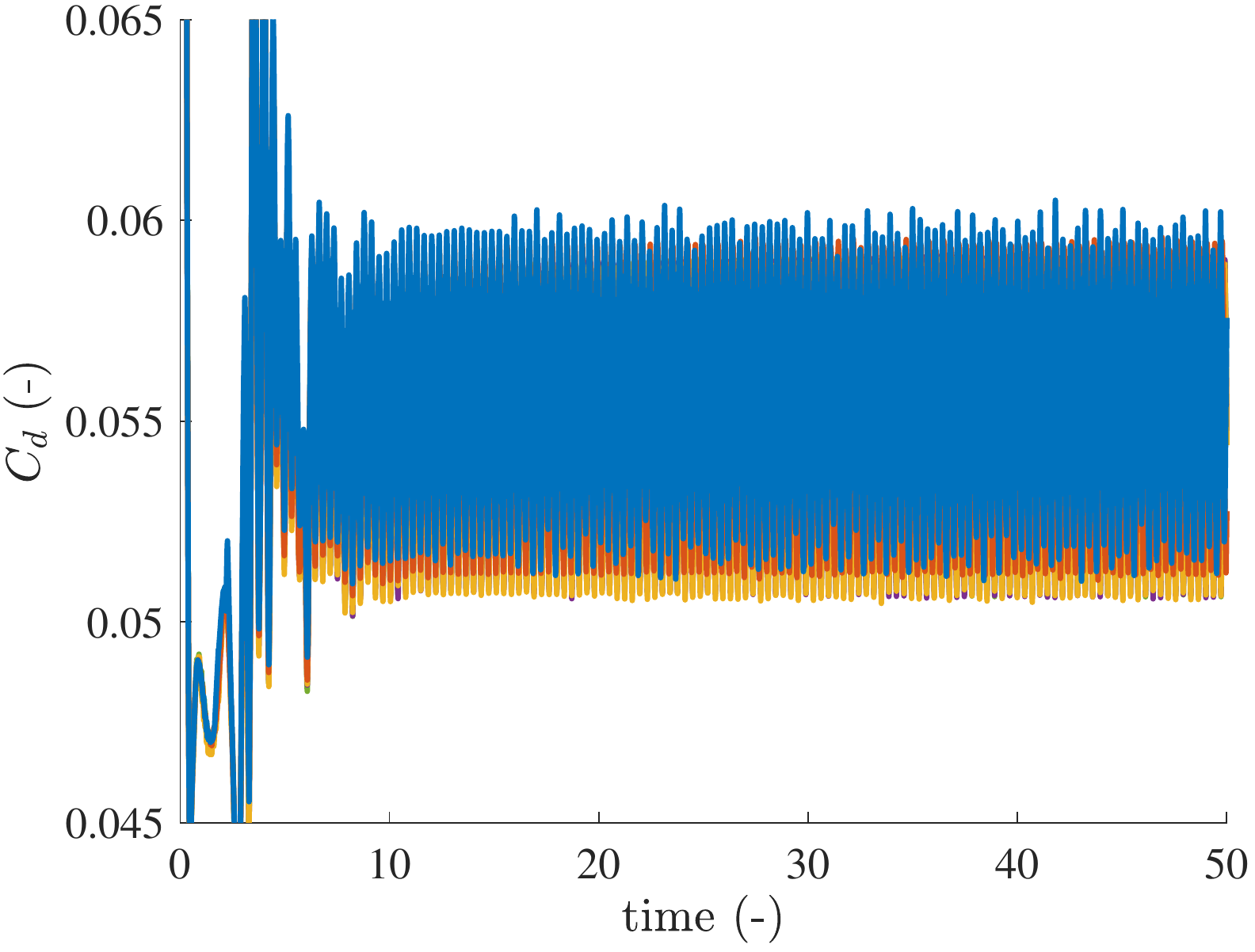}
    }
    \makebox[\textwidth][c]{
    \makebox[0.45\textwidth][c]{(a) Lift coefficient}
    \hfill
    \makebox[0.45\textwidth][c]{(b) Drag coefficient}
    }
	\caption{Lift and drag coefficients over time for different computational domain sizes and $\alpha$\,=\,4$^\circ$, $M$\,=\,0.3.}
	\label{fig:clcd_domain_4deg}
\end{figure}

\begin{figure}
	\makebox[\textwidth][c]{
    \includegraphics[width=0.45\textwidth]{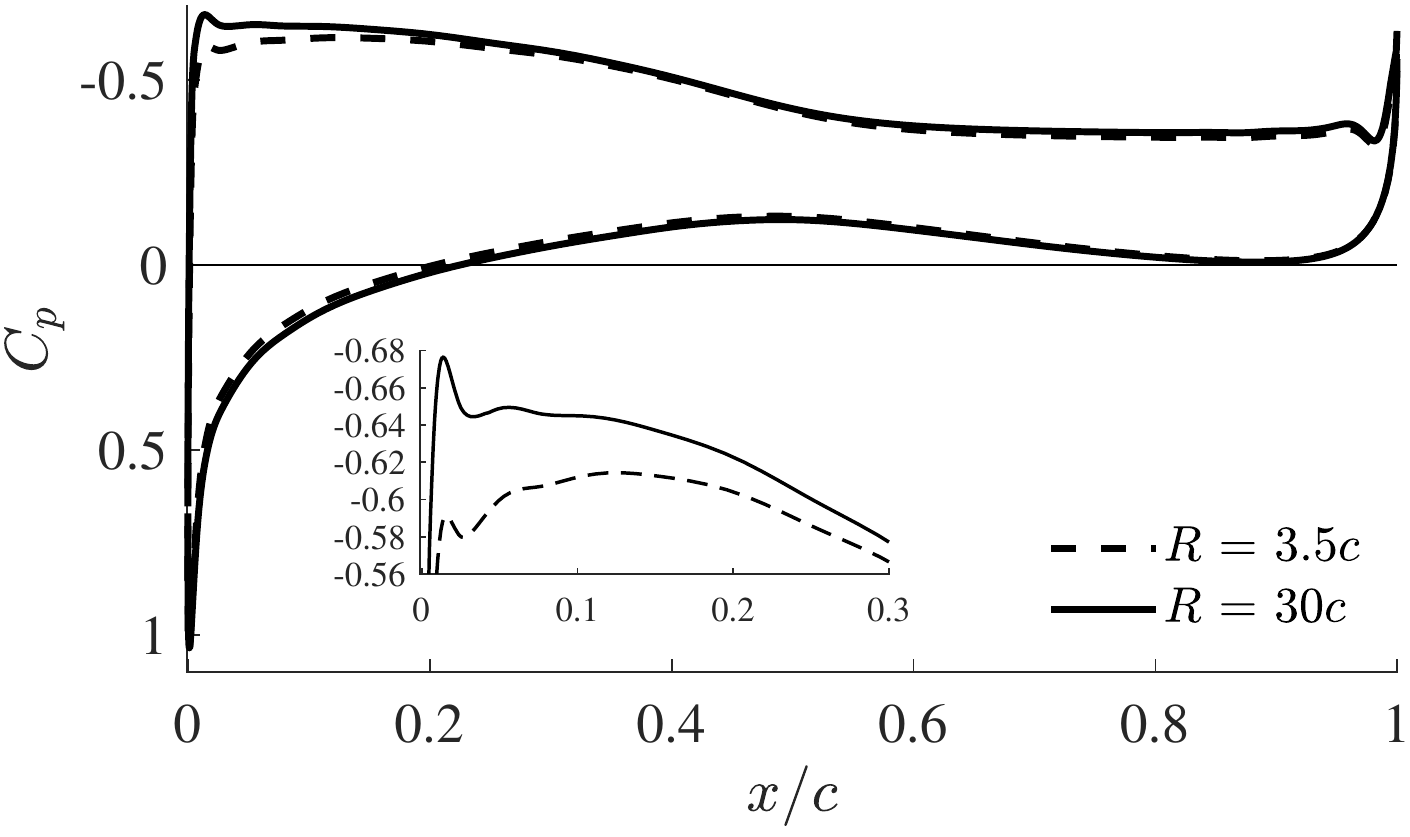}
    \hfill
    \includegraphics[width=0.45\textwidth]{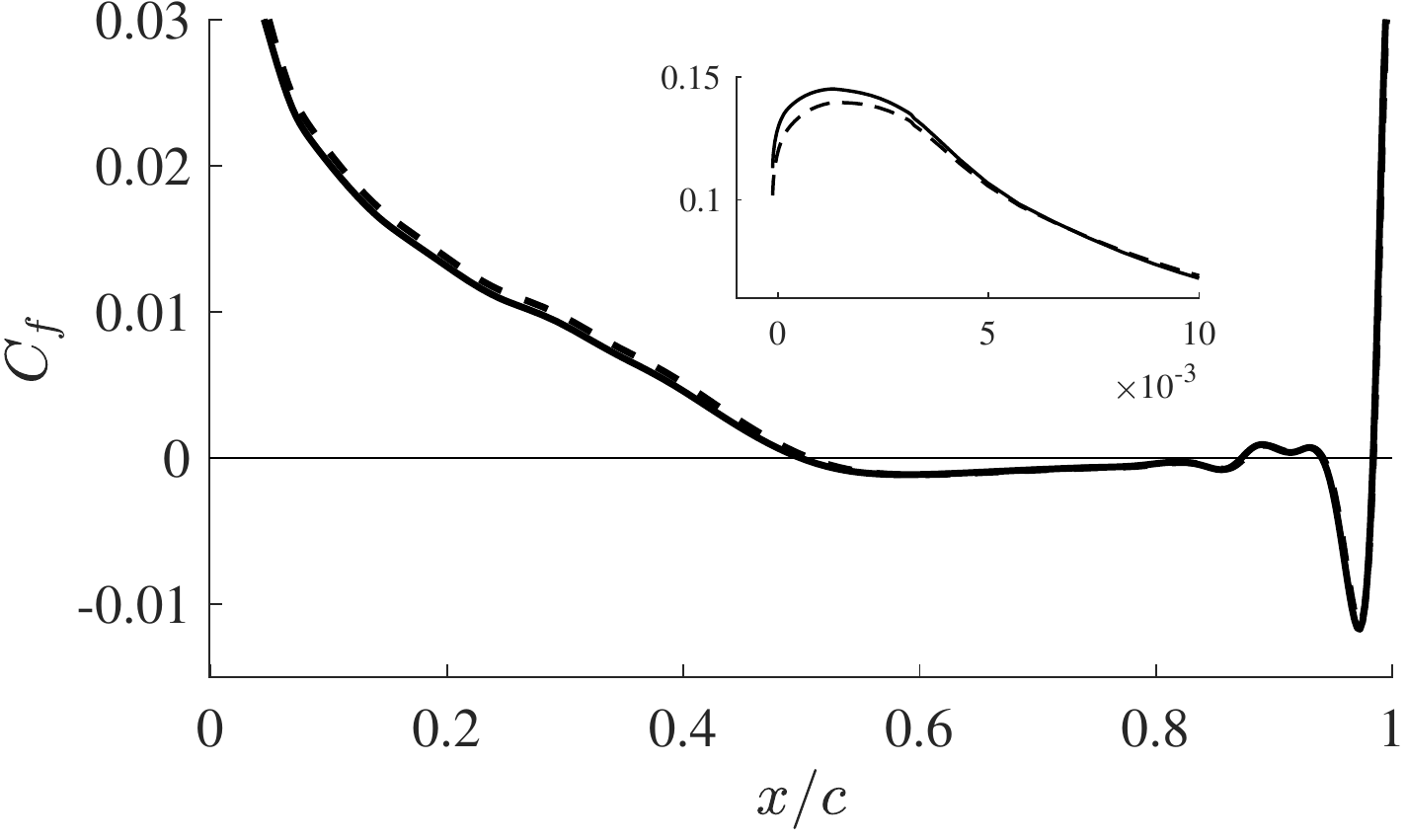}
    }
    \makebox[\textwidth][c]{
    \makebox[0.45\textwidth][c]{(a) Pressure coefficient}
    \hfill
    \makebox[0.45\textwidth][c]{(b) Skin friction coefficient}
    }
	\caption{Time-averaged pressure and skin friction coefficients for $R$\,=\,3.5$c$ and $R$\,=\,30$c$ at $\alpha$\,=\,4$^\circ$, $M$\,=\,0.3.}
	\label{fig:cpcf_domain_4deg}
\end{figure}

Figure \ref{fig:clcd_domain_4deg} illustrates the lift and drag coefficient for the flow at 4$^\circ$ incidence and domain radii from $R$\,=\,3.5$c$ to $R$\,=\,50$c$.
Corresponding pressure and skin friction distributions over the wing are plotted in figure \ref{fig:cpcf_domain_4deg} for  $R$\,=\,3.5$c$ and 30$c$. 
The free-stream boundaries show a strong impact on the pressure coefficient at the leading edge, which is significantly lower for the larger domain and indicates that the proximity of the boundaries for $R$\,=\,3.5$c$ forces the flow in this region.
The pressure deviation is reflected in the trend of the lifting force with deviations of the time-averaged solution of 6\% between small and large domains (see table \ref{tab:clcd_domain_4deg}). 
As the discrepancies are mainly caused by the differences in the pressure distribution, the drag force shows only minor variations between the cases and converges more quickly.
The strongly sinusoidal time dependency of the forces is maintained for all domain radii.

\begin{table}
    \begin{center}
    \begin{tabular}{lcccccc}
        Domain radius   & $C_l$     & $C_{l,p}$     & $C_{l,f}$     & $C_d$    & $C_{d,p}$    & $C_{d,f}$ \\
        \hline
        3.5$c$          & 0.434     & 0.433         & 0.001         & 0.055    & 0.036        & 0.019 \\
        30$c$           & 0.463     & 0.462         & 0.001         & 0.054    & 0.036        & 0.019 \\
    \end{tabular}
    \caption{Lift and drag coefficients for $\alpha$\,=\,4$^\circ$ and different domain sizes.}
    \label{tab:clcd_domain_4deg}
    \end{center}
\end{table}

\begin{figure}
    \centerline{\includegraphics[width=0.65\textwidth]{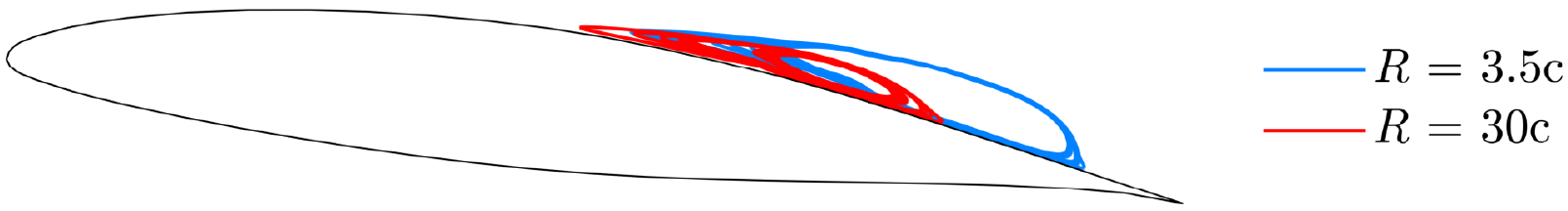}}
	\caption{Laminar separation bubble for domain radii $R$\,=\,3.5$c$ and $R$\,=\,30$c$ at $\alpha$\,=\,7$^\circ$, $M$\,=\,0.3.}
	\label{fig:lsb_domain_7deg}
\end{figure}

\begin{figure}
	\makebox[\textwidth][c]{
    \includegraphics[width=0.45\textwidth]{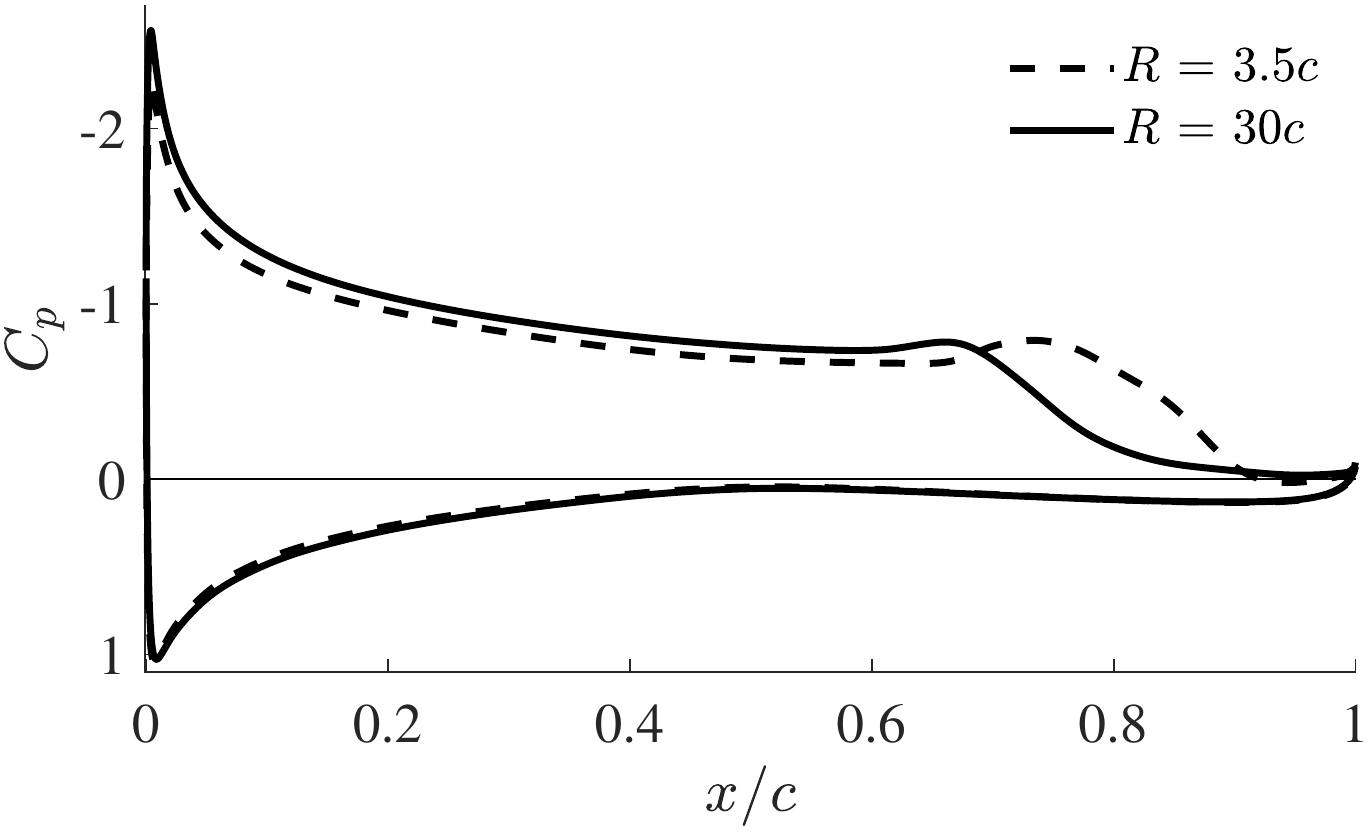}
    \hfill
    \includegraphics[width=0.45\textwidth]{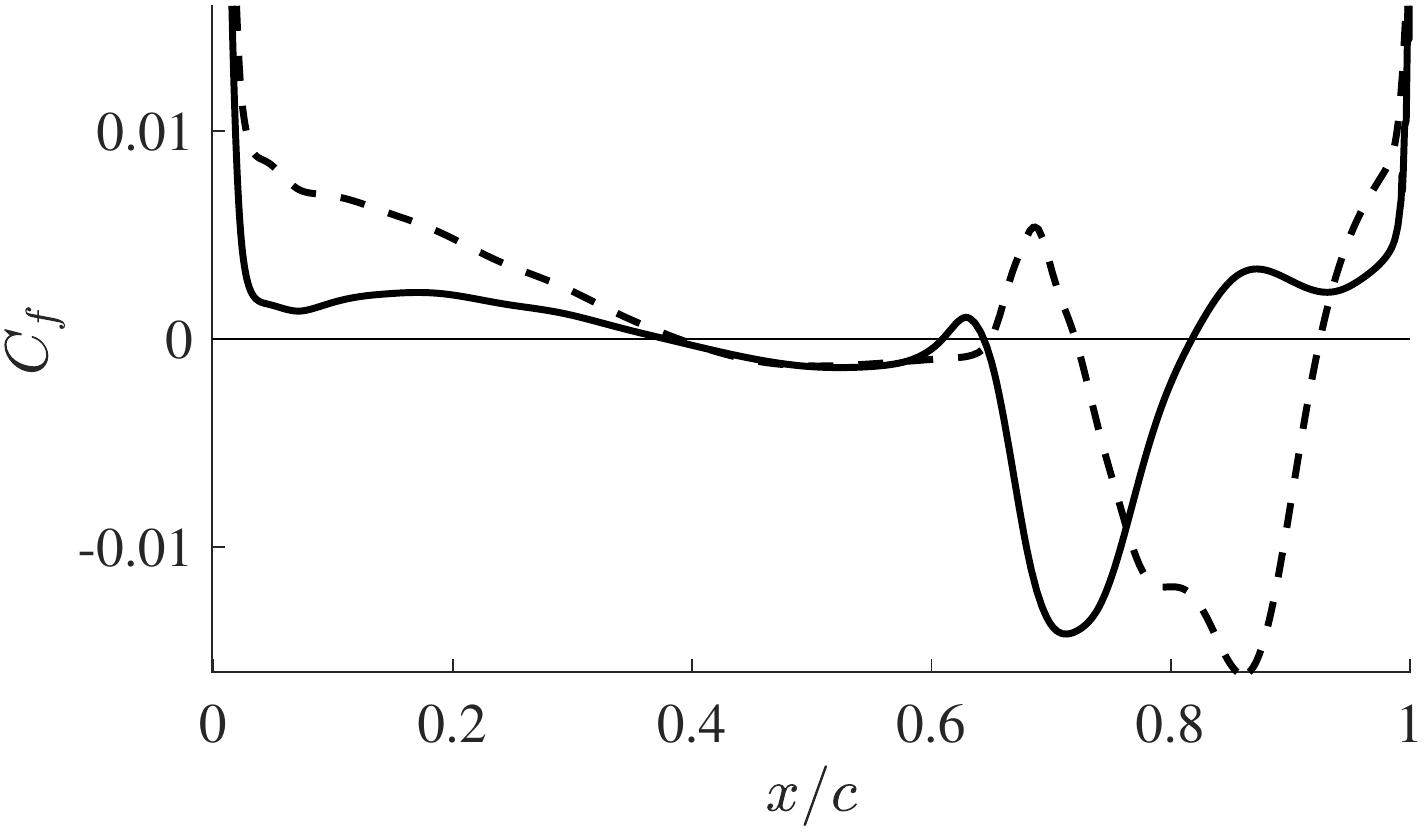}
    }
    \makebox[\textwidth][c]{
    \makebox[0.45\textwidth][c]{(a) Pressure coefficient}
    \hfill
    \makebox[0.45\textwidth][c]{(b) Skin friction coefficient}
    }
	\caption{Time-averaged pressure and skin friction coefficients for $R$\,=\,3.5$c$ and $R$\,=\,30$c$ at $\alpha$\,=\,7$^\circ$, $M$\,=\,0.3.}
	\label{fig:cpcf_domain_7deg}
\end{figure}

Because the magnitude of the pressure and friction forces increases with the flow angle, the influence of the free-stream boundaries also becomes more distinct. 
For 7$^\circ$ incidence, the separated boundary layer reattaches at the rear of the airfoil and forms a local LSB.
Streamlines of the time-averaged solution within the separation bubble are plotted in figure \ref{fig:lsb_domain_7deg} for domain sizes of $R$\,=\,3.5$c$ and $R$\,=\,30$c$. 
The LSB is significantly larger in the smaller domain where the free-stream boundaries impact the solution stronger by forcing the flow.
The difference in LSB sizes is distinctly visible in the time-averaged profiles of the surface pressure and skin friction coefficients, where both, $C_p$ and $C_f$, show the shift of the reattachment point of the LSB (see figure \ref{fig:cpcf_domain_7deg}).
Despite these significant differences in the flow topology, the time-averaged lift coefficient deviates only by 1.5\%, while the drag force differs by more than 40\% (see table \ref{tab:clcd_domain_7deg}). 
Note that the magnitude of the drag is only about 5\% of the lift force and hence is more susceptible to such changes.

\begin{table}
    \begin{center}
    \begin{tabular}{lcccccc}
        Domain radius   & $C_l$     & $C_{l,p}$     & $C_{l,f}$     & $C_d$    & $C_{d,p}$    & $C_{d,f}$ \\
        \hline
        3.5$c$          & 0.931     & 0.928         & 0.002         & 0.074    & 0.062        & 0.012 \\
        30$c$           & 0.946     & 0.944         & 0.002         & 0.052    & 0.041        & 0.011 \\
    \end{tabular}
    \caption{Lift and drag coefficients for $\alpha$\,=\,7$^\circ$ and different domain sizes.}
    \label{tab:clcd_domain_7deg}
    \end{center}
\end{table}

\begin{figure}
    \centerline{\includegraphics[width=0.65\textwidth]{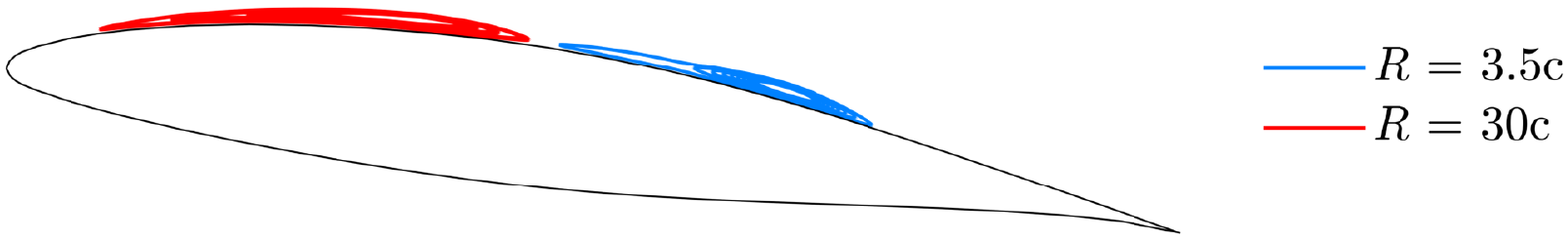}}
	\caption{Laminar separation bubble for domain radii $R$\,=\,3.5$c$ and $R$\,=\,30$c$ at $\alpha$\,=\,8$^\circ$, $M$\,=\,0.3.}
	\label{fig:lsb_domain_8deg}
\end{figure}

\begin{figure}
	\makebox[\textwidth][c]{
    \includegraphics[width=0.45\textwidth]{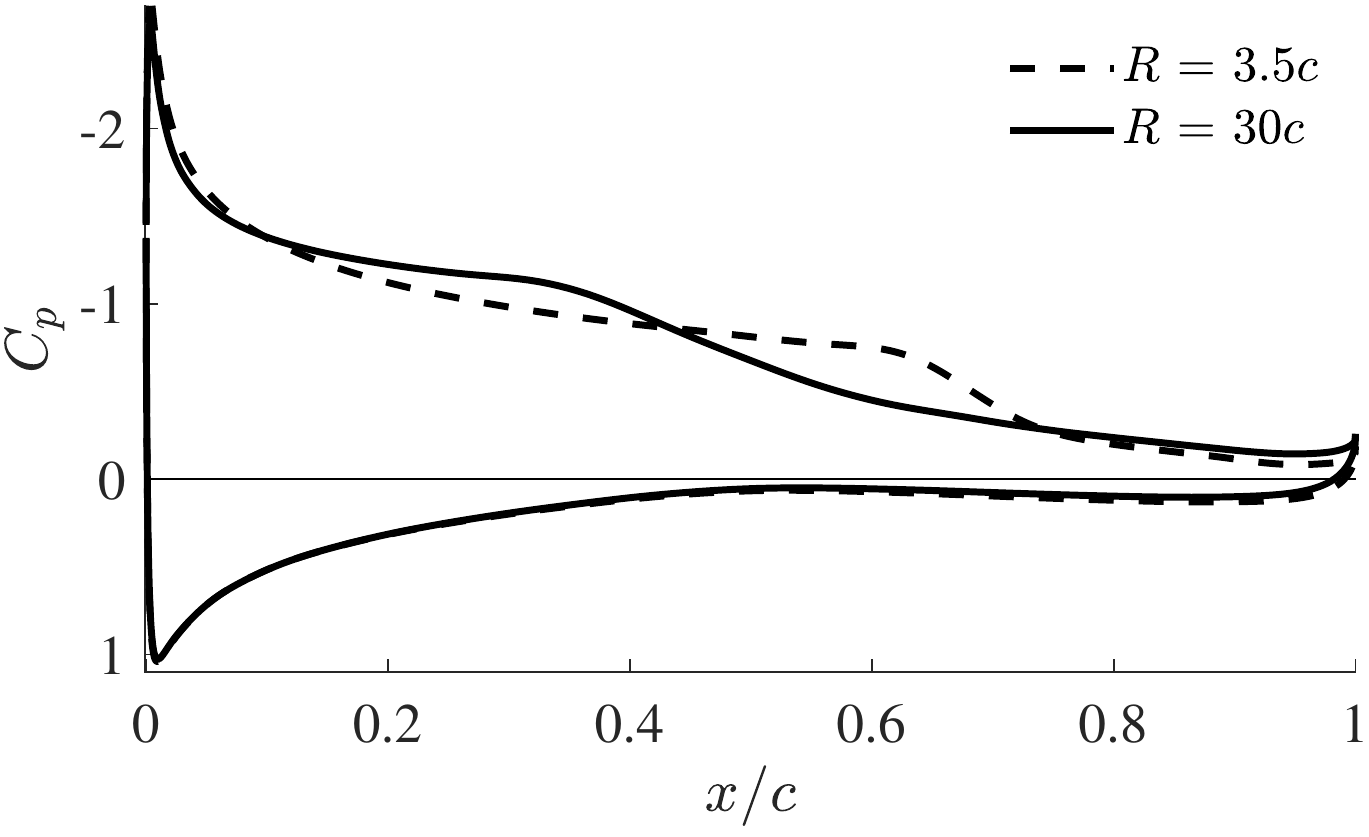}
    \hfill
    \includegraphics[width=0.45\textwidth]{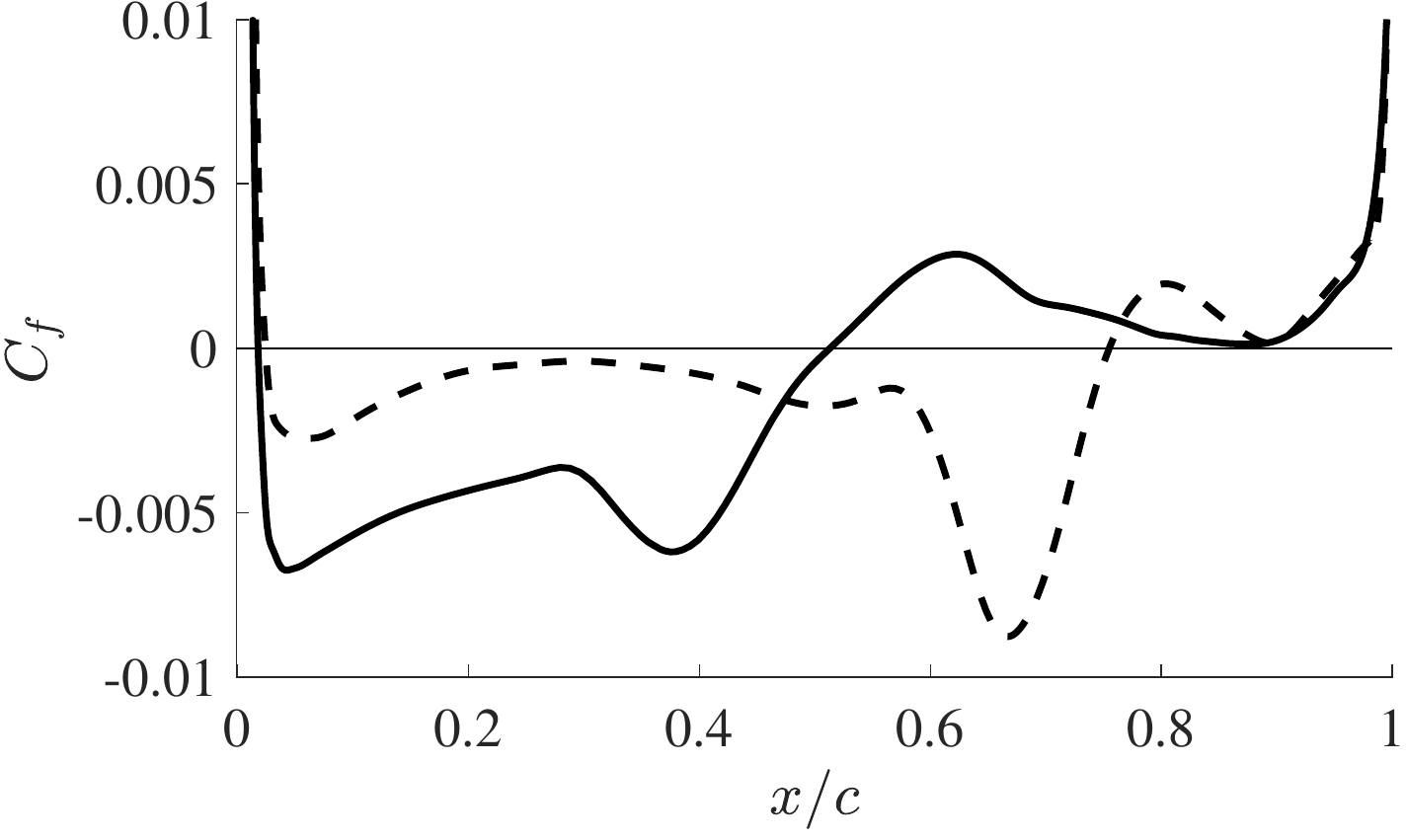}
    }
    \makebox[\textwidth][c]{
    \makebox[0.45\textwidth][c]{(a) Pressure coefficient}
    \hfill
    \makebox[0.45\textwidth][c]{(b) Skin friction coefficient}
    }
	\caption{Time-averaged pressure and skin friction coefficients for $R$\,=\,3.5$c$ and $R$\,=\,30$c$ at $\alpha$\,=\,8$^\circ$, $M$\,=\,0.3.}
	\label{fig:cpcf_domain_8deg}
\end{figure}

The effect of free-stream boundaries on the flow topology is even more pronounced at 8$^\circ$ incidence, where the location of the LSB completely shifts between the front and the rear side of the airfoil (see figure \ref{fig:lsb_domain_8deg}).
This, again, is reflected in the surface pressure and the skin friction coefficient (see figure \ref{fig:cpcf_domain_8deg}), but curiously does not translate into a significant change in the integrated lift or the drag force, as summarized in table \ref{tab:clcd_domain_8deg}.
The reason is that the bubble height is small and hence only slightly changes the pressure distribution, which remains approximately constant throughout separated flow regions. 
Given that both lift and drag coefficients are dominated by the pressure force (see table \ref{tab:clcd_domain_8deg}), the location of the LSB has only a limited affect on the lift as long as the bubble remains slender.

\begin{table}
    \begin{center}
    \begin{tabular}{lcccccc}
        Domain radius   & $C_l$     & $C_{l,p}$     & $C_{l,f}$     & $C_d$    & $C_{d,p}$    & $C_{d,f}$ \\
        \hline
        3.5$c$          & 0.989     & 0.987         & 0.002         & 0.064    & 0.053        & 0.011 \\
        30$c$           & 0.962     & 0.961         & 0.002         & 0.058    & 0.047        & 0.010 \\
    \end{tabular}
    \caption{Lift and drag coefficients for $\alpha$\,=\,8$^\circ$ and different domain sizes.}
    \label{tab:clcd_domain_8deg}
    \end{center}
\end{table}

The parametric study of two-dimensional Navier-Stokes simulations show that although the LSB location can be notably affected by changes in domain size, resolution, and Mach number, the results do not indicate that any of the tested parameters move the critical angle of attack to higher values.
Particularly, the good agreement between DGSEM and \textit{FLUENT} simulations confirm that the flow at 8$^\circ$ incidence has converged to a reasonable level across different numerical solvers.
We therefore consider the high-order results presented in this paper to be high-fidelity DNS.

%% file: main.bbl
\def\cprime{$'$}
\begin{thebibliography}{72}
\providecommand{\natexlab}[1]{#1}
\providecommand{\url}[1]{\texttt{#1}}
\expandafter\ifx\csname urlstyle\endcsname\relax
  \providecommand{\doi}[1]{doi: #1}\else
  \providecommand{\doi}{doi: \begingroup \urlstyle{rm}\Url}\fi

\bibitem[Anderson(2010)]{anderson:10}
J.~D. Anderson.
\newblock \emph{Fundamentals of aerodynamics}.
\newblock Tata McGraw-Hill Education, 2010.

\bibitem[Destarac and {van der Vooren}(2004)]{destarac:04}
D.~Destarac and J.~{van der Vooren}.
\newblock Drag/thrust analysis of jet-propelled transonic transport aircraft;
  definition of physical drag components.
\newblock \emph{Aerospace Science and Technology}, 8\penalty0 (6):\penalty0
  545--556, 2004.
\newblock ISSN 1270-9638.
\newblock \doi{https://doi.org/10.1016/j.ast.2004.03.004}.

\bibitem[Lissaman(1983)]{Lissaman83}
P.~B.~S. Lissaman.
\newblock Low-reynolds-number airfoils.
\newblock \emph{Annual Review of Fluid Mechanics}, 15\penalty0 (1):\penalty0
  223--239, 1983.
\newblock \doi{10.1146/annurev.fl.15.010183.001255}.

\bibitem[Horton(1968)]{Horton68}
H.~P. Horton.
\newblock \emph{Laminar separation in two and three-dimensional incompressible
  flow}.
\newblock Phd dissertation, Univeristy of London, 1968.

\bibitem[Stewartson(1970)]{stewartson:70}
K.~Stewartson.
\newblock Is the singularity at separation removable?
\newblock \emph{J. Fluid Mech.}, 44:\penalty0 347--364, 1970.

\bibitem[Smith(1979)]{smith:79}
F.~T. Smith.
\newblock Laminar flow of an incompressible fluid past a bluff body, the
  separation, reattachment, eddy properties and drag.
\newblock \emph{J. Fluid Mech.}, 92:\penalty0 171--205, 1979.

\bibitem[Alam and Sandham(2000)]{AS00}
M.~Alam and N.~D. Sandham.
\newblock Direct numerical simulation of ‘short’ laminar separation bubbles
  with turbulent reattachment.
\newblock \emph{Journal of Fluid Mechanics}, 410:\penalty0 1--28, 2000.
\newblock \doi{10.1017/S0022112099008976}.

\bibitem[Scheichel et~al.(2008)Scheichel, Braun, and Kluwick]{scheichl:08}
S.~Scheichel, S.~Braun, and A.~Kluwick.
\newblock On a similarity solution in the theory of unsteady marginal
  separation.
\newblock \emph{Acta Mech.}, 201:\penalty0 153--170, 2008.

\bibitem[Jones et~al.(2008)Jones, Sandberg, and Sandham]{JSS08}
L.~E. Jones, R.~D. Sandberg, and N.~D. Sandham.
\newblock Direct numerical simulations of forced and unforced separation
  bubbles on an airfoil at incidence.
\newblock \emph{Journal of Fluid Mechanics}, 602:\penalty0 175--207, 2008.

\bibitem[Burgmann et~al.(2008)Burgmann, Dannemann, and
  Schr\"{o}der]{burgmann:08a}
S.~Burgmann, J.~Dannemann, and W.~Schr\"{o}der.
\newblock Time-resolved and volumetric {PIV} measurements of a transitional
  separation bubble on an {SD}7003 airfoil.
\newblock \emph{Exp. Fluids}, 44:\penalty0 609--622, 2008.

\bibitem[Burgmann and Schr\"{o}der(2008)]{burgmann:08b}
S.~Burgmann and W.~Schr\"{o}der.
\newblock Investigation of the vortex induced unsteadiness of a separation
  bubble via time-resolved and scanning {PIV} measurements.
\newblock \emph{Exp. Fluids}, 45:\penalty0 675--691, 2008.

\bibitem[Wei and Smith(1986)]{WS86}
T.~Wei and C.~R. Smith.
\newblock Secondary vortices in the wake of circular cylinders.
\newblock \emph{Journal of Fluid Mechanics}, 169:\penalty0 513--533, 1986.
\newblock \doi{10.1017/S0022112086000733}.

\bibitem[Williamson(1996{\natexlab{a}})]{Williamson96}
C.~Williamson.
\newblock Vortex dynamics in the cylinder wake.
\newblock \emph{Annual Review of Fluid Mechanics}, 28:\penalty0 477--539,
  1996{\natexlab{a}}.

\bibitem[Williamson(1996{\natexlab{b}})]{Williamson96a}
C.~H.~K. Williamson.
\newblock Three-dimensional wake transition.
\newblock \emph{Journal of Fluid Mechanics}, 328:\penalty0 345--407,
  1996{\natexlab{b}}.
\newblock \doi{10.1017/S0022112096008750}.

\bibitem[Kerswell(2002)]{Kerswell02}
R.~R. Kerswell.
\newblock Elliptical instability.
\newblock \emph{Annual Review of Fluid Mechanics}, 34\penalty0 (1):\penalty0
  83--113, 2002.
\newblock \doi{10.1146/annurev.fluid.34.081701.171829}.

\bibitem[Crow(1970)]{Crow70}
S.~C. Crow.
\newblock Stability theory for a pair of trailing vortices.
\newblock \emph{AIAA Journal}, 8\penalty0 (12):\penalty0 2172--2179, 1970.
\newblock \doi{10.2514/3.6083}.

\bibitem[Leweke et~al.(2016)Leweke, Le~Dizès, and Williamson]{LLW16}
T.~Leweke, S.~Le~Dizès, and C.~H.~K. Williamson.
\newblock Dynamics and instabilities of vortex pairs.
\newblock \emph{Annual Review of Fluid Mechanics}, 48\penalty0 (1):\penalty0
  507--541, 2016.
\newblock \doi{10.1146/annurev-fluid-122414-034558}.

\bibitem[Marxen et~al.(2013)Marxen, Lang, and Rist]{MLR13}
Olaf Marxen, Matthias Lang, and Ulrich Rist.
\newblock Vortex formation and vortex breakup in a laminar separation bubble.
\newblock \emph{Journal of Fluid Mechanics}, 728:\penalty0 58--90, 2013.
\newblock \doi{10.1017/jfm.2013.222}.

\bibitem[Theofilis(2011)]{Theofilis11}
V.~Theofilis.
\newblock Global linear instability.
\newblock \emph{Annual Review of Fluid Mechanics}, 43:\penalty0 319--352, 2011.

\bibitem[Ducoin et~al.(2016)Ducoin, Loiseau, and Robinet]{DLR16}
A.~Ducoin, J.-Ch. Loiseau, and J.-Ch. Robinet.
\newblock Numerical investigation of the interaction between laminar to
  turbulent transition and the wake of an airfoil.
\newblock \emph{European Journal of Mechanics - B/Fluids}, 57:\penalty0 231 --
  248, 2016.
\newblock ISSN 0997-7546.
\newblock \doi{https://doi.org/10.1016/j.euromechflu.2016.01.005}.
\newblock URL
  \url{http://www.sciencedirect.com/science/article/pii/S0997754615302685}.

\bibitem[Jones et~al.(2010)Jones, Sandberg, and Sandham]{JSS10}
L.~E. Jones, R.~D. Sandberg, and N.~D. Sandham.
\newblock Stability and receptivity characteristics of a laminar separation
  bubble on an aerofoil.
\newblock \emph{Journal of Fluid Mechanics}, 648:\penalty0 257--296, 2010.

\bibitem[Glezer and Amitay(2002)]{glezer02}
Ari Glezer and Michael Amitay.
\newblock Synthetic jets.
\newblock \emph{Annual Review of Fluid Mechanics}, 34\penalty0 (1):\penalty0
  503--529, 2002.
\newblock \doi{10.1146/annurev.fluid.34.090501.094913}.

\bibitem[Suzuki et~al.(2004)Suzuki, Colonius, and Pirozzoli]{suzuki04}
Takao Suzuki, Tim Colonius, and Sergio Pirozzoli.
\newblock {Vortex shedding in a two-dimensional diffuser: theory and simulation
  of separation control by periodic mass injection}.
\newblock \emph{Journal of Fluid Mechanics}, 520:\penalty0 187--213, December
  10 2004.

\bibitem[Visbal(2011)]{visbal11}
Miguel~R. Visbal.
\newblock Numerical investigation of deep dynamic stall of a plunging airfoil.
\newblock \emph{AIAA Journal}, 49\penalty0 (10):\penalty0 2152--2170, 2011.

\bibitem[Bhattacharjee et~al.(2020)Bhattacharjee, Klose, Jacobs, and
  Hemati]{BKJH20}
Debraj Bhattacharjee, Bjoern Klose, Gustaaf~B. Jacobs, and Maziar~S. Hemati.
\newblock Data-driven selection of actuators for optimal control of airfoil
  separation.
\newblock \emph{Theoretical and Computational Fluid Dynamics}, 2020.

\bibitem[Yang and Spedding(2013)]{yang:13a}
S.L. Yang and G.R. Spedding.
\newblock Separation control by external acoustic excitation on a finite wing
  at low reynolds numbers.
\newblock \emph{AIAA J.}, 51:\penalty0 1506 -- 1515, 2013.
\newblock \doi{10.2514/1.J052191}.

\bibitem[Yang and Spedding(2014)]{yang:14}
S.L. Yang and G.R. Spedding.
\newblock Local acoustic forcing of a wing at low reynolds numbers.
\newblock \emph{AIAA J.}, 52:\penalty0 2867 -- 2876, 2014.
\newblock \doi{10.2514/1.J052984}.

\bibitem[Cattafesta and Sheplak(2011)]{CS11}
L.~N. Cattafesta and M.~Sheplak.
\newblock Actuators for active flow control.
\newblock \emph{Annual Review of Fluid Mechanics}, 43\penalty0 (1):\penalty0
  247--272, 2011.
\newblock \doi{10.1146/annurev-fluid-122109-160634}.

\bibitem[Selig et~al.(1995)Selig, Guglielmo, Broeren, and Giguere]{selig:95}
M.~S. Selig, J.~J. Guglielmo, A.~P. Broeren, and P.~Giguere.
\newblock \emph{Summary of Low-Speed Airfoil Data}, volume~1.
\newblock SoarTech Publications, Virginia Beach, Virginia, 1995.

\bibitem[Tank et~al.(2017)Tank, Smith, and Spedding]{TSS17}
J.~Tank, L.~Smith, and G.~R. Spedding.
\newblock On the possibility (or lack thereof) of agreement between experiment
  and computation of flows over wings at moderate reynolds number.
\newblock \emph{Interface Focus}, 7\penalty0 (1):\penalty0 20160076, 2017.

\bibitem[Selig et~al.(1989)Selig, Donovan, and Fraser]{SD89}
M.~S. Selig, S.~F. Donovan, and D.~B. Fraser.
\newblock \emph{Airfoils at Low Speeds}.
\newblock H. A. Stokely, Virginia Beach, VA, 1989.

\bibitem[Shan et~al.(2005)Shan, Jiang, and Chaoqun]{SJL05}
H.~Shan, L.~Jiang, and L.~Chaoqun.
\newblock Direct simulation of flow separation around a {NACA} 0012 airfoil.
\newblock \emph{Computers \& Fluids}, 34:\penalty0 1096--1114, 2005.

\bibitem[Jones and Sandberg(2011)]{JS11}
L.~E. Jones and R.~D. Sandberg.
\newblock Numerical analysis of tonal airfoil self-noise and acoustic
  feedback-loops.
\newblock \emph{Journal of Sound and Vibration}, 330\penalty0 (25):\penalty0
  6137 -- 6152, 2011.
\newblock \doi{https://doi.org/10.1016/j.jsv.2011.07.009}.

\bibitem[Almutairi et~al.(2010)Almutairi, Jones, and Sandham]{AJS10}
J.~H. Almutairi, L.~E. Jones, and N.~D. Sandham.
\newblock Intermittent bursting of a laminar separation bubble on an airfoil.
\newblock \emph{AIAA Journal}, 48\penalty0 (2):\penalty0 414--426, 2010.

\bibitem[Lee et~al.(2015)Lee, Nonomura, Oyama, and Fujii]{LNOF15}
D.~Lee, T.~Nonomura, A.~Oyama, and K.~Fujii.
\newblock Comparison of numerical methods evaluating airfoil aerodynamic
  characteristics at low reynolds number.
\newblock \emph{Journal of Aircraft}, 52\penalty0 (1):\penalty0 296--306, 2015.
\newblock \doi{10.2514/1.C032721}.

\bibitem[Cadieux and Domaradzki(2016)]{cadieux:16}
F.~Cadieux and J.~A. Domaradzki.
\newblock Periodic filtering as a subgrid-scale model for les of laminar
  separation bubble flows.
\newblock \emph{J. Turbulence}, 2016.
\newblock \doi{10.1080/14685248.2016.1208825}.

\bibitem[Galbraith and Visbal(2010)]{GV10}
M.~Galbraith and M.~Visbal.
\newblock \emph{Implicit Large Eddy Simulation of Low-{R}eynolds-Number
  Transitional Flow Past the {SD7003} Airfoil}.
\newblock 2010.
\newblock \doi{10.2514/6.2010-4737}.

\bibitem[Uranga et~al.(2011)Uranga, Persson, Drela, and Peraire]{uranga11}
A.~Uranga, P.-O. Persson, M.~Drela, and J.~Peraire.
\newblock Implicit large eddy simulation of transition to turbulence at low
  reynolds numbers using a discontinuous galerkin method.
\newblock \emph{International Journal for Numerical Methods in Engineering},
  87:\penalty0 232--261, 2011.

\bibitem[Beck et~al.(2014)Beck, Bolemann, Flad, Frank, Gassner, Hindenlang, and
  Munz]{BBFFGHM14}
A.~D. Beck, T.~Bolemann, D.~Flad, H.~Frank, G.~J. Gassner, F.~Hindenlang, and
  C.-D. Munz.
\newblock High-order discontinuous {G}alerkin spectral element methods for
  transitional and turbulent flow simulations.
\newblock \emph{International Journal for Numerical Methods in Fluids},
  76\penalty0 (8):\penalty0 522--548, 2014.

\bibitem[Drela(1989)]{xfoil}
M.~Drela.
\newblock {XFOIL}: An analysis and design system for low {R}eynolds number
  airfoils.
\newblock In T.~J. Mueller, editor, \emph{Low {R}eynolds Number Aerodynamics},
  pages 1--12, Berlin, Heidelberg, 1989. Springer Berlin Heidelberg.
\newblock ISBN 978-3-642-84010-4.

\bibitem[Durbin(2018)]{Durbin18}
P.~A. Durbin.
\newblock Some recent developments in turbulence closure modeling.
\newblock \emph{Annual Review of Fluid Mechanics}, 50\penalty0 (1):\penalty0
  77--103, 2018.
\newblock \doi{10.1146/annurev-fluid-122316-045020}.

\bibitem[Rogallo and Moin(1984)]{RM84}
R.~S. Rogallo and P.~Moin.
\newblock Numerical simulation of turbulent flow.
\newblock \emph{Ann. Rev. Fluid Mech.}, 16:\penalty0 99--137, 1984.

\bibitem[Sengupta et~al.(2007)Sengupta, Mashayek, and
  Jacobs]{sengupta2007large}
K.~Sengupta, F.~Mashayek, and G.~Jacobs.
\newblock Large-eddy simulation using a discontinuous {G}alerkin spectral
  element method.
\newblock In \emph{45th AIAA Aerospace Sciences Meeting and Exhibit}, page 402,
  2007.

\bibitem[Spalart(2009)]{Spalart09}
P.~R. Spalart.
\newblock Detached-eddy simulation.
\newblock \emph{Annual Review of Fluid Mechanics}, 41\penalty0 (1):\penalty0
  181--202, 2009.
\newblock \doi{10.1146/annurev.fluid.010908.165130}.

\bibitem[Herrig et~al.(1951)Herrig, Emery, and Erwin]{HEE51}
L.~Joseph Herrig, James~C. Emery, and John~R. Erwin.
\newblock Systematic two-dimensional cascade tests of naca 65-series compressor
  blades at low speeds.
\newblock Naca technical note 3916, 1951.

\bibitem[Wright(1974)]{Wright74}
L.~C. Wright.
\newblock Blade selection for a modern axial-flow compressor.
\newblock Nasa conference proceedings, 1974.

\bibitem[Haller(2004)]{Haller04}
G~Haller.
\newblock Exact theory of unsteady separation for two-dimensional flows.
\newblock \emph{J. Fluid Mech.}, 512:\penalty0 357--311, 2004.

\bibitem[Nelson(2015)]{nelson}
Daniel~Alan Nelson.
\newblock \emph{High-fidelty {L}agrangian coherent structures analysis and
  {DNS} with discontinuous-Galerkin methods}.
\newblock {PhD Thesis}, University of California, San Diego in conjuction with
  San Diego State University, San Diego, {CA}, Month unknown 2015.
\newblock http://escholarship.org/uc/item/2cv4f732.

\bibitem[Kopriva(2009)]{kopriva}
David~A. Kopriva.
\newblock \emph{Implementing Spectral Methods for Partial Differential
  Equations}.
\newblock Springer, New York, 2009.

\bibitem[Klose et~al.(2020)Klose, Jacobs, and Kopriva]{KJK20}
Bjoern~F. Klose, Gustaaf~B. Jacobs, and David~A. Kopriva.
\newblock Assessing standard and kinetic energy conserving volume fluxes in
  discontinuous galerkin formulations for marginally resolved navier-stokes
  flows.
\newblock \emph{Computers \& Fluids}, 205:\penalty0 104557, 2020.
\newblock ISSN 0045-7930.
\newblock \doi{https://doi.org/10.1016/j.compfluid.2020.104557}.

\bibitem[Jacobs et~al.(2003)Jacobs, Kopriva, and Mashayek]{JKM03}
G.~B. Jacobs, D.~A. Kopriva, and F.~Mashayek.
\newblock A comparison of outflow boundary conditions for the multidomain
  staggered-grid spectral method.
\newblock \emph{Numerical Heat Transfer, Part B: Fundamentals}, 44:\penalty0
  225--251, 2003.

\bibitem[Nelson et~al.(2016)Nelson, Jacobs, and Kopriva]{nelson16}
D.~A. Nelson, G.~B. Jacobs, and D.~A. Kopriva.
\newblock {Effect of Boundary Representation on Viscous, Separated Flows in a
  Discontinuous-Galerkin Navier-Stokes Solver}.
\newblock \emph{Theoretical Computational Fluid Dynamics}, 30:\penalty0
  363--385, March 30 2016.

\bibitem[Deng et~al.(2007)Deng, Jiang, and Liu]{DJL07}
Shutian Deng, Li~Jiang, and Chaoqun Liu.
\newblock {DNS} for flow separation control around an airfoil by pulsed jets.
\newblock \emph{Computers \& Fluids}, 36\penalty0 (6):\penalty0 1040--1060,
  July 2007.

\bibitem[Zhang et~al.(2015)Zhang, Cheng, Gao, Qamar, and Samtaney]{ZCGQS15}
Wei Zhang, Wan Cheng, Wei Gao, Adnan Qamar, and Ravi Samtaney.
\newblock Geometrical effects on the airfoil flow separation and transition.
\newblock \emph{Computers \& Fluids}, 116:\penalty0 60 -- 73, 2015.

\bibitem[Balakumar(2017)]{balakumar}
Ponnampalam Balakumar.
\newblock Direct numerical simulation of flows over an naca-0012 airfoil at low
  and moderate reynolds numbers.
\newblock \emph{AIAA Fluid Dynamics Conference}, 47, June 2017.

\bibitem[Serson et~al.(2017)Serson, Meneghini, and Sherwin]{SMS17}
Douglas Serson, Julio~R. Meneghini, and Spencer~J. Sherwin.
\newblock Direct numerical simulations of the flow around wings with spanwise
  waviness.
\newblock \emph{Journal of Fluid Mechanics}, 826:\penalty0 714–731, 2017.
\newblock \doi{10.1017/jfm.2017.475}.

\bibitem[Pirozzoli(2011)]{pirozzoli}
Sergio Pirozzoli.
\newblock {Numerical Methods for High-Speed Flows}.
\newblock \emph{Annual Review of Fluid Mechanics}, 43:\penalty0 163--194, 2011.

\bibitem[Georgiadis et~al.(2010)Georgiadis, Rizzetta, and Fureby]{GRF10}
Nicholas~J. Georgiadis, Donald~P. Rizzetta, and Christer Fureby.
\newblock Large-eddy simulation: Current capabilities, recommended practices,
  and future research.
\newblock \emph{AIAA Journal}, 48\penalty0 (8), August 2010.

\bibitem[Robinson(1991)]{Robinson91}
S~K Robinson.
\newblock Coherent motions in the turbulent boundary layer.
\newblock \emph{Annual Review of Fluid Mechanics}, 23\penalty0 (1):\penalty0
  601--639, 1991.
\newblock \doi{10.1146/annurev.fl.23.010191.003125}.

\bibitem[Jeong and Hussain(1995)]{JH95}
J.~Jeong and F.~Hussain.
\newblock On the identification of a vortex.
\newblock \emph{Journal of Fluid Mechanics}, 285:\penalty0 69–94, 1995.

\bibitem[Lasheras and Choi(1988)]{LC88}
J.~C. Lasheras and H.~Choi.
\newblock Three-dimensional instability of plane free shear layer: {A}n
  experimental study of the formation and evolution of streamwise vortices.
\newblock \emph{Journal of Fluid Mechanics}, 189:\penalty0 51--86, 1988.

\bibitem[Chatelain et~al.(2008)Chatelain, Curioni, Bergdorf, Rossinelli,
  Andreoni, and Koumoutsakos]{CCBRAK08}
P.~Chatelain, A.~Curioni, M.~Bergdorf, D.~Rossinelli, W.~Andreoni, and
  P.~Koumoutsakos.
\newblock Billion vortex particle direct numerical simulations of aircraft
  wakes.
\newblock \emph{Computer Methods in Applied Mechanics and Engineering},
  197\penalty0 (13):\penalty0 1296--1304, 2008.
\newblock ISSN 0045-7825.
\newblock \doi{https://doi.org/10.1016/j.cma.2007.11.016}.

\bibitem[Leweke and Williamson(1998{\natexlab{a}})]{LW98}
T.~Leweke and C.~H.~K. Williamson.
\newblock Cooperative elliptic instability of a vortex pair.
\newblock \emph{Journal of Fluid Mechanics}, 360:\penalty0 85--119,
  1998{\natexlab{a}}.
\newblock \doi{10.1017/S0022112097008331}.

\bibitem[Leweke and Williamson(1998{\natexlab{b}})]{LW98a}
T.~Leweke and C.~H.~K. Williamson.
\newblock Three-dimensional instabilities in wake transition.
\newblock \emph{European Journal of Mechanics - B/Fluids}, 17\penalty0
  (4):\penalty0 571 -- 586, 1998{\natexlab{b}}.
\newblock ISSN 0997-7546.
\newblock \doi{https://doi.org/10.1016/S0997-7546(98)80012-5}.
\newblock Special Issue Dynamics and Statistics of Concentrated Vortices in
  Turbulent Flow (Euromech Colloquium 364).

\bibitem[Sakai et~al.(2020)Sakai, Diamessis, and Jacobs]{SJD20}
T.~Sakai, P.~J. Diamessis, and G.~B. Jacobs.
\newblock Self-sustained instability, transition, and turbulence induced by a
  long separation bubble in the footprint of an internal solitary wave. {I}.
  {F}low topology.
\newblock \emph{Physical Review Fluids}, 2020.
\newblock accepted for publication.

\bibitem[Spurk and Aksel(2008)]{SA08}
J.~Spurk and N.~Aksel.
\newblock \emph{Fluid Mechanics}.
\newblock Springer-Verlag, Berlin Heidelberg, 2008.

\bibitem[Chaudhuri et~al.(2017)Chaudhuri, Jacobs, Don, Abbassi, and
  Mashayek]{CJDAM16}
A.~Chaudhuri, G.B. Jacobs, W.S. Don, H.~Abbassi, and F.~Mashayek.
\newblock Explicit discontinuous spectral element method with entropy
  generation based artificial viscosity for shocked viscous flows.
\newblock \emph{J. Comp. Phys.}, 32:\penalty0 99--117, 2017.

\bibitem[Tank et~al.(2019)Tank, Klose, Jacobs, and Spedding]{TKJS19a}
J.~Tank, B.~F. Klose, G.~Jacobs, and G.~R. Spedding.
\newblock Computer and laboratory studies on the aerodynamics of the naca
  65(1)-412 at reynolds number 20 000.
\newblock {AIAA} scitech 2019 forum, 2019.

\bibitem[Choi(2020)]{Choi20}
D.~A. Choi.
\newblock {Wind tunnel experiments on the flow over a NACA 65(1)-412 airfoil at
  a Reynolds number of 20,000}.
\newblock Master thesis, San Diego State University, San Diego, CA, Fall 2020.

\bibitem[Fluent.Inc.(2018)]{Fluent19}
Fluent.Inc.
\newblock Fluent 19.2 user manual.
\newblock \emph{Fluent.Inc.}, 2018.

\bibitem[Pelletier and Mueller(2001)]{PMJA01}
Alain Pelletier and Thomas~J. Mueller.
\newblock Effect of endplates on two-dimensional airfoil testing at low
  reynolds number.
\newblock \emph{Journal of Aircraft}, 38\penalty0 (6):\penalty0 1056--1059,
  2001.

\bibitem[Klose et~al.(2021)Klose, Spedding, and Jacobs]{KSJ21}
B.~F. Klose, G.~R. Spedding, and G.~B. Jacobs.
\newblock \emph{What is the effect of self-induced pressure waves and their
  wall reflections on low Reynolds number airfoil flow in wind tunnels?}
\newblock 2021.
\newblock \doi{10.2514/6.2021-1195}.

\end{thebibliography}
